\documentclass[a4paper,11pt]{article}

\usepackage{imakeidx}
\makeindex[columns=2, title=Alphabetical Index, intoc]
\usepackage{jheppub} % for details on the use of the package, please see the JINST-author-manual
\usepackage{lineno}
\usepackage{amsmath}
\usepackage{amssymb}
\usepackage{nicefrac}
\usepackage[dvipsnames]{xcolor}
\usepackage{mathtools}
\usepackage{cleveref}
\usepackage{slashed}
\usepackage{multirow}
\usepackage{subfigure}
\usepackage{placeins}
\usepackage{stmaryrd}
\usepackage{capt-of}
\usepackage{tikz-feynman,contour}

\usepackage[export]{adjustbox}

\usepackage{rotating}

\newcommand{\ie}{i.e.~}

\newcommand{\wrt}{w.r.t.~}

\newcommand{\z}{{\boldsymbol \zeta}}

\newcommand{\as}{\alpha_s}

\newcommand{\mathd}{\mathrm{d}}
\newcommand{\ep}{\epsilon}
\newcommand{\vep}{\varepsilon}
\newcommand{\vepT}{\widetilde{\varepsilon}}

\newcommand{\Li}{\mathrm{Li}}

\newcommand{\Th}{\hat T}

\newcommand{\zbar}{\bar z}

\renewcommand{\Im}{{\rm Im}}
\renewcommand{\Re}{{\rm Re}}

\newcommand{\pp}{{\bf p}}
\newcommand{\qp}{{\bf q}}
\newcommand{\kp}{{\bf k}}
\newcommand{\zp}{{\bf z}}
\newcommand{\pP}[1]{{p_{#1,\perp}}}

\newcommand{\pPc}[1]{{\bar p_{#1,\perp}}}

\newcommand{\Wb}{\widetilde{W}}
\newcommand{\meas}[1]{[ \mathrm{d} #1 ] }
\newcommand{\measF}[1]{\{ \mathrm{d} #1 \} }
\newcommand{\measJ}[1]{ \mathfrak{D} #1 }
\newcommand{\LL}{\text{LL}}
\newcommand{\NLL}{\text{NLL}}
\newcommand{\NNLL}{{\text{NNLL}}}
\newcommand{\timeOrd}{ \mathbb{T} }
\newcommand{\wilsonV}{\mathcal{W}}
\newcommand{\wilsonU}{\mathcal{U}}

\newcommand\MSbar{\overline{\rm MS}}

\newcommand{\IFpre}{\mathcal{J}}

\newcommand{\IF}{\mathcal{I}}
\newcommand{\IFfin}{\hat{\mathcal{I}}}
\newcommand{\IFU}{\mathcal{I}}
\newcommand{\twoDint}{\mathcal{K}}
\newcommand{\taufin}{\hat{\tau}}
\newcommand{\tauU}{\tau}
\newcommand{\gcdot}{\!\cdot\!}
\newcommand{\tIep}{\tilde I^{(\ep)}}
\newcommand{\tDep}{\tilde \Delta^{(\ep)}}
\newcommand{\tBep}{\tilde B^{(\ep)}}

\keywords{QCD, scattering amplitudes, Regge limit, BFKL}

\makeatletter
\newcommand\ps@indexpagestyle{
  \renewcommand\@oddfoot{\hfill-- \thepage\ --\hfill}
  \renewcommand\@oddhead{}
}

\preprint{
\begin{flushright}
OUTP-24-06P  \\
SLAC-PUB-241120 \\
TUM-HEP-1537/24
\end{flushright}
}
\allowdisplaybreaks

\makeatother
\indexsetup{firstpagestyle=indexpagestyle}

\def\OX{Rudolf Peierls Centre for Theoretical Physics, University of Oxford, Clarendon Laboratory, Parks
Road, Oxford OX1 3PU}
\def\TUM{Physik Department,
Technische Universit\"at M\"unchen,
James-Franck-Straße 1, 
D–85748 Garching, Germany}
\def\WDM{Wadham College, University of Oxford, Parks Road, Oxford OX1 3PN, UK}

\def\SLAC{SLAC National Accelerator Laboratory, Stanford University, Stanford, CA 94039, USA}

\author[a]{Federico Buccioni,}
\author[b,c]{Fabrizio Caola,}
\author[b,d]{Federica Devoto,}
\author[b]{Giulio Gambuti}

\emailAdd{federico.buccioni@tum.de}
\emailAdd{fabrizio.caola@physics.ox.ac.uk}
\emailAdd{federica@slac.stanford.edu}
\emailAdd{giulio.gambuti@physics.ox.ac.uk}

\affiliation[a]{\TUM}
\affiliation[b]{\OX}
\affiliation[c]{\WDM}
\affiliation[d]{\SLAC}

\title{
  Investigating the universality of five-point QCD scattering amplitudes
  at high energy}
\abstract{
  We
  investigate $2\to3$ QCD scattering amplitudes in multi-Regge kinematics,
  i.e. where the final partons are strongly ordered in rapidity.
  In this regime amplitudes exhibit intriguing factorisation properties
  which can be understood in terms of
  effective degrees of freedom called \emph{reggeons}.
  Working within the Balitsky/JIMWLK framework, we predict these
  amplitudes for the first time to next-to-next-to-leading logarithmic
  order, and compare against the limit of QCD scattering amplitudes in
  full colour and kinematics. 
  We find that the latter can be described in terms of universal
  objects, and that the apparent non-universality arising at NNLL
  comes from well-defined and under-control contributions that we can
  predict.
  Thanks to this observation, we  extract for the first time
  the universal vertex that controls the emission of the
  central-rapidity gluon, both in QCD and $\mathcal N=4$ super Yang-Mills. 
}

\begin{document}

\maketitle
\raggedbottom
\allowdisplaybreaks

%%%%%% -------
%%%%%% -------
%%%%%% -------
%%%%%% -------
%%%%%% -------
%%%%%%%%%%%%%%%%%%%%%%%%%%%%%%%%%%%%%%%%%%%%%%%%%%%%%%%%%%%%%%%%%%%%%%%%%%%%%%%%%%%%%%%%%%%%%%%%%%%%%%%%%%%%%%%%%%%%%%%%%%%%%%%%%%%%%%%%%%%%%%%%%%%%%%%%%%%%%%%%%%%%%%%%%%%%%%%%%%%%%%%%%%
%%%%%% -------
%%%%%% -------
%%%%%% -------
%%%%%% -------
%%%%%% -------

\section{Introduction}
\label{se:introduction}

Scattering processes in the high-energy -- or \emph{Regge} -- limit
offer a rich laboratory to explore properties of gauge theories, both
at amplitude and cross-section level. In this regime, the invariant
mass of any pair of final-state particles grows with the scattering
energy $\sqrt{s}$ and it is much larger than their individual
transverse momenta that are instead held fixed.
This can equivalently be expressed by requiring that the final-state particles are strongly ordered in rapidity while having comparable transverse momenta. Such a configuration for $2\to n$ scattering is referred to as
multi-Regge kinematics (MRK).
In this scenario, scattering amplitudes feature interesting properties,
the most remarkable one being the phenomenon of \textit{reggeisation}:
they naturally organise in terms of $t$-channel exchanges of
effective degrees of freedom called $reggeons$, whose propagator is
dressed with a power-law behaviour $s^\tau$. In perturbative QCD, or more
generally in non-abelian gauge theories, the dominant contribution is
given by a reggeised gluon, whose power-law behaviour is controlled
by the gluon Regge trajectory $\tau_g$. 

Following the seminal work by Fadin, Kuraev and Lipatov on the
reggeisation of gauge bosons in non-abelian gauge theories with broken
symmetries~\cite{Fadin:1975cb,Lipatov:1976zz}, the investigation of
QCD as a massless theory came shortly
after~\cite{Kuraev:1976ge,Kuraev:1977fs,Balitsky:1978ic}.  These
studies led to the conception of the celebrated
Balitsky-Fadin-Kuraev-Lipatov (BFKL) formalism and its related
evolution equation. The latter allows for the resummation of large
terms of type $\ln (s/|t|)$ arising in the Regge limit at any order in
perturbative QCD, both at leading-logarithmic
(LL)~\cite{Fadin:1975cb,Lipatov:1976zz,Kuraev:1976ge,Kuraev:1977fs,Balitsky:1978ic},
\ie $\left[\alpha_s \ln(s/|t|)\right]^n$, and
next-to-leading-logarithmic (NLL)
accuracy~\cite{Fadin:1998py,Ciafaloni:1998gs,Kotikov:2000pm}, \ie
$\alpha_s\left[\alpha_s \ln(s/|t|)\right]^n$. Apart from its formal
interest, the BFKL formalism has a broad spectrum of applications,
ranging from the physics of small-$x$ parton distribution functions,
see
e.g. refs~\cite{Jaroszewicz:1982gr,Catani:1990eg,Altarelli:2001ji,Ciafaloni:2003ek},
to the phenomenology of processes with large rapidity gaps in hadronic
collisions, see
e.g. refs~\cite{Mueller:1986ey,DelDuca:1993mn,Stirling:1994he,Andersen:2001kta,Colferai:2010wu,Ducloue:2013hia,Caporale:2014gpa,Andersen:2011hs,Andersen:2023kuj,Golec_Biernat_2018,Celiberto_2021a,Celiberto_2021b}. Improving the BFKL approach beyond the current frontier to reach
next-to-next-to-leading-logarithmic (NNLL) accuracy,
\ie $\alpha_s^2\left[\alpha_s \ln(s/|t|)\right]^n$, would 
significantly enhance those studies and our understanding of QCD in
extreme regimes. 

The BFKL formalism relies on fundamental properties of scattering
amplitudes in the high-energy regime. At LL accuracy, amplitudes
are described to all-orders in perturbation theory as
a tree-level exchange of reggeised gluons
in the $t$-channel, whose power-law behaviour is fully determined
by the one-loop Regge trajectory.
In MRK, the
interaction between these reggeons and the gluons emitted centrally in
the large rapidity gap is described by an effective vertex known as
the Lipatov or central-emission
vertex (CEV)~\cite{Lipatov:1976zz}. Owing to single-reggeon exchanges
in the $t$-channels, scattering amplitudes have a simple pole in the
complex-angular momentum
plane~\cite{Collins:1977jy,Ioffe:2010zz}. Thus, at LL accuracy, the
iterated structure of reggeon propagators and CEVs goes under the name
of Regge-pole factorisation.
Starting at NLL, multi-reggeon exchanges appear and they give rise to
cuts in the complex-angular momentum plane. However, these contribute
solely to the absorptive (imaginary) part of the amplitude, whereas
the dispersive (real) part is still described as a single reggeon
exchange, thus showing a Regge-pole factorisation
behaviour~\cite{Fadin:1998py}. At this logarithmic order, the
factorised structure also entails the interaction between a single
reggeon and the particles that sit at the edges of the large rapidity
gap, defining the so-called impact factors. The latter are flavour-dependent, and in QCD, there are impact factors for both quarks and gluons. It is worth stressing that these are the only
process-dependent ingredients in Regge-pole factorisation. Once
they are accounted for, this factorisation reflects into a statement
about Regge-pole universality in (M)RK.

Several ingredients required for predictions in the BFKL formalism
beyond LL accuracy are known. Two-loop corrections to the gluon Regge
trajectory were computed long ago~\cite{Fadin:1996tb}, and more
recently three-loop ones in both $\mathcal{N}=4$ super
Yang-Mills (sYM)~\cite{Henn:2016jdu} and full
QCD~\cite{DelDuca:2021vjq,Caola:2021izf,Falcioni:2021dgr} became
available. One-loop QCD corrections to the quark and gluon impact
factors were computed in
refs.~\cite{Fadin:1992zt,Fadin:1993wh,Fadin:1993qb,Bern:1998sc,DelDuca:1998kx,Fadin:1999de}
and two-loop ones appeared in ref.~\cite{Caron-Huot:2017fxr}.  Finally, the
CEV is known with one-loop accuracy in
QCD~\cite{Fadin:1993wh,Fadin:1994fj,Fadin:1996yv,DelDuca:1998cx}, and
it was recently presented in dimensional regularisation up to second
order in the regulator~\cite{Fadin:2023roz}. The only missing
components required for computing the Regge-pole contribution to
scattering amplitudes at NNLL are the one-loop corrections to the
central two-gluon emission vertex, and the two-loop ones to the CEV.
While the former has been recently presented in $\mathcal N=4$
sYM~\cite{Byrne:2022wzk}, the latter is unknown.

Alongside the evaluation of these contributions, an outstanding issue
in QCD that prevents a robust generalisation of the BFKL framework
beyond NLL is the appearance of cuts in the complex angular momentum
plane.  These are understood as multi-reggeon exchanges. Starting from
NNLL, such cuts also appear in the dispersive part of the result,
making the identification of the Regge pole contribution
problematic.\footnote{This scenario is very different in planar
$\mathcal N=4$ sYM, leading to a much better understanding, see
e.g. section 6 in ref.~\cite{Caron-Huot:2013fea} and references
therein.}
High-energy factorisation breaking at $\NNLL$ in the real part of
two-loop $2\to2$ QCD scattering amplitudes was first reported in
ref.~\cite{DelDuca:2001gu}. The observation that
factorisation-violating terms are infrared (IR) divergent motivated
investigations into their contributions to the IR poles of scattering
amplitudes at two- and three-loop orders in
QCD~\cite{DelDuca:2013ara,DelDuca:2014cya}.
In the recent past, several approaches have appeared to address this
problem in a systematic way. Ref.~\cite{Caron-Huot:2013fea} developed
an effective theory based on the Balitsky/JIMWLK
formalism~\cite{Balitsky:1995ub,Jalilian-Marian:1997qno,Jalilian-Marian:1997jhx,Jalilian-Marian:1997ubg,Kovner:2000pt,Weigert:2000gi,Iancu:2000hn,Iancu:2001ad,Ferreiro:2001qy,Caron-Huot:2013fea}
that paved the way for many amplitude-level investigations, see
e.g.~\cite{Caron-Huot:2017fxr,Caron-Huot:2020grv,Falcioni:2020lvv,Falcioni:2021buo,Falcioni:2021dgr}.
Fadin and Lipatov studied instead the complete contribution of three-reggeon cuts to the $2\to 2$ scattering amplitude, using a diagrammatic
approach~\cite{Fadin:2017nka,Fadin:2019tdt,Fadin:2021csi,Fadin:2024eyf}.
More recently, a SCET-based formalism based on Glauber exchanges has
also been
developed~\cite{Rothstein:2016bsq,Rothstein:2024fpx,Moult:2022lfy,Gao:2024qsg}.

Thanks to impressive progress on the calculation of multi-loop
multi-leg scattering
amplitudes~\cite{Abreu:2018aqd,Caron-Huot:2020vlo,Caola:2021izf,Caola:2022dfa,Agarwal:2023suw,DeLaurentis:2023izi,DeLaurentis:2023nss},
we now have analytic data that allow us to $a)$ validate the
approaches described above, $b)$ gain direct insight into the
high-energy structure of perturbative QCD, $c)$ extract the universal
building blocks required to extend the BFKL programme beyond NLL
accuracy.  In this paper, we take an important step in this direction
by considering the high-energy limit of the $2\to 3$ QCD scattering
amplitudes in full colour and
kinematics~\cite{Agarwal:2023suw,DeLaurentis:2023izi,DeLaurentis:2023nss}
and comparing them against predictions that we obtained from the framework
of ref.~\cite{Caron-Huot:2013fea}. By matching against the
EFT~\cite{Caron-Huot:2013fea}, we both validate this approach at NNLL
in MRK at the two-loop level, and extract for the first time the universal
two-loop vertex that describes the emission of a central-rapidity gluon.
This can be seen as a crucial step towards a robust definition of the
Lipatov vertex at two loops and beyond. 

The remainder of this paper is organised as follows: in
\cref{se:generalaspects} we discuss the $2 \to 3$ MRK,
list the contributing partonic channels, describe
aspects related to signature and colour and illustrate how we expanded the
five-point QCD scattering  amplitudes~\cite{Agarwal:2023suw} in MRK.
Section \ref{se:shockwave} is devoted to a review of the
Balitsky/JIMWLK formalism and to a discussion of how it can be used to
predict the form of our scattering amplitudes in MRK. We do this
by extending the results of ref.~\cite{Caron-Huot:2013fea} and expressing the
two-loop scattering amplitude in terms of fully-predicted quantities
and one unknown two-loop universal vertex that describes the emission
of the central-rapidity gluon.
In \cref{se:results} we compare the results from
\cref{se:generalaspects,se:shockwave}. After finding full agreement
between the two for all the terms that we predict unambiguously at NNLL, we
leverage the knowledge of the full $2 \to 3$ QCD amplitude to extract
for the first time the two-loop universal vertex, both in QCD and
$\mathcal{N}=4$ sYM. Moreover, we exploit the well-known IR structure of
two-loop gauge-theory amplitudes to define finite remainders for their
universal building blocks in the high-energy regime. We also document
the various checks that we have performed to validate our
calculations.  We present our conclusions, final remarks and outlook
in \cref{se:conclusions}.

%%%%%% -------
%%%%%% -------
%%%%%% -------
%%%%%% -------
%%%%%% -------
%%%%%%%%%%%%%%%%%%%%%%%%%%%%%%%%%%%%%%%%%%%%%%%%%%%%%%%%%%%%%%%%%%%%%%%%%%%%%%%%%%%%%%%%%%%%%%%%%%%%%%%%%%%%%%%%%%%%%%%%%%%%%%%%%%%%%%%%%%%%%%%%%%%%%%%%%%%%%%%%%%%%%%%%%%%%%%%%%%%%%%%%%%
%%%%%% -------
%%%%%% -------
%%%%%% -------
%%%%%% -------
%%%%%% -------

\section{Five-point scattering amplitudes in MRK}
\label{se:generalaspects}
%%%------------------------------------------------------------------------------------
In this section we discuss the defining features of five-point scattering amplitudes in MRK. We begin with a precise description of the kinematics, and then turn to the discussion of signature eigenstates as well as the choice of appropriate colour bases for the various partonic channels. In the second half we describe how we obtain the high-energy limit of $2\to3$ amplitudes up to two loops starting from their known expressions in general kinematics~\cite{Agarwal:2023suw}, 
with an emphasis on the expansion of the transcendental functions.
We conclude by presenting our results for the infrared-subtracted scattering amplitudes.
\subsection{Kinematics}
\label{se:multireggekin}
%%%---------------------------------------------------------------------
\begin{figure}[t!]
    \centering
    \includegraphics[scale=0.67]{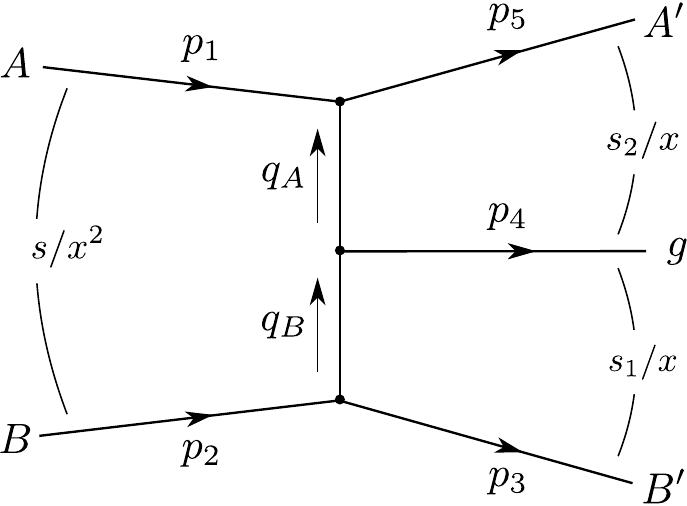}
    \caption{Schematic representation of the kinematics for the scattering process in~\cref{eq:scattering}}
    \label{fig:kinematics}
\end{figure}
We consider the scattering process
\begin{equation}
\label{eq:scattering}
    A^{\lambda_A}(p_1)\;B^{\lambda_B}(p_2) \to B'^{\lambda_{B'}}(p_3) \;
    g^{\lambda_4}(p_4) \; A'^{\lambda_{A'}}(p_5)
\end{equation}
where $A,A',B,B'$ are flavour indices and can either be
$q(\bar{q})$ for quarks(anti-quarks) or $g$ for gluons, $p_i$ label
the momenta of the scattering partons and $\lambda_i$ refer
to their helicities.
The MRK regime is defined as the configuration where the final-state
partons are strongly ordered in rapidity, while their transverse
components are commensurate and much smaller than the centre-of-mass
energy $s_{12} = 2p_1\cdot p_2$:
\begin{equation}
  p_1^+\sim p_5^+ \gg p_4^+ \gg p_3^+ \,, \quad
  p_2^- \sim p_3^- \gg p_4^- \gg p_5^-\,, \quad
  p_4^\pm \sim |{\bf q}_A| \sim |{\bf p}_4| \sim |{\bf q}_B|\,,
\end{equation}
with $p_1^+=p_2^-=\sqrt{s_{12}}$,
where we have introduced the light-cone coordinates in the notation $p^{\pm} = p^0 \pm p^z$, 
${\bf p} = \{p^x,p^y\}$, and introduced the $t$-channel momenta
\begin{equation}
    \label{eq:tchannelmomenta}
    q_A = p_5 - p_1, \quad\quad\quad
    q_B = p_4 + p_5 - p_1 = p_2 - p_3\,,
\end{equation}
see~\cref{fig:kinematics}.
The helicity amplitudes for the process~\eqref{eq:scattering}
can be described in terms of the five Mandelstam invariants
\begin{equation}
  \label{eq:invariants}
    s_{12} = 2 p_1\cdot p_2, \quad s_{23} = -2 p_2\cdot p_3, 
    \quad s_{34} = 2 p_3\cdot p_4, \quad s_{45} = 2 p_4\cdot p_5, \quad s_{51} = -2 p_1\cdot p_5\,,
\end{equation}
and the parity-odd quantity
\begin{equation}
\label{eq:trfive}
\mathrm{tr}_5 \equiv {\rm Tr}[\gamma_5 \slashed p_1 \slashed p_2
  \slashed p_3 \slashed p_4] = 4 i \epsilon_{\mu \nu \rho \sigma}
p_1^\mu p_2^\nu p_3^\rho p_4^\sigma ,
\end{equation}
where $\epsilon_{\mu \nu \rho \sigma}$ is the totally anti-symmetric
Levi-Civita symbol. In MRK, they parametrically scale as
\begin{equation}
  \label{eq:scaling}
  s_{12} \sim \mathrm{tr}_5 \sim 1/x^2,
  \quad
  s_{34}\sim s_{45} \sim 1/x,
  \quad
  s_{23}\sim s_{51} \sim 1,
\end{equation}
with the scaling parameter $x\ll 1$. Also, in this limit
\begin{equation}
  \label{eq:quad}
  \mathrm{tr}^2_5 = \Delta = s_{34}^2 s_{45}^2 (1-2\alpha-2\beta - 2 \alpha
  \beta + \alpha^2+\beta^2) + \mathcal O\left(1/x^3\right),
\end{equation}
where $\Delta = \mathrm{det}(G_{ij})$ and $G_{ij} = 2 p_i\cdot p_j$ with $i,j=1...4$ is the Gram matrix. This implies $\alpha = -s_{12}s_{23}/(s_{34} s_{45})$ and $\beta = -s_{12}s_{51}/(s_{34} s_{45})$.

To make the scaling~\cref{eq:scaling} manifest and simplify the quadratic relation~\eqref{eq:quad},
we follow ref.~\cite{Caron-Huot:2020vlo} and parametrise the kinematics in terms of $\mathcal O(1)$
invariants $\{s,s_1,s_2\}$, a small dimensionless parameter $x$, and a complex variable $z$ such that
\begin{equation}
\label{eq:sij_mrk_scaling}
    s_{12} = \frac{s}{x^2}, \quad s_{23} = -\frac{s_1 s_2}{s} z \zbar,
    \quad s_{34} = \frac{s_1}{x}, \quad s_{45} = \frac{s_2}{x}, \quad
    s_{51} = -\frac{s_1 s_2}{s} (1-z) (1-\zbar)\,
\end{equation}
with $\bar z = z^*$.
Thanks to \cref{eq:sij_mrk_scaling}, the parity odd invariant $\mathrm{tr}_5$ 
is then simply given by
\footnote{Although the quadratic relation \cref{eq:trf_mrk} allows for two solutions,
we are free to pick one. Once such choice is made, given a set of independent $s_{ij}$,
the kinematics of the process and the components of the individual momenta are entirely determined.}
\begin{equation}
\label{eq:trf_mrk}
\mathrm{tr}_5 = \frac{s_1 s_2}{x^2} (z- \zbar) +\mathcal O (1/x) \,.
\end{equation}
As we explicitly show in \cref{se:spinorproducts}, 
it follows that the complex variable $z$ is related to the transverse momenta $q_{A,B}$ via
\begin{equation}
\label{eq:zzb_to_qab}
    z   = -\frac{q_B^x-i q_B^y}{|{\bf{p}}_4|}, \quad\quad\quad
    1-\zbar = \frac{q_A^x +i q_A^y}{|{\bf{p}}_4|}\,.
\end{equation}
Let us stress that the role of the scaling parameter $x$ is
effectively to ensure that the invariants $s$, $s_1$, and $s_2$ are
of the same order, and that large-rapidity logarithms manifest
themselves through the large quantity $\ln(x)$.

In the $2\to3$ physical scattering region, the five Mandelstam invariants fulfil the following set of conditions~\cite{Gehrmann:2018yef}
\begin{equation}
\label{eq:physicalregion}
    s_{12} \ge s_{34}, \quad\; s_{12}-s_{34}\ge s_{45}, \quad\;
    s_{45}-s_{12} \le s_{23} \le 0, \quad\; s^{-}_{51} \le s_{51} \le s^{+}_{51},
\end{equation}
with
\begin{align}
\label{eq:s51pm}
s_{51}^\pm = \frac{1}{(s_{12}-s_{45})^2} &\Big[ s_{12}^2 s_{23} + s_{34} s_{45} (s_{45} - s_{23}) - s_{12} (s_{34} s_{45} + s_{23} s_{34} + s_{23} s_{45}) \notag \\
&\pm \sqrt{s_{12} s_{23} s_{34} s_{45} (s_{12} + s_{23} - s_{45})(s_{34} + s_{45} - s_{12})}\, 
\Big]\,.
\end{align}
\begin{figure}[t!]
    \centering
    \includegraphics[scale=0.5]{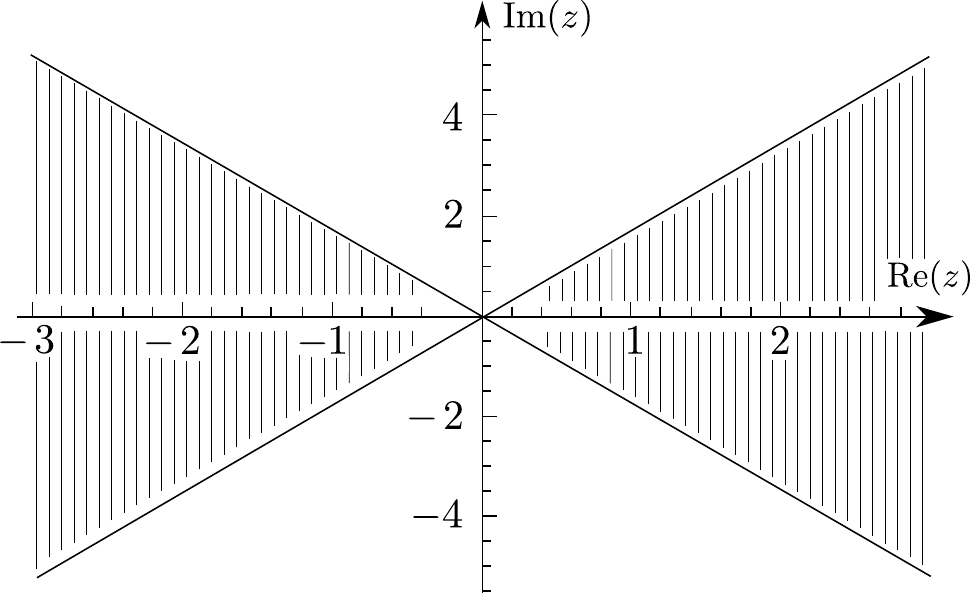}
    \caption{Representation of the $z$-complex plane in MRK in the strict $x\to 0$ limit. 
    The shaded regions are forbidden whereas the lower and upper triangles are permitted.}
    \label{fig:z_complex_plane}
\end{figure}
Considering the $x\to 0$ behaviour of \cref{eq:sij_mrk_scaling}, the first three conditions in
\cref{eq:physicalregion} are trivially satisfied. 
The last one, at fixed values of $s$, $s_{1}$ and $s_2$, 
defines an $x$-dependent exclusion region in the $z$-complex plane. 
In the strict $x \to 0$ limit, for any value of $s$ and $s_{1,2}$, this reduces simply to
\begin{equation}
 |\mathrm{Im}(z)| \ge \sqrt{3} |\Re(z)|,
\end{equation}
which is pictorially represented in~\cref{fig:z_complex_plane}. 

Finally we consider the scenario where the two light-cones are exchanged, \ie we apply the $p_1 \leftrightarrow p_2$ and $p_3 \leftrightarrow p_5$ permutations. This allows one to investigate the $BA$ scattering channel starting from the $AB$ one of \cref{eq:scattering}. 
As for the invariants, see \cref{eq:sij_mrk_scaling,eq:trfive}, 
this leads to
\begin{equation}
\label{eq:permutation_invariants}
s_{1} \leftrightarrow s_{2} \quad\quad z\zbar 
\leftrightarrow (1-z)(1-\zbar), \quad\quad
\mathrm{tr}_5 \to \mathrm{tr}_5, 
\end{equation}
where the last relation follows from the fact that we are considering an
even permutation of momenta.
In particular, since $\mathrm{tr}_5$ in MRK is given by \cref{eq:trf_mrk},
this implies that $(z-\zbar)\leftrightarrow (z-\zbar)$. 
The latter, combined with the second relation in \cref{eq:permutation_invariants} implies $z \leftrightarrow 1-\zbar$. 
Therefore, exchanging the two light-cones amounts simply to the transformations $s_{1} \leftrightarrow s_{2}$ and $z \leftrightarrow 1-\zbar$. Amplitudes where particles $A(B)$ and $A'(B')$ have the same
flavour and helicity are then symmetric under this transformation.

%%%---------------------------------------------------------------------
\subsection{Partonic channels, colour bases, and signature symmetry}
\label{se:colour}
%%%---------------------------------------------------------------------
In MRK, the only partonic configurations that contribute to
\cref{eq:scattering} at leading power in $x$ are those where both flavour
and helicity are conserved along the large light-cone momenta lines, \ie
\begin{equation}\label{eq:partchann}
  A^{\lambda_A} (p_1) B^{\lambda_B}(p_2) \to
  B^{\lambda_B} (p_3) g^{\lambda_4}(p_4) A^{\lambda_A}(p_5).
\end{equation}
To fix the notation, we define the scattering amplitude $\mathcal{A}$ in terms of the connected component of the $\mathcal{S}$-matrix as
\begin{equation} \label{eq:S_to_A}
    \mathcal{S}_{\text{connected}} \equiv (2 \pi)^{d} \delta^{d}(\textstyle\sum_i p_i) (i\mathcal{A}) .
\end{equation}
We then write the ultra-violet (UV) renormalised scattering amplitude for the
process~\eqref{eq:partchann} as
\begin{equation}
\label{eq:uv_ren_ampls}
  \mathcal A^{[AB]}_{\boldsymbol{\lambda}} =
  g_s^3 \,2 \sqrt{2} \, \Phi^{[AB]}_{\boldsymbol{\lambda}}
  \sum_{n} \mathcal C_n^{[AB]}\mathcal B_{n}^{[AB]}(\{s_{ij}\},
  \mathrm{tr}_5;\boldsymbol{\lambda}),
\end{equation}
where the pair $[AB]$ identifies the flavour configuration in the initial state, \ie $[A B]$ can be $[gg],\,[qg]$ or $[qQ]$.
In \cref{eq:uv_ren_ampls} $g_s$ is the strong
coupling, $\boldsymbol{\lambda}=\{\lambda_B,\lambda_4,\lambda_A\}$ are
the final-state helicities, $\Phi^{[AB]}$ are helicity- and
channel-dependent spinor factors, $\mathcal C_n$ are a set of colour
tensors that span the colour space of the process, and
$\mathcal B_n^{[AB]}$ are spinor- and colour-stripped
scalar functions that contain
the non-trivial perturbative information about the scattering amplitude.
We expand them in terms of the strong coupling constant $\alpha_s$ as
\begin{equation}\label{eq:perturbative_amplitude}
  \mathcal B_n^{[AB]} = 
  \mathcal B_n^{[AB],(0)}+
  \left(\frac{\alpha_s}{4\pi}\right) \mathcal B_n^{[AB],(1)}+
  \left(\frac{\alpha_s}{4\pi}\right)^2 \mathcal B_n^{[AB],(2)}
  +\mathcal O(\alpha_s^3),
\end{equation}
where $\alpha_s = \alpha_s(\mu_R)$ with $\mu_R$ the renormalisation scale.
Since we are working with UV-renormalised amplitudes, infrared (IR) divergences
manifest themselves as poles in the dimensional regulator $\epsilon = (4-d)/2$. 
We keep the $\epsilon$-dependence in $\mathcal B_n^{[AB],(l)}$ implicit. Also,
note that throughout this paper we will use the superscript $(l)$
to denote the coefficient of $(\as\big/4\pi)^l$ in the perturbative
expansion of the corresponding quantity (possibly ignoring an overall
$g_s^3$ tree-level coupling factor, such that $\mathcal A^{[AB],(0)}$ denotes the tree-level amplitude). 

We now discuss the various terms in \cref{eq:partchann},
starting with the spinor factors $\Phi^{[AB]}_{\boldsymbol{\lambda}}$.
We choose them to be the tree-level (MHV) spinor amplitudes
for the process \cref{eq:partchann}. Specifically, for the
$\boldsymbol{\lambda} = \{+,+,+\}$ configuration, we then have
\begin{align}
\label{eq:spinorfactors}
    \Phi^{[gg]}_{\lbrace+++\rbrace} & = \frac{\langle12\rangle^4}{\langle12\rangle \langle23\rangle \langle34\rangle\langle45\rangle\langle51\rangle}, \\
    \Phi^{[qg]}_{\lbrace+++\rbrace} & = \frac{\langle12\rangle^3 \langle25\rangle}{\langle15\rangle \langle52\rangle \langle23\rangle\langle34\rangle\langle41\rangle},\\
    \Phi^{[qQ]}_{\lbrace+++\rbrace} & = \frac{\langle12\rangle^2\langle13\rangle\langle25\rangle}{\langle15\rangle \langle52\rangle \langle23\rangle\langle34\rangle\langle41\rangle}\,.
\end{align}
Results for other helicites (as well as for anti-quark channels) can
be obtained from these via parity transformations, permutation of
external momenta, and crossing symmetry.
For a detailed discussion regarding our spinor products conventions in 
MRK we refer the reader to \cref{se:spinorproducts}.

In order to discuss colour, we introduce the standard operators
${\bf T}_i$ defined as
\begin{equation}\label{eq:colour_insertion_operators}
    \left(\mathbf{T}_{k}\right)^c_{ab} = i f^{acb} \;\; \mathrm{if} \; k=g,\quad\;
    \left(\mathbf{T}_{k}\right)^c_{ij} = T^{c}_{ij}  \;\; \mathrm{if} \; k=q,\quad\;
    \left(\mathbf{T}_{k}\right)^c_{ij} = - T^{c}_{ji}  \;\; \mathrm{if} \; k=\bar{q}\,.
\end{equation}
Here, both quarks and anti-quarks are taken to be outgoing, and $T^c_{ij}$ are generators of $SU(N_c)$  in the fundamental representation satisfying $\mathrm{Tr}(T^a T^b) = \delta^{ab}/2$.
For future convenience, we also introduce the combinations
\begin{equation}
\label{eq:titjpm}
    {\bf T}^{\pm}_{15} = {\bf T}_{1} \pm {\bf T}_{5}, \quad\quad
    {\bf T}^{\pm}_{23} = {\bf T}_{2} \pm {\bf T}_{3}.
\end{equation}
As it will become apparent later on, in MRK it is natural to work in a
colour basis $\mathcal C_i$ where each element is an eigenstate of
both $({\bf T}^{+}_{15})^2$ and $({\bf T}^{+}_{23})^2$. We achieve
this by choosing a basis of irreducible $SU(N_c)$ representations in
both the $s_{51}$ and $s_{23}$ channels.  Each basis
element can be labelled by a pair $(r_1,r_2)$, where $r_1$ and $r_2$
correspond to the representations in the $q_A\, (s_{51})$ and
$q_B\,(s_{23})$ channels respectively. Specifically, we define
$r_{1,2}$ through
\begin{equation}\label{eq:colrep}
    R_1 \otimes R_5 = \oplus\, r_1
    \qquad 
    R_2 \otimes R_3 = \oplus\, r_2\,,
\end{equation}
with $R_i$ the representation of the $i$-th parton. The relevant
decompositions we need are
\begin{equation}
\label{eq:ggdecomp}
8 \otimes 8 = \overbrace{0 \oplus 1 \oplus 8_s \oplus 27}^\mathrm{symmetric} \; \oplus \; \overbrace{8_a \oplus 10 \oplus \overline{10}}^{\mathrm{anti-symmetric}},
\end{equation}
for a gluon line, and
\begin{equation}
3 \otimes \bar 3 = 1 \oplus 8\,,
\end{equation}
for a quark line. The subscripts $a$ and $s$ in \cref{eq:ggdecomp}
refer to the anti-symmetric and symmetric adjoint representations
respectively, while ``0'' denotes an irreducible representation that is
present in $SU(N_c)$ for $N_c>3$, but not in $SU(3)$. We report more
details about the colour decomposition in \cref{se:appendixcolour}, 
including the explicit expressions for the $\mathcal C_n$ tensors (listed in \cref{tab:colorGG,tab:colorQG}).
Here we only mention that we choose orthogonal bases, \ie
\begin{equation}
    \sum_{\mathrm{colour}} \left(\mathcal{C}^{[AB]}_n\right)^\dagger
    \mathcal{C}^{[AB]}_m = \mathcal{N}^{[AB]}_m \delta_{mn}\,.
\end{equation}
Also, we note that in the gluon case this colour basis naturally
splits into tensors which have well-defined symmetry properties under
exchange of the colour labels for the two external gluons of the
relevant channel, as shown by the braces in \cref{eq:ggdecomp}.

Finally, in MRK it is also convenient to work with signature
eigenstates, i.e. states which have well-defined symmetry properties
under the exchange of initial and final states. Indeed, only one
signature eigenstate captures the leading behaviour of the cross
section in the high-energy limit (see e.g.~\cite{Caron-Huot:2017fxr}). 
We then define the following
operations on the scalar functions $\mathcal B$~\footnote{Note that the
helicity of particle 1(2) and 5(3) are the same, cf. \cref{eq:partchann}.}
\begin{align}
  &\mathcal B_n^{[AB],(l)}(s_{12},s_{23},s_{34},s_{45},s_{51},\mathrm{tr}_5;\boldsymbol{\lambda})\big|_{1\leftrightarrow5}
  = \mathcal B_n^{[AB],(l)}(s_{52},s_{23},s_{34},s_{41},s_{51},-\mathrm{tr}_5;\boldsymbol{\lambda}), \nonumber
  \\
    &\mathcal B_n^{[AB],(l)}(s_{12},s_{23},s_{34},s_{45},s_{51},\mathrm{tr}_5;\boldsymbol{\lambda})\big|_{2\leftrightarrow3}
  = \mathcal B_n^{[AB],(l)}(s_{13},s_{23},s_{24},s_{45},s_{51},-\mathrm{tr}_5;\boldsymbol{\lambda}), \\
    &\mathcal B_n^{[AB],(l)}(s_{12},s_{23},s_{34},s_{45},s_{51},\mathrm{tr}_5;\boldsymbol{\lambda})\big|_{\substack{ 1\leftrightarrow5 \\  2\leftrightarrow3}}
  = \mathcal B_n^{[AB],(l)}(s_{53},s_{23},s_{24},s_{41},s_{51},+\mathrm{tr}_5;\boldsymbol{\lambda}). \nonumber
\end{align}
Crucially, these transformations exchange positive with negative
Mandelstam invariants and thus require a non-trivial analytic
continuation of the amplitudes. 
To achieve the latter, we exploit the explicit expressions of the ensuing transcendental
functions in all 5! kinematic regions of $2\to3$ scattering~\cite{Chicherin:2020oor} and follow the procedure
described in refs.~\cite{Agarwal:2021vdh,Agarwal:2023suw} to cross those functions from a given region to another one.

We define the following eigenstates~\footnote{Note the apparently different signs in front of
$\sigma_i$ with respect to the standard definition. This is because we have
factored out from our amplitudes an antisymmetric spinor factor $\Phi_{\mathbf{\lambda}}$.}
\begin{equation}
\label{eq:signatureampls}
\begin{aligned}
    \mathcal{B}_n^{[AB],(l),(\sigma_1,\sigma_2)} = \frac{1}{4} \Big[ &
      \mathcal{B}_n^{[AB],(l)}-
      \sigma_1\mathcal{B}_n^{[AB],(l)}\big|_{1\leftrightarrow 5} \,\,-
      \\ &\sigma_2\mathcal{B}_n^{[AB],(l)}\big|_{2\leftrightarrow3}\,\,+
      \sigma_1\sigma_2\mathcal{B}_n^{[AB],(l)}\big|_{\substack{ 1\leftrightarrow5 \\  2\leftrightarrow3}}
      \Big] \, .
\end{aligned}
\end{equation}
In the same spirit, we introduce the definite-signature colour operators
\begin{equation}
\label{eq:colcombinations}
\boldsymbol{\mathcal{T}}_{\sigma_1 \sigma_2} = {\bf T}_{15}^{\sigma_1} \cdot {\bf T}_{23}^{\sigma_2}\,,
\end{equation}
and note that all our colour basis elements are by construction eigenstates of
$\boldsymbol{\mathcal{T}}_{++}$ since
\begin{equation}
\label{eq:tpp_diag}
   \boldsymbol{\mathcal{T}}_{++} = \frac{1}{2}\left(
{{\bf T}_4}^2 - ({\bf T}^{+}_{15})^2 - ({\bf T}^{+}_{23})^2\right)\,.
\end{equation}
We point out that thanks to Bose statistics the (anti)symmetrisation over a gluon
line is equivalent to (anti)symmetrising its colour indices. Therefore,
selecting a signature eigenstate corresponds to selecting a symmetric or
antisymmetric representation in the $8 \otimes 8$ decomposition, see
\cref{eq:ggdecomp}. The same is not true in general for quarks.

We conclude this section by stressing that all the manipulations and
definitions described above do not require the MRK limit. However, they make the study of this kinematic configuration
particularly transparent. Indeed, as we will review later on, in
the MRK regime the amplitude naturally organises into contributions coming from
$t$-channel exchanges of effective degrees of freedom -- reggeons -- 
which have well-defined colour and signature symmetry. Writing the amplitude in the
way described in this section helps uncover such a structure.

%%%---------------------------------------------------------------------
\subsection{MRK expansion of the full scattering amplitudes}
\label{se:amplexpansion}
Having set our notation, we can now expand in MRK the full two-loop
scattering amplitudes that some of us
computed~\cite{Agarwal:2023suw}. To do so, we first note that the
spinor factors defined in \cref{eq:spinorfactors} capture the
leading-power multi-Regge behaviour,
$\Phi^{[AB]}_{\boldsymbol{\lambda}} \sim 1/x^2$. This implies
that at leading power
\begin{equation}
  \mathcal B^{[AB]}_{n} \sim {\rm const} + \mathcal O(x).
\end{equation}
The scattering amplitudes are schematically written as~\cite{Agarwal:2023suw}
\begin{equation}
  \mathcal B^{[AB]}_{n} = \sum \left[R_{a,k}(\{s_{ij}\};\ep_5)
    +\mathrm{tr}_5 R_{b,k}(\{s_{ij}\};\ep_5)\right] \times f_k(\{W\}),
\end{equation}
where $R$ are rational functions of the external kinematics, while $f_k$
are transcendental functions, referred to as pentagon
functions~\cite{Gehrmann:2018yef,Abreu:2018aqd,Chicherin:2018old,Chicherin:2020oor}.
The latter are pure functions of uniform transcendentality, 
that depend on the 31 letters of the pentagon alphabet 
$\{W\}$~\cite{Abreu:2018aqd,Chicherin:2018old,Chicherin:2020oor}.
Finally, we introduced a further dependence on $\ep_5$, which is related to the Gram determinant via $\ep_5 = i \sqrt{|\Delta|}$. 
Note that for $\ep_5$ we adopt the same convention of
ref.~\cite{Chicherin:2020oor}, i.e. we define it with a positive
imaginary part\footnote{We have explicitly checked the correctness of this choice by evaluating one-loop 
and two-loop planar pentagon integrals in a dozen of kinematic points using the numerical implementation of ref.~\cite{Chicherin:2020oor} 
and the \texttt{AMFlow} program~\cite{Liu:2022chg} and found agreement.}. 
This implies $\ep_5 = +i |{\rm tr}_5|$, and therefore in MRK
\begin{equation}
\label{eq:signepsfive}
    \epsilon_5 \simeq \frac{s_1 s_2}{x^2} \times 
     \begin{cases}
          \left(z-\bar{z}\right) \quad \mathrm{if} \; \mathrm{Im}(z) > 0,\\
          \left(\bar{z}-z\right) \quad \mathrm{if} \; \mathrm{Im}(z) < 0\, .
     \end{cases}
\end{equation}
The expansion of the rational
functions is straightforward. However, due to their large number
and sheer size, this task can be computationally intensive. For this
purpose we use the computer algebra program
\texttt{FORM}~\cite{Vermaseren:2000nd,Ruijl:2017dtg}.  Furthermore,
when the amplitude is expanded in terms of the pentagon functions, the
rational coefficients $R$ develop spurious higher poles in $x=0$.  To
obtain the leading MRK behaviour of the amplitude, the transcendental
functions then need to be expanded around $x=0$ beyond leading power.
In particular, we require the expansion of $f_k$ up to second order in
$x$.  Fortunately, the pentagon functions are written as iterated
integrals, which makes such an expansion straightforward.

Before describing how we proceed with the expansion, we comment on \cref{eq:signepsfive}. As discussed in detail in ref.~\cite{Caron-Huot:2020vlo}, amplitudes in MRK are non real analytic when crossing the $\epsilon_5=0$ ($\mathrm{Im}(z)=0$) hyper-surface. This would 
require to analytically continue the ensuing results from the upper to the lower half of the complex plane, or vice-versa. Similarly to ref.~\cite{Caron-Huot:2020vlo}, we prefer instead to expand the pentagon functions separately in both regions. The approach is indeed identical, and care just needs to be taken when fixing an initial boundary condition.

\paragraph{Leading-power expansion}
We now discuss how to obtain the leading-power MRK behaviour
of a generic weight-$w$ pentagon function $f_i^{(w)}(\{W\})$ of
ref.~\cite{Chicherin:2020oor}\footnote{See also ref.~\cite{Caron-Huot:2020vlo} for an analogous discussion.}. 
By construction, those obey a differential equation
of the form
\begin{equation}
  \mathrm{d} f_i^{(w)} = \sum_{k=1}^{31} {\mathrm{d}}\ln (W_k)
  \times\bigg[
    \mathrm{linear~combinations~of~lower~weight~} f_{j}^{(w')}
    \bigg].
  \label{eq:diff_equations}
\end{equation}
Crucially, $a)$ the bracket on the r.h.s. involve linear combinations of $f^{(w')}_j$
whose coefficients are rational numbers, 
and $b)$ the total transcendental weight of each term in the bracket is exactly $w-1$. 
Assuming that the weight-$(w-1)$ functions in the MRK expansion are known, 
one can then readily obtain the desired result at weight $w$ by expanding the 
pentagon alphabet $\{W\}$ at leading power in $x$ in the differential equation
in~\cref{eq:diff_equations}, and integrate it back. Although all of this is
quite standard, we now provide additional details on the procedure for the sake of completeness.

First, we point out that the pentagon alphabet drastically simplifies in the MRK limit.
Indeed, the 31 $\mathrm{d}\ln(W_k)$ can all be expressed in terms of 12
$\mathrm{d}\ln v_i$, where the letters $v_i$ are~\cite{Caron-Huot:2020vlo}
\begin{equation}
\label{eq:lettersmrk}
\lbrace x \rbrace, \;\;\;
\left\lbrace \frac{s_1 s_2}{s} \right\rbrace, \;\;\;
\lbrace s_1, s_2, s_1 - s_2, s_1 + s_2\rbrace, \;\;\;
\lbrace z, \bar{z}, 1-z,1-\bar{z}, 1-z-\bar{z}, z-\bar{z} \rbrace\, .
\end{equation}
A few comments on the alphabet~\cref{eq:lettersmrk} are in order. First, note the
separation between longitudinal ($s_i$) and transverse ($z,\bar z)$
variables.  Second, this alphabet implies that all our results can be
expressed in terms of $\ln(x)$, and
2dHPLs~\cite{Gehrmann:2000zt,Gehrmann:2001ck,Gehrmann:2001jv} of
$\{s,s_1,s_2\}$ and of $\{z,\bar z\}$.
Given this, obtaining the desired result is straightforward. 
We assume that the result at weight $w-1$ is both known and expressed in terms of a minimal basis of 2dHPLs $\{G^{(w-1)}\}$. Then, all one
has to do is
\begin{itemize}
\item write the most general weight$-w$ ansatz starting
  from a minimal 2dHPLs basis $\{G^{(w)}\}$ (and products of lower-weight
  2dHPLs and constants such that the total weight is $w$)~\footnote{
  In practice, given the simplicity of the alphabet \cref{eq:lettersmrk}
  one does not need a full basis of 2dHPLs but only a subset of it.};
\item
  take the differential of the ansatz,
  and express it in terms of $\{G^{(w-1)}\}$;
\item match this result against~\cref{eq:diff_equations} to fix all the
  unknown coefficients;
\item fix a missing overall constant by computing the result in a
  specific kinematic point (which must be in the exact MRK limit).
\end{itemize}
We now provide some extra detail on the practical implementation of
this procedure. First, we note that in principle one could read off
the differential of the pentagon functions directly from their
iterated form given in ref.~\cite{Chicherin:2020oor}. However, this
would require dealing with a very large number of different weight-3
functions. We then decided to re-derive the differential equations in
\cref{eq:diff_equations} starting from the canonical basis of one- and two-loop 
five-point master integrals provided in ref.~\cite{Chicherin:2020oor}, 
exploiting their expressions in terms of pentagon functions. 
To do so, we used the programs
\texttt{Reduze}~\cite{Studerus:2009ye,vonManteuffel:2012np} to obtain
the derivatives \wrt the $s_{ij}$, and
\texttt{Kira}~\cite{Klappert:2019emp,Klappert:2020nbg} to perform the
necessary integration-by-parts reduction to master integrals.
We then reconstruct the differential equations for the canonical master integrals $G_i$ in the form 
\begin{equation} \label{eq:full_diff}
    \mathrm{d} G_i = \epsilon \, \sum_{jk}   A_{ijk} {\mathrm{d}}\ln (W_j) \; G_k
\end{equation}   
by numerically fitting the coefficients $A_{ijk}$. 
Finally, substituting the solutions for the master integrals, expressed in terms of pentagon functions $f^{(w)}_i$~\cite{Chicherin:2020oor}, into \cref{eq:full_diff} and collecting the different powers of $\ep$ on the l.h.s.~and r.h.s.~, we obtain the differential equations of \cref{eq:diff_equations}.

We also note that the alphabet~ \eqref{eq:lettersmrk} leads to the following spurious
singularities:
\begin{itemize}
    \item  $s_1 = \pm s_2$, corresponding to $s_{34} = \pm s_{45}$,
    \item $z=\zbar$, corresponding to $\epsilon_5 = 0$,
    \item $1-z-\zbar=0 \Leftrightarrow {\rm Re}(z) = 1/2$,  corresponding to $s_{23}=s_{51}$.
\end{itemize}
Although they cannot be present in the physical
result, they appear in individual pentagon functions and the fate of
their cancellation is different. Indeed, the $s_1=\pm s_{2}$
singularity explicitly drops out from the UV-renormalised amplitudes.
Functions involving the $z-\zbar$ letter are present in the
UV-renormalised amplitudes, but drop out from four-dimensional finite
remainders (see~\cref{se:finite_remainders}). Finally, ${\rm Re}(z)=1/2$ remains as a spurious singularity if the amplitude
is expressed in terms of 2dHPLs, or in terms of the single-valued polylogarithms 
described in~\cref{se:finite_remainders}.\footnote{Upon expanding around $z_0$ with ${\rm Re}(z_0)=1/2$,
we have explicitly checked that the amplitude is regular at this point.} 

To conclude our discussion about the leading-power expansion of the
pentagon functions, we now illustrate how we obtained a boundary
condition in MRK. By construction, in the point
$X_0=\{s_{12},s_{23},s_{34},s_{45},s_{51}\} = \{3,-1,1,1,-1\}$,
the pentagon functions either vanish or can be expressed in terms of
few simple transcendental constants~\cite{Chicherin:2020oor}.
We point out that in principle both $\mathrm{Im}(z)>0$ and $\mathrm{Im}(z)<0$ are allowed when translating this point from the $s_{ij}$ to the complex variables of \cref{eq:sij_mrk_scaling}. We thus solve the system of differential equations in both the upper and lower halves of the complex plane choosing $\mathrm{Im}(z)$ accordingly in the boundary value.
We then consider the family of kinematic points $Y = \{3/x^2,-1,1/x,1/x,-1\}$ and use the differential equation in $x$ to transport the $X_0$ boundary 
($x=1$) to the $x=0$ MRK point. 
The differential equation in $x$ contains square roots, 
but they are easily rationalisable, allowing us to obtain an analytic result. 

We express the result in terms of Goncharov polylogarithms, that we numerically 
evaluate with very high precision using \texttt{GiNaC}~\cite{Bauer:2000cp} 
in the boundary points $x=1$ and $x=0$. 
We then use the \texttt{PSLQ} algorithm~\cite{Ferguson:PSLQ} to express the
final boundary points for the expanded pentagon functions in terms of
simple transcendental constants. A complete set of such constants was already discussed in ref.~\cite{Caron-Huot:2020vlo} (see tab.~2 therein), so we do not repeat them here. 
Throughout the whole procedure we made use of the program \texttt{PolyLogTools}~\cite{Duhr:2019tlz} to deal with the differential equations and for manipulations involving the Goncharov polylogarithms.

\paragraph{Sub-leading-power expansion}
If the leading-power expansion of the pentagon functions is known,
obtaining higher powers is straightforward. Indeed, all one
needs to do is to write a generalised power series in $x$ of the form
\begin{equation}
\label{eq:generalisedseries}
  f^{(w)}_i = \sum_{\ell=0}\sum_{m=0}^w x^\ell \ln^m(x) 
  g^{(w)}_{i,\ell m}\left(\lbrace s,s_1,s_2\rbrace,\lbrace z,\bar{z}\rbrace\right)\,,
\end{equation}
insert it in the differential equation~\eqref{eq:diff_equations}, expand
the $\mathrm{d}\ln \left(W_k\right)$ forms accordingly up to desired
power in $x$, and solve the differential equation term by term in $x$ and $\ln x$.
Since the resulting differential equation has the schematic form
\begin{equation}
  {\mathrm d}f_i^{(w)} = \left[\frac{A_{ij}^{(-1)}}{x} + A_{ij}^{(0)} +
    x \,A_{ij}^{(1)} + ...\right] f_i^{(w)},
\end{equation}
it is easy to see that beyond leading power, at any
transcendental weight $w$ and power in $x$, the problem is turned into solving a linear system of equations
for the coefficient functions $g^{(w)}_{i,\ell m}$. For this purpose,
we make use of the program \texttt{FiniteFlow}~\cite{Peraro:2019svx},
that allows us to efficiently reach high powers in $x$.
In the ancillary files provided alongside this publication, we present expansions of 
all the pentagon functions in MRK up to $\mathcal{O}(x^4)$. 
Following the discussion on the non-real analyticity property of the pentagon functions in crossing $\mathrm{Im}(z)=0$, we provide results in both the lower and upper halves of the complex $z$ plane.

We checked the correctness
of such expansions by comparing the results in a dozen of kinematic points against a numerical
evaluation of the complete pentagon functions in quadruple precision~\cite{Chicherin:2020oor} 
for small-$x$ values. We found excellent agreement within the expected accuracy of the expansion.

\subsection{IR subtraction and finite remainders}
\label{se:finite_remainders}
The procedures described in the previous sections lead to UV-renormalised amplitudes in MRK which still contain (universal) IR
singularities. The factorisation of IR divergences in scattering amplitudes allows us to define a finite remainder (or hard amplitude), which
encodes the non-trivial physical information. As we noted in the
previous section, its structure is simpler than the one
of the IR-divergent amplitude.

The IR structure of two-loop QCD amplitudes is well known~\cite{Catani:1998bh,Aybat:2006mz,Becher:2009qa,Gardi:2009zv}.
Here we follow ref.~\cite{Becher:2009qa} and define ($\MSbar$) finite
remainders $\mathbf{\mathcal H}^{[AB]}_n$ as
\begin{equation}
  \label{eq:bfin}
  \boldsymbol{\mathcal H}^{[AB]} = \lim_{\ep\to0} \,
  \mathbf{Z}_{IR}^{-1}\,
  \boldsymbol{\mathcal B}^{[AB]},
\end{equation}
where we used the vector notation $\boldsymbol{\mathcal H}^{[AB]} =
\big\{\mathcal H^{[AB]}_n\big\}$, $\mathbf{Z}_{IR}$ is a matrix in colour
space and the perturbative expansion of $\mathbf{\mathcal H}^{[AB]}_n$
is defined in the same way as in \cref{eq:perturbative_amplitude}. 
From now on, we will drop the $[AB]$ superscript whenever this is not ambiguous to avoid cluttering the notation.

Up to two loops, the $\mathbf{Z}_{IR}$ matrix can be written as
\begin{equation}\label{eq:Z_exponentiation}
    \mathbf{Z}_{IR} (\epsilon,\{p\},\mu) = \exp \left[\int^\infty_\mu
      \frac{\mathrm{d} \mu'}{\mu'} {\bf
        \Gamma}_{IR}(\{p\},\mu')\right] \; ,
\end{equation}
where $\mu$ is an arbitrary IR scale, and 
\begin{align}
\label{eq:soft_anomalous_dimension}
    {\bf \Gamma}_{IR}(\{p\},\mu) &= \gamma_K(\as) \sum_{\substack{i,j=1 \\ i>j }}^5 \mathbf{T}_i \cdot \mathbf{T}_j  \ln\left( \frac{\mu^2}{-s_{ij}-i\delta}\right)  + \sum_{i=1}^5 \gamma_i(\as) \, .
\end{align}
In the equation above, 
$\as = \as(\mu)$, $\gamma_K$ is the QCD cusp
anomalous dimension~\cite{Korchemsky:1987wg,Moch:2004pa} and
$\gamma_{i}$ are the collinear anomalous
dimensions~\cite{Ravindran:2004mb,Moch:2005id,Moch:2005tm}. They are
given explicitly in \cref{se:beta_and_anomalous} up to the relevant
perturbative order. In MRK, the soft anomalous dimension
\cref{eq:soft_anomalous_dimension} can be written as
\begin{align}
  \label{eq:softad_mrk}
    {\bf \Gamma}_{IR} =& 
    \frac{\gamma_K}{2}\bigg\{
    \left({\bf T}^{15}_+\right)^2
    \left[
      \ln\frac{s_{45}}{-s_{51} } + \ln\frac{s_{45}}{{\bf p}_4^2} - i \pi
      \right]
    +
    \left({\bf T}^{23}_+\right)^2
    \left[
     \ln\frac{s_{34}}{-s_{23}} + \ln\frac{s_{34}}{{\bf p}_4^2} - i \pi
      \right]
    \\
    - & 2\, \mathcal C_A \ln \frac{\mu^2}{-s_{51}}
      -2\, \mathcal C_B \ln \frac{\mu^2}{-s_{23}}
    -  \mathcal C_g \ln \frac{\mu^2}{{\bf p}_4^2}
    + i \pi \left[
      \boldsymbol{\mathcal T}_{+-} + \boldsymbol{\mathcal T}_{-+} + 
      \boldsymbol{\mathcal T}_{--}
      -\boldsymbol{\mathcal T}_{++}
      \right]
    \bigg\} \nonumber \\
    +& 2 \gamma_A + 2\gamma_B + \gamma_g, \nonumber
\end{align}
where we assume that the pairs of particles $(1,5)$ and $(2,3)$ have the same flavour within each pair, and where $\mathcal{C}_{A(B)}$ is the quadratic Casimir of the colour representation of particle $A(B)$, \ie $\mathcal{C}_g = C_A$ and $\mathcal C_q = \mathcal C_{\bar q} = C_F$.
Note that, in our colour basis, the soft anomalous dimension is diagonal
except for the signature-changing term
\begin{equation}
  \frac{\gamma_K}{2}\times i\pi\left[
    \boldsymbol{\mathcal T}_{+-} + \boldsymbol{\mathcal T}_{-+} + \boldsymbol{\mathcal T}_{--}\right].
\end{equation}

As we will discuss in \cref{se:hard_functions}, the soft anomalous dimension~\eqref{eq:softad_mrk}
can be organised in a form that makes MRK factorisation manifest. For now, we
limit ourselves to discussing the properties of the finite remainders
defined through \cref{eq:bfin}. For a given signature, we write them as 
\begin{equation}
  \boldsymbol{\mathcal H}^{(l),(\sigma_1,\sigma_2)} =
  \sum_{k=0}^l\boldsymbol{\mathcal H}^{(l),(\sigma_1,\sigma_2)}_{(k)} L^k,
  \label{eq:logexp}
\end{equation}
where for convenience we have expanded in $L=-\ln(x) - i\pi/2$. We will
justify this form in \cref{se:shockwave}.
At LO, only the odd-odd signature featuring a
double antisymmetric colour-octet exchange is present in MRK.
Specifically, within our choice of spinor factors $\Phi_{\boldsymbol{\lambda}}$
we have
\begin{equation}
  \mathcal H^{[gg],(0),(--)}_{(8_a, 8_a)} = 
  \mathcal H^{[qg],(0),(--)}_{(8, 8_a)_a} = 
  \mathcal H^{[qQ],(0),(--)}_{(8, 8)_a} = 1 + \mathcal O(x),
  \label{eq:LOH}
\end{equation}
while all other contributions are power suppressed. The subscripts in the finite remainders
$\mathcal{H}$ label the colour-basis element, see~\cref{tab:colorGG,tab:colorQG} for details.

As we will summarise later on, this corresponds to single-reggeon exchanges in both the
$s_{51}$ and $s_{23}$ channels.

\paragraph{Basis of transcendental functions}
Beyond leading order (LO), we find that the
finite remainders can be expressed entirely in terms of simple logarithms
of longitudinal variables and single-valued functions of $z$ and $\zbar$. 
The relevant basis has already been discussed in ref.~\cite{Caron-Huot:2020vlo}, thus we borrow it from there.
The one-loop finite remainder can be written in terms of
\begin{equation}
 \ln\left( \frac{\mu^2}{{\bf p}_4^2}\right), \quad\quad
 \ln\left( \frac{\mu^2}{s_1}\right), \quad\quad
 \ln\left( \frac{\mu^2}{s_2}\right), \quad\quad
\end{equation}
plus the following weight-one single-valued functions~\cite{Caron-Huot:2020vlo}
\begin{equation}
  \begin{gathered}
    g_{1,4} = \ln(z\zbar),\quad
    g_{1,5} = \ln\big((1-z)(1-\zbar)\big),
    \\
    g_{1,6} = \ln(z)-\ln(\zbar),\quad
    g_{1,7} = \ln(1-z)-\ln(1-\zbar),
  \end{gathered}
  \label{eq:sv1}
\end{equation}
and, at weight two,
\begin{align}
  \label{eq:sv2}
  g_{2,1} &= D_2(z,\zbar), \\
  g_{2,2} &= \Li_2(z)+\Li_2(\zbar), \\ 
  g_{2,3} &= \Li_2\left(\frac{z}{1-\zbar}\right) + \Li_2\left(\frac{\zbar}{1-z}\right) + \left(g_{1,4}-g_{1,5}\right)\ln(|1-z-\zbar|)  \notag \\
  & + i\pi \left(g_{1,6}+g_{1,7}\right){\rm sgn}[{\rm Im}(z)]\Theta\left({\rm Re}(z)-\frac{1}{2}\right),
\end{align}
where $\Theta$ is the Heaviside step function, and $D_2(z,\zbar)$ is the Bloch-Wigner dilogarithm defined as
\begin{equation}
   D_2(z,\bar{z}) =
   \mathrm{Li}_2(z)-\mathrm{Li}_2(\zbar)+
   \frac{\ln(z\zbar)}{2}\left(\ln(1-z)-\ln(1-\zbar)\right),
\end{equation}
which enjoys the property $D_2(z,\zbar) = D_2(1-\zbar,1-z)$.

For the two-loop finite remainder, weight 3 and 4 functions are also required. In
ref.~\cite{Caron-Huot:2020vlo} it was observed that in $\mathcal N=4$ sYM
the weight-four term can always be written in terms of product of functions
of lower weight. We find that the same property also holds in QCD. 
More explicitly, at weight four there are only products of simple logarithms and 
$D_2(z,\zbar)$ multiplied by $\zeta_2$. 
To write the two-loop finite remainder, we then only need to supplement
\cref{eq:sv1,eq:sv2} with the weight-three functions~\cite{Caron-Huot:2020vlo}
\begin{align}
  g_{3,1} &= D_3(z,\zbar), \\
  g_{3,2} &= D_3(1-z,1-\zbar), \\
  g_{3,3} &= \Li_3(z)-\Li_3(\zbar), \\
  g_{3,4} &= \Li_3(1-z)-\Li_3(1-\zbar) \\
  g_{3,5} &= \Li_3\left(\frac{z\zbar}{(1-z)(1-\zbar)} \right)
  + \frac{1}{2}\ln(1-z-\zbar)\ln^2
  \left(\frac{z\zbar}{(1-z)(1-\zbar)} \right), \\
  g_{3,6} &= 2\,\Li_3\left(\frac{z}{1-\zbar}\right)
    -2\Li_3\left(\frac{\zbar}{1-z}\right)
    -\ln\left(\frac{z\zbar}{(1-z)(1-\zbar)}
    \right) D_2\left(\frac{z}{1-\zbar},\frac{\zbar}{1-z}\right) \notag \\
    & + \frac{i\pi}{2}
    \left[(g_{1,4}-g_{1,5})^2+(g_{1,6}+g_{1,7})^2
      \right]{\rm sgn}[{\rm Im}(z)]
    \,\Theta\left({\rm Re}(z)-\frac{1}{2}\right), \\
  g_{3,9} & = \Li_3\left( \frac{1-z-\zbar}{(1-z)(1-\zbar)}\right),
\end{align}
where we introduced
\begin{equation}
  D_3(z,\zbar) =
  \Li_3(z)+\Li_3(\zbar)-\frac{1}{2}\ln(z\zbar)
  \big(\Li_2(z)+\Li_2(\zbar)\big)
  -\frac{1}{4}\ln^2(z\zbar)
  \ln\left((1-z)(1-\zbar)\right).
\end{equation}
In passing, we note that our QCD results share the same set of single-valued functions of the $\mathcal{N}=4$ sYM results, except
for $g_{2,3}$ and $g_{3,9}$ which instead appear in $\mathcal{N}=8$ supergravity~\cite{Caron-Huot:2020vlo}.

As discussed at the end of \cref{se:multireggekin}, the $s_1 \leftrightarrow s_2$ and $z\leftrightarrow 1-\zbar$ transformations
corresponds to interchanging the two light cones. We thus find it convenient to introduce a basis of functions with definite symmetry under these transformations, either even or odd.
At weight one and two we define,
\begin{equation}
\begin{gathered}
    h_{1,1} = g_{1,4} + g_{1,5}, \quad
    h_{1,2} = g_{1,4} - g_{1,5}, \quad
    h_{1,3} = g_{1,6} + g_{1,7}, \quad
    h_{1,4} = g_{1,6} - g_{1,7}, \quad \\
    h_{2,1} = g_{2,1} \quad \\
    h_{2,2} = g_{2,2} + \frac{h^2_{1,1}-h^2_{1,2}+h^2_{1,3}-h^2_{1,4}}{16} - \zeta_2, \quad
    h_{2,3} = g_{2,3} + \frac{h_{1,2}^2+h_{1,3}^2}{8}- 2\zeta_2\, ,
\end{gathered}
\end{equation}
and at weight three,
\begin{equation}
\begin{gathered}
    h_{3,1} = g_{3,1} + g_{3,2}, \quad
    h_{3,2} = g_{3,1} - g_{3,2}, \quad
    h_{3,3} = g_{3,3} + g_{3,4}, \quad
    h_{3,4} = g_{3,3} - g_{3,4}, \\
    h_{3,5} = g_{3,5} + \frac{h^3_{1,2}}{12} - \zeta_2 \,h_{1,2}, \quad
    h_{3,6} = g_{3,9} + \frac{g_{3,5}}{2} - \frac{h_{1,1}h^2_{1,2}}{8}-\frac{\zeta_3}{2}, \quad \\
    h_{3,7} = g_{3,6} + \frac{h^3_{1,3}}{24} + \frac{h^2_{1,2}h_{1,3}}{8}
    -2\zeta_2 \,h_{1,3}\,.
\end{gathered}
\end{equation}
Given that the $h_{w,i}$ functions are built in terms of the $g_{w,j}$ ones, they also are single valued. In particular, the functions 
$h_{1,1}$, $h_{1,4}$, $h_{2,1}$, $h_{3,1}$, $h_{3,4}$, $h_{3,5}$ and $h_{3,7}$ are symmetric under $z\leftrightarrow 1-\zbar$, 
while the other ones are odd.
Furthermore, they are also either even or odd under the transformation $z \to \zbar$, a property inherited from the $g_{i,j}$ functions above. 

\paragraph{Final results}
The complete results for the finite remainders for all partonic channels can be found in the ancillary files.
Here we only report a few examples to illustrate their simplicity. 

We focus on $gg$ scattering, and consider antisymmetric colour-octet exchanges in both the $s_{51}$ and $s_{23}$ channels. 
As explained in \cref{se:colour}, for gluon scattering this automatically selects the $(--)$ signature component.
This amplitude is symmetric under the exchange of the plus and minus light-cones, thus any even(odd) function $h_{w,i}$ needs to be paired to an even(odd) rational function of $(z,\zbar)$. Up to two loops we find only six independent rational functions, 
\begin{equation}
\begin{gathered}
    r_1 = \frac{z^3 + (1-\zbar)^3}{(1-z-\zbar)^3}, \quad
    r_{2} = \frac{z(1-\zbar)}{(1-z-\zbar)^2}\left(\frac{1}{1-z}+\frac{1}{\zbar}\right),\quad
    r_{3} = \frac{1+z-\zbar}{1-z-\zbar}, \\
    r_{4} = \frac{z(1-\zbar)}{(1-z)\zbar},\quad
    r_{5} = \frac{z(1-\zbar)}{(1-z-\zbar)^2}, \quad
    r_{6} = \frac{z(1-\zbar)(z-\zbar)}{\zbar(1-z)(1-z-\zbar)}\,,
\end{gathered}
\end{equation}
with $r_{1}$, $r_{3}$ and $r_{6}$ odd, and $r_{2}$, $r_{4}$ and $r_5$ even.
For the centrally-emitted gluon with positive helicity, at one loop we have
\begin{align}
  \label{eq:ggfin_mrk_1L}
  \mathcal H_{(8_a, 8_a)}^{[gg],(1),(--)} &= N_c \Bigg[
  \frac{17\pi^2}{12}-2 h_{1,1} L -\frac{h^2_{1,2}}{2} + 
  \left( \ln\left(\frac{s_1 s_2}{s^2}\right) -\frac{i \pi}{2}\right)h_{1,1}
  -\ln\left(\frac{s_1}{s_2}\right)h_{1,2}\Bigg] \notag \\
  &+ \left(N_c-n_f\right)\left[\frac{h_{1,1}-r_1 h_{1,2}-r_2}{3}\right]
  +\frac{3}{2}N_c \left(h_{1,1}-r_3 h_{1,2}\right) - \frac{\gamma^{(2)}_K}{4},
\end{align}
where the explicit expression of the cusp anomalous dimension coefficient $\gamma^{(2)}_K$ is given in \cref{eq:cusp_anomalous_coefficients}. 
The equivalent result
for the negative helicity case can be obtained by the simple replacement $z\leftrightarrow\zbar$.
For the same helicity configuration, at two loops we find
  \begin{align}
   &\mathcal H_{(8_a, 8_a)}^{[gg],(2),(--)} =
   N_c^2 \Bigg[2 L^2 h_{1,1}^2 +L \bigg(-2
   \log \left(\frac{s_1 s_2}{s^2}\right) h_{1,1}^2+i \pi  h_{1,1}^2+h_{1,2}^2 h_{1,1}-17 \zeta _2 h_{1,1} 
   \notag\\
   & \quad\quad
   +2 \log
   \left(\frac{s_1}{s_2}\right) h_{1,2} h_{1,1}+3 r_3 h_{1,2} h_{1,1}-4 \zeta _3+\frac{232}{9}\bigg)+ \frac{h_{1,2}^4}{8}+\frac{1}{2} \log ^2\left(\frac{s_1}{s_2}\right)
   h_{1,2}^2
   \notag\\
   & \quad\quad
   -\frac{9}{2} h_{2,2} h_{1,2}+\frac{1}{2} \log ^2\left(\frac{s_1 s_2}{s^2}\right)
   h_{1,1}^2 +\frac{1677 \zeta _4}{16}
    -\zeta _3 h_{1,1}+\frac{58 h_{1,1}}{9}-3 \big(r_2+r_1 h_{1,2}\big)
   \notag\\
   & \quad\quad
    +\log
   \left(\frac{s_1}{s_2}\right) \bigg(\frac{1}{2} \big(3 r_3+h_{1,2}-3\big) h_{1,2}^2
   -\frac{17}{2} \zeta _2
   h_{1,2}-\log \left(\frac{s_1 s_2}{s^2}\right) h_{1,1} h_{1,2}
   \notag\\
   & \quad\quad
   +\frac{i \pi}{2}  h_{1,1} h_{1,2}\bigg)-\frac{\zeta _2}{4}
   \bigg( 6 h_{1,1}^2+23 h_{1,2}^2+\frac{209}{3} + 39 r_3 h_{1,2}\bigg)
   \notag\\
   & \quad\quad
   +\log
   \left(\frac{s_1 s_2}{s^2}\right) \bigg(-\frac{1}{2} i \pi  h_{1,1}^2+\frac{17}{2} \zeta _2 h_{1,1}-\frac{3}{2} r_3
   h_{1,2} h_{1,1}+2 \zeta _3-\frac{1}{18} \big(9 h_{1,1} h_{1,2}^2+232\big)\bigg)
   \notag\\
   & \quad\quad
   +i \pi  \bigg(\frac{1}{4}
   h_{1,1} h_{1,2}^2-\frac{\zeta _3}{2}-\frac{17}{4} \zeta _2 h_{1,1}+\frac{3}{16} r_3 \big(h_{1,1} h_{1,2}-h_{1,3}
   h_{1,4}-8 h_{2,2}+8 h_{2,3}\big)+\frac{29}{9}\bigg)
   \notag\\
   & \quad\quad
   -\frac{3}{16} r_3 \big(2 h_{1,2}^3 +\big(3
   h_{1,1}^2-16\big) h_{1,2}+h_{1,1} \big(h_{1,3} h_{1,4}+8 h_{2,2}-8 h_{2,3}\big)+64
   h_{3,6}\big)-\frac{1717}{54}\Bigg]
   \notag\\
   & \quad\quad
   +N_c \big(N_c-N_f\big) \Bigg[+\frac{2}{27} L \big(9 h_{1,1} \big(r_2+r_1 h_{1,2}\big)+56\big) + \frac{1}{3} \log \left(\frac{s_1}{s_2}\right)
   h_{1,2} \big(r_2+\big(r_1-1\big) h_{1,2}\big)
   \notag\\
   & \quad\quad
   +\frac{1}{27} \log \left(\frac{s_1 s_2}{s^2}\right) \big(-9 h_{1,1} \big(r_2+r_1
   h_{1,2}\big)-56\big)
   +\frac{1}{18} \zeta _2 \big(-63 r_2-39 r_1
   h_{1,2}-55\big)
   \notag \\
   & \quad\quad
   +\frac{1}{162}
   \big(33 h_{1,1}-162 h_{1,2} h_{2,2}-445\big)
   +i \pi  \bigg(\frac{1}{24} \big(4 \big(r_4+r_2
   \big(h_{1,1}-1\big)\big)
   \notag\\
   & \quad\quad
   +\big(12 r_3+r_1 \big(h_{1,1}-12\big)\big) h_{1,2}-r_1 \big(h_{1,3} h_{1,4}+8
   h_{2,2}-8 h_{2,3}\big)\big) +\frac{14}{27}\bigg)+\frac{1}{24} \big(-2 r_1 h_{1,2}^3
   \notag\\
   & \quad\quad
   +4 \big(r_2-r_5\big)
   h_{1,2}^2+\big(4 \big(r_6+r_3 \big(41-6 h_{1,1}\big)\big)-r_1 \big(3 \big(h_{1,1}-8\big)
   h_{1,1}+152\big)\big) h_{1,2}
   \notag\\
   & \quad\quad
   +4 r_4 h_{1,1}-4 r_2 \big(h_{1,1}+46\big)-\big(12 r_3+r_1
   \big(h_{1,1}-12\big)\big) h_{1,3} h_{1,4}
   \notag\\
   & \quad\quad
   -8 \big(12 r_3+r_1 \big(h_{1,1}-12\big)\big) h_{2,2}+8
   \big(12 r_3
   +r_1 \big(h_{1,1}-12\big)\big) h_{2,3}-64 r_1 h_{3,6}\big)\Bigg]
   \notag\\
   & \quad\quad
    + N_c\beta ^{\text{(0)}} \Bigg[\frac{1}{2} L
   \big(h_{1,2}^2-h_{1,1}^2\big) + \frac{1}{2} \log \left(\frac{s_1}{s_2}\right) h_{1,2}^2+\frac{73 \zeta _3}{6}+\frac{1}{4} \log
   \left(\frac{s_1 s_2}{s^2}\right) 
   \big(h_{1,1}^2-h_{1,2}^2\big)
   \notag\\
   & \quad\quad
   +\frac{1}{4} \zeta _2 \big(h_{1,1}-8 h_{1,4}\big)+i \pi  \bigg(\frac{\zeta
   _2}{2}+\frac{1}{8} \big(-h_{1,1}^2+2 h_{1,2}^2-8 h_{2,1}\big)\bigg) 
   +\frac{1}{48} \big(-2 h_{1,1}^3
   \notag\\
   & \quad\quad
  -3 h_{1,4}
   h_{1,1}^2-3 \big(h_{1,2}^2-2 h_{1,3} h_{1,2}+16 h_{2,1}\big) h_{1,1}
   -h_{1,4}^3+3 h_{1,2}^2 h_{1,4}+3 h_{1,3}^2
   h_{1,4}
   \notag\\
   & \quad\quad
   -9 h_{1,2} \big(h_{1,3} h_{1,4}-8 h_{2,3}\big)-48 \big(-2 h_{3,4}+2
   h_{3,5}+h_{3,7}\big)\big)\Bigg] -\frac{4}{27} \big(N_c-N_f\big){}^2
   \notag\\
   & \quad\quad
   +\frac{\beta ^{\text{(1)}}}{2} \bigg[r_2+h_{1,1}+
   \big(r_1-2 r_3\big)
   h_{1,2}\bigg] 
   -6 \zeta _2 \big(3 h_{1,1}^2+5
   h_{1,2}^2+8 h_{2,1}\big) -\frac{\gamma _K^{(3)}}{4} \label{eq:ggfin_mrk_2L}\,.
  \end{align}
In both \cref{eq:ggfin_mrk_1L} and \cref{eq:ggfin_mrk_2L} we set
$\mu=|{\bf p}_4|$. 
Results with full $\mu$ dependence for all partonic channels and all signatures can be found in the ancillary files. 
As for the two-loop case, we provide separate expressions for both the
lower and upper half of the $z$-complex plane. Here, we just limit ourselves to point out that the two 
differ only for those amplitudes with even-odd or odd-even signature, 
whereas they are the identical when the signature is the same in the two $t$-channels.

%%%%%% -------
%%%%%% -------
%%%%%% -------
%%%%%% -------
%%%%%% -------
%%%%%%%%%%%%%%%%%%%%%%%%%%%%%%%%%%%%%%%%%%%%%%%%%%%%%%%%%%%%%%%%%%%%%%%%%%%%%%%%%%%%%%%%%%%%%%%%%%%%%%%%%%%%%%%%%%%%%%%%%%%%%%%%%%%%%%%%%%%%%%%%%%%%%%%%%%%%%%%%%%%%%%%%%%%%%%%%%%%%%%%%%%
%%%%%% -------
%%%%%% -------
%%%%%% -------
%%%%%% -------
%%%%%% -------

\section{Predictions from the Balitsky/JIMWLK formalism}
\label{se:shockwave}
In this section, we will briefly summarise how one can obtain
predictions for scattering amplitudes in MRK from the Balitsky/JIMWLK
formalism~\cite{Balitsky:1995ub,Jalilian-Marian:1997qno,Jalilian-Marian:1997jhx,Jalilian-Marian:1997ubg,Kovner:2000pt,Weigert:2000gi,Iancu:2000hn,Iancu:2001ad,Ferreiro:2001qy,Caron-Huot:2013fea}.
Although nothing presented in this section is conceptually new, to the
best of our knowledge many of the required details have not been
spelled out explicitly before, so we report them here. We start from a
brief summary of the formalism itself, but we refer the reader
to the vast literature on the subject, and in particular to
ref.~\cite{Caron-Huot:2013fea}, for a more in-depth presentation.

\begin{figure}[t!]
    \centering
    \scalebox{1}{\includegraphics[valign=m]{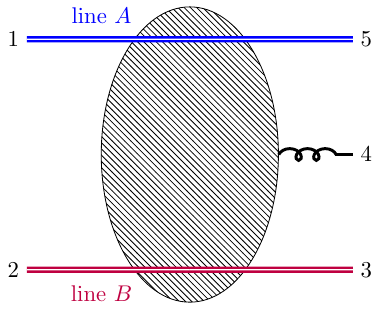}}
    \caption{Graphical notation for MRK scattering. The upper double line denotes the projectile boosted in the $+$ lightcone direction, while the lower one denotes the projectile boosted along the $-$ lightcone direction. 
    The shaded blob represents the interaction among the projectiles and the emitted gluon.}
    \label{fig:MRK_notation}
\end{figure}

Before moving on, we take the opportunity to introduce the graphical notation and some nomenclature that we will use in this section. 
As already explained, the process involves two highly boosted ``projectiles" which retain their identities in the interaction. We will refer to these as $A$ and $B$ lines, in accordance with the notation of \cref{eq:partchann}. We will also use the $A/B$ subscripts to denote quantities associated to the respective lines.
As depicted in \cref{fig:MRK_notation}, we will use blue double lines to draw projectile $A$ and red double lines to draw projectile $B$.
The double lines can indicate either gluons or (anti-)quarks. 
In \cref{fig:MRK_notation} the dashed blob represents the full interaction among the projectiles $A$ and $B$ and with the central-rapidity gluon. 

\subsection{The MRK amplitude from the Balitsky/JIMWLK formalism}
\label{se:review}
The Balitsky/JIMWLK formalism is a convenient framework to analyse systems
with large rapidity gaps. 
In this formalism, in the strict high-energy limit ($x=0$),
the line $A$ is represented by
a single infinite Wilson line $U^{\{r\}}$, which in position space is
localised on the $x_+=0$ lightcone and only depends non trivially on
transverse coordinates $\mathbf z$:
\begin{equation}
  U^{\{r\}}(z) = \mathcal P e^{ig_s\int_{-\infty}^\infty dx^- A^a_-
    (x^+=0,\, x^-, \,\mathbf{z}) T_r^{a}},
\end{equation}
where $A$ is the gluon field, $\mathcal P$ is the usual path-ordering
in colour space, and $r$ stands for the colour
representation.\footnote{In what follows, we will omit $r$ whenever
this does not create ambiguities.}  Such infinite Wilson lines are
divergent, and can be regulated by working at finite rapidity.
Their evolution in rapidity -- which generates the large $\ln x$ logs -- is
also known, and at LO is given by the celebrated Balitsky/JIMWLK
equation. The latter is a non-linear equation and, in particular, the
evolution in rapidity generates additional Wilson lines.

At finite rapidity, 
line $A$ can be represented as
\begin{equation}\label{eq:projectile_expansion_wilson}  
\begin{aligned}
  a_{\lambda_1}^{a_1,\dagger}(p_1) & a_{\lambda_5}^{a_5}(p_5) \sim
  2\pi \delta(p_1^+ - p_5^+) \delta_{\lambda_1\lambda_5} \times
  2 p_1^+ \times\\ & \Big[ \IFpre(\qp_A)
    \overline{U}_{\eta_A}(\qp_A) + \frac{1}{2}\int \measF{\qp} \,
    \IFpre'(\qp_A,\qp)
    \overline{U}_{\eta_A}(\qp_A-\qp)\overline{U}_{\eta_A}(\qp) + \dots
    \Big]^{a_1 a_5} ,
\end{aligned}
\end{equation}
where $a_{\lambda_1}^{a_1,\dagger}(p_1)$ and $ a_{\lambda_5}^{a_5}(p_5) $ are creation and annihilation operators of external states.
Moreover, in \cref{eq:projectile_expansion_wilson}, $2 p_1^+$ is the standard eikonal vertex (with our definition
of the lightcone coordinates),
the subscript $\eta_A$ indicates that we have regulated the
Wilson line by tilting it by a fixed rapidity $\eta_A = 1/2 \ln
p_5^+/p_5^-$, and we have defined
\begin{equation}
\label{eq:fourierdef}
  \overline{U}_\eta(\pp)\equiv U_\eta(\pp)-\mathbb{I}
  \quad \text{and} \quad 
  U_\eta(\pp) = \int \meas{\zp} \, e^{-i \qp \cdot \zp}  U_\eta(\zp) .
\end{equation}
We also introduced the notations 
\begin{equation}
\meas{\zp}  = \mathrm{d}^{2-2\ep}\zp,
\quad\quad 
\measF{\pp} = \mathrm{d}^{2-2\epsilon} \pp / (2\pi)^{2-2\epsilon}.
\end{equation}
The coefficients $\IFpre$ and $\IFpre'$ in 
\cref{eq:projectile_expansion_wilson} are radiatively-generated impact
factors.  We note that $\IFpre'$ is a colour operator so that the
product $\IFpre' \overline{U}\, \overline{U}$ is in the same
representation of the $A$ projectile. In our normalisation, $\IFpre
\sim 1 + \mathcal O(\as)$ and $\IFpre' \sim \mathcal
O(\as)$. An analogous construction holds for line $B$, with the
role of the $+$ and $-$ lightcones exchanged.

In the absence of a central gluon, to compute the amplitude one would
need to evolve \cref{eq:projectile_expansion_wilson} down to the
rapidity $\eta_B = 1/2 \ln p_3^+/p_3^-$ -- thus generating the large
rapidity logs --  and then compute a same-rapidity
correlator with the equivalent of
\cref{eq:projectile_expansion_wilson} for the $B$
line, see ref.~\cite{Caron-Huot:2013fea}. In our
case however, one first needs to evolve down to the rapidity of the
central gluon ($\eta_4=1/2 \ln p_4^+/p_4^-$), and consider the
interaction of the resulting Wilson lines with the gluon 4. To the
accuracy needed for this paper, it is sufficient to consider the
interaction of the gluon with a single Wilson line. At LO, one can
compute such interaction using the shockwave formalism, where gluon 4
interacts with the Wilson line in the background
generated by projectile $B$. 
In terms of the annihilation operator of the emitted gluon $a_{\lambda_4}^{a}$, 
the result reads~\cite{Caron-Huot:2013fea}
\begin{equation}
  \begin{split}
  U_\eta(\pp) a^{a}_{\lambda_4}(p_4) & \sim -2g_s \int \meas{\zp_1}
  \meas{\zp_2} e^{-i \pp \cdot \zp_1 - i \pp_4 \cdot \zp_2} \:
  \big[U^{ab}_{\eta,\rm adj}(\zp_2) \Th^b_{1,R}-\Th^{a}_{1,L}\big]
  U_\eta(\zp_1) \times \\ &\times \int \measF{\kp} e^{i\kp \cdot
    (\zp_2-\zp_1)} \:\frac{{\boldsymbol{\vep}}^*_{\lambda_4} \cdot
    \kp}{\kp^2},
  \end{split}
  \label{eq:OPE}
\end{equation}
where we used the standard notation
\begin{equation}
  \begin{split}
    T^a_{i,L} \, U(\zp_1)... U(\zp_i)... U(\zp_n)\equiv
    U(\zp_1)... \, T^a U(\zp_i)... U(\zp_n),
    \\
    T^a_{i,R} \, U(\zp_1)... U(\zp_i)... U(\zp_n)\equiv
    U(\zp_1)... U(\zp_i) T^a \, ... U(\zp_n).
  \end{split}
\end{equation}
In \cref{eq:OPE}, $\boldsymbol{\varepsilon}^{*}_\lambda$ is the
two-dimensional polarisation vector of the gluon 4, with the implicit
choice $\varepsilon_\lambda\cdot p_2 = 0$.\footnote{Beyond LO,
\cref{eq:OPE} gets corrected in multiple ways. As it will become clear
later however, for the purposes of this paper one only needs to
consider multiplicative corrections to it.} After \cref{eq:OPE} has
been used, one can evolve the resulting Wilson lines down to rapidity
$\eta_B$.

In practice, the rapidity evolution of the Wilson lines is non
trivial. Fortunately, for any perturbative application, one can use
the dilute-field approximation
\begin{equation}
\label{eq:W_def}
  U_\eta(\zp) \equiv \exp\left\{i g_s T^a W_{\eta}^a(\zp) \right\}=
  1+(ig_s)W_{\eta}^a T^a + \frac{1}{2}(ig_s)^2 W_{\eta}^a W_{\eta}^b T^aT^b + \dots,
\end{equation}
with $g_s W_{\eta} \ll 1$, see ref.~\cite{Caron-Huot:2013fea} for details.
This allows us to perturbatively expand the shock-wave expression
\cref{eq:projectile_expansion_wilson} for the $A$ line, the equivalent
formula for the $B$ line, and the gluon-Wilson line interaction~\cref{eq:OPE}.
\begin{figure}
    \centering
    \scalebox{1.1}{\includegraphics[valign=m]{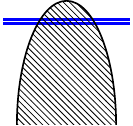}}
    $\quad\sim\quad$ 
    \scalebox{1}{\includegraphics[valign=m]{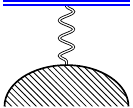}}
    $+$
    \scalebox{1}{\includegraphics[valign=m]{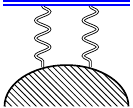}}
    $+$
    \scalebox{1}{\includegraphics[valign=m]{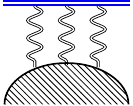}}
    $+ \quad \dots$
    \caption{Graphical representation of the expansion of the Wilson line $A$ in terms of $W$ fields, represented by double wavy lines. The dashed blob stands for the background generated by the other projectile. 
    A similar expansion holds for line $B$.}
    \label{fig:Wilson_exp}
\end{figure}

Introducing the Fourier conjugate of the $W$ field
\begin{equation}
     W^a_{\eta}(\zp) = \int \measF{\qp}  e^{i \qp \cdot \zp} W^a_{\eta}(\qp),
\end{equation}
we find that the projectile expansion
\cref{eq:projectile_expansion_wilson} becomes
\begin{align}
\label{eq:projectile_W_exp}
a^{a_1,\dagger}_{\lambda_1}(p_1) a_{\lambda_5}^{a_5}(p_5) \sim &
\,2\pi \delta(p_1^+ - p_5^+) \delta_{\lambda_1\lambda_5} \times
2p_1^+ \times
\Bigg\{ (i g_s) \IFpre(\qp_A)\llbracket W(\qp_A)
\rrbracket_A +
\notag \\ &+ \frac{ (i g_s)^2}{2!} \int \measF{\qp}
\left[ 1+ \IFpre'(\qp_A,\qp) \right]\, \llbracket W(\qp_A-\qp) W(\qp)
\rrbracket_A + \\ &+ \frac{ (i g_s)^3}{3!} \int \measF{\qp_1}\measF{\qp_2} \,
\llbracket W(\qp_A-\qp_1) W(\qp_1-\qp_2)W(\qp_2) \rrbracket_A + \dots
\Bigg\}_{\eta=\eta_A}, \nonumber
\end{align}
where we have only kept terms that contribute up to NNLL accuracy and 
where we have introduced the notation
\begin{equation}
\begin{aligned}    
    \llbracket O_1 O_2 \dots O_n \rrbracket_r^{ab} &\equiv 
 (T_r^{c_1})_{aa_1} (T_r^{c_2})_{a_1 a_2} \dots (T_r^{c_n})_{a_{n-1}  b}
 \: O^{c_1}_1 O^{c_2}_2 \dots O^{c_n}_n  ,\\
 \llbracket O_1 O_2 \dots O_n \rrbracket^{ab} &\equiv \llbracket O_1 O_2 \dots O_n \rrbracket^{ab}_{\mathrm{adj}}\, ,
\end{aligned}
\end{equation}
to refer to fields contracted with products of $SU(N_c)$ generators.
Graphically, \cref{eq:projectile_W_exp} corresponds to the expansion in \cref{fig:Wilson_exp}.

We now consider the interaction of a Wilson line with a gluon,
and expand both sides of \cref{eq:OPE} using
\cref{eq:W_def}. After some algebra, comparing left-
and right-hand sides of the equation we find that \cref{eq:OPE} provides
the interaction of a gluon with a single $W$ in the form
\begin{equation}
  \begin{split}
    W(\pp&)^b \, a^a_\lambda(q) 
    \sim \\ 
    &\quad 2 g_s  \llbracket W\rrbracket^{a b} (\qp+\pp) \left[\frac{\boldsymbol{\vep}^*_\lambda\cdot \pp}{\pp^2}+\frac{\boldsymbol{\vep}^*_\lambda\cdot \qp}{\qp^2}\right]  +
    \\
    &+ i g_s^2  \int \measF{\kp_1}
    \llbracket W(\qp+\pp-\kp_1) W(\kp_1) \rrbracket^{a b}\left[
      \frac{\boldsymbol{\vep}^*_\lambda\cdot \pp}{\pp^2} + \frac{\boldsymbol{\vep}^*_\lambda\cdot (\kp_1-\pp)}{(\kp_1-\pp)^2}
      \right]+
    \\
    &+ g_s^3 \int \measF{\kp_1}\measF{\kp_2}
    \llbracket W(\qp+\pp-\kp_1) W(\kp_1-\kp_2)W(\kp_2)\rrbracket^{a b}
    \times\\
    &\times\bigg[
      \frac{1}{6}\left(\frac{\boldsymbol{\vep}^*_\lambda\cdot (\kp_1-\pp)}{(\kp_1-\pp)^2}\right)
      -\frac{1}{2}\left(\frac{\boldsymbol{\vep}^*_\lambda\cdot (\kp_2-\pp)}{(\kp_2-\pp)^2}\right)
      -\frac{1}{3}\left(\frac{\boldsymbol{\vep}^*_\lambda\cdot \pp}{\pp^2}\right)
      \bigg]+\dots,
  \end{split}
  \label{eq:ope1_0}
\end{equation}
\begin{figure}
    \centering
    \scalebox{1}{\includegraphics[valign=m]{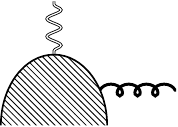}}
    $\quad \sim \quad$
    \scalebox{1}{\includegraphics[valign=m]{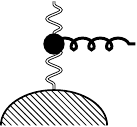}}
    $+$
    \scalebox{1}{\includegraphics[valign=m]{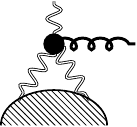}}
    $+$
    \scalebox{1}{\includegraphics[valign=m]{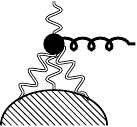}}
    $+ \quad \dots$
    \caption{Graphical representation of \cref{eq:ope1_0}. The shaded area represents the background generated by line $B$, while the black dots correspond to interactions ``vertices".
    Similar diagrams can be drawn for \cref{eq:ope2,eq:ope3} by simply replacing the upper $W$ field with two and three $W$ fields respectively.}
    \label{fig:Wa_expansion}
\end{figure}
for which we provide a cartoon in \cref{fig:Wa_expansion}.
We also find the interactions with multiple $W$ fields to be 
\begin{align}
      [W^b \otimes W^c](\pp) \, a_\lambda^{a}(q) &\sim  
      2 g_s 
      \int \measF{\kp_1} \llbracket W (\qp+\pp-\kp_1) \rrbracket^{ab}W^{c}(\kp_1) \times \notag\\
      &\times\left[
        \frac{\boldsymbol{\vep}^*_\lambda\cdot \qp}{\qp^2} + \frac{\boldsymbol{\vep}^*_\lambda\cdot(\pp-\kp_1)}{(\pp-\kp_1)^2}
        \right]
      + (b\leftrightarrow c) + \dots, \label{eq:ope2}\\
  [W^b \otimes W^c \otimes W^d](\pp)  \, a^{a}_\lambda(q) &\sim   2 g_s
  \int \measF{\kp_1}\measF{\kp_2}
   \llbracket W(\qp+\pp-\kp_1) \rrbracket^{ab} W^{c}(\kp_1-\kp_2)W^{d}(\kp_2)\times
  \notag\\
  &\times\left[
        \frac{\boldsymbol{\vep}^*_\lambda\cdot \qp}{\qp^2} + \frac{\boldsymbol{\vep}^*_\lambda\cdot(\pp-\kp_1)}{(\pp-\kp_1)^2}
        \right]
  + (b\leftrightarrow c) + (b\leftrightarrow d) + \dots, \label{eq:ope3}
\end{align}
where $[f\otimes g](\qp) = \int \measF{\kp} f(\kp) g(\qp-\kp)$ is the
standard Fourier convolution and where we refrain from writing
expressions for interactions with more than
3 $W$s as well as higher order terms
since they will not play any role in our analysis.
As we have mentioned, \cref{eq:ope1_0,eq:ope2,eq:ope3} receive
perturbative corrections. However, in the next subsection we will see
that for the purposes of this paper the only relevant ones are those
affecting the first term on the right-hand side of
\cref{eq:ope1_0}, which gets dressed with a radiative 
vertex correction $\wilsonV_\lambda = 1 + \mathcal O(\as)$ yielding

\begin{equation}
  \begin{split}
    2 g_s \llbracket W\rrbracket^{a b} (\qp+\pp)
    &\left[\frac{\boldsymbol{\vep}^*_\lambda\cdot
        \pp}{\pp^2}+\frac{\boldsymbol{\vep}^*_\lambda\cdot
        \qp}{\qp^2}\right] \longrightarrow
    \\
    &2 g_s \llbracket W\rrbracket^{a b} (\qp+\pp)
     \left[\frac{\boldsymbol{\vep}^*_\lambda\cdot
        \pp}{\pp^2}+\frac{\boldsymbol{\vep}^*_\lambda\cdot
         \qp}{\qp^2}\right]
     \wilsonV_{\lambda}(\pp,\qp).
     \label{eq:cevW}
    \end{split}
\end{equation}
We conclude by reporting some of the main features of the $W$ field,
once again referring the reader to ref.~\cite{Caron-Huot:2013fea} for additional details.  Up to two loops, $W$ is an eigenstate of the Balitsky/JIMWLK
rapidity evolution, and its eigenvalue coincides with (minus) the gluon Regge
trajectory $\tau_g$:
\begin{equation}
  \label{eq:full_regge_trajectory_def}
  -\frac{\mathd}{\mathd\eta} W_{\eta}({\pp}) = \tau_g(\pp) W_{\eta}(\pp).
\end{equation}
Explicit results for $\tau_g$ up to two loops are reported in
\cref{se:hard_functions}. We note that the rapidity evolution
also induces transitions between $n$ and $n+2$, $n+4$ etc. states, but
such transitions are suppressed by at least one power of $\as$ compared
to the $n\to n$ ones, and do not enter in our analysis.

At LO the $W$ fields are free, so that their same-rapidity
correlator is
\begin{equation}\label{eq:vev_1}
\begin{aligned}
    \Big\langle  
    \timeOrd 
    \big[W(\pp_1) \cdots & W(\pp_n)\big]_\eta
    \big[\Wb(\qp_1) \cdots \Wb(\qp_m)\big]_\eta
    \Big\rangle = \\
    &\delta_{nm}\sum_{\sigma \in S_n} G(\pp_1,\qp_{\sigma(1)}) \dots G(\pp_n,\qp_{\sigma(n)})  + \mathcal{O}(\as) ,
\end{aligned}
\end{equation}
where $\timeOrd$ is the standard time ordering,
$\Wb$ is associated with projectile $B$, 
and the 2-point correlator reads
\begin{equation}\label{eq:vev_0}
G(\pp,\qp)  =
\Big\langle   \timeOrd\,
W^a_\eta(\pp) \Wb^b_\eta(\qp) 
\Big\rangle  = 
(2 \pi)^{2-2\epsilon} \delta^{2-2\epsilon}(\pp-\qp) \frac{i \delta^{ab}}{\pp^2} + \mathcal{O}(\as).
\end{equation}
Due to CPT invariance, the correlator between even and odd numbers of
$W$ fields is zero at any order ($W$ is a signature-odd
field)~\cite{Caron-Huot:2013fea}. Beyond LO, it is convenient to
redefine the $W$ field
\begin{equation}    
W \to (1+\alpha_s r_1 +  \dots )W + (\alpha_s^2 s_1 + \dots )WWW + \dots ,
\end{equation}
such that $a)$ correlators with different numbers of $W$ fields on the two
lightcones vanish,
\begin{equation}\label{eq:vev_2}
    \Big\langle  
    \timeOrd 
    \big[W(\pp_1) \cdots  W(\pp_n)\big]_\eta
    \big[\Wb(\qp_1) \cdots \Wb(\qp_m)\big]_\eta
    \Big\rangle = 0 \quad \quad  \text{for} \quad n \neq m,
\end{equation}
and $b)$ the two-point correlator is equal to the free one
\begin{equation}\label{eq:vev_3}
G(\pp,\qp) = \Big\langle \timeOrd\, W^a_\eta(\pp) \Wb^b_\eta(\qp)
\Big\rangle = (2 \pi)^{2-2\epsilon} \delta^{2-2\epsilon}(\pp-\qp)
\frac{i \delta^{ab}}{\pp^2},
\end{equation}
at any order in $\as$. 
Note that after this redefinition \cref{eq:W_def} would not hold anymore, but
\cref{eq:projectile_W_exp,eq:full_regge_trajectory_def} still hold
(with modified impact factors, but unchanged gluon Regge trajectory)
and our computation remains identical to the required order.

These are all the details that we need to predict the MRK behaviour of
the 5-point scattering amplitude up to two loops. We study this in the
next subsections. We start by discussing the $(--)$ signature, which is
the only one that contributes at LL and NLL accuracy at the cross-section
level, see e.g. ref.~\cite{BARTELS1980365}.

\subsection{LL and NLL predictions for the MRK amplitudes}
\label{se:LL&NLL}
\subsubsection{LL and NLL predictions for the $(--)$ signature}
To obtain a prediction for the LL amplitude in the formalism
summarised in the previous section, we recall that large logarithms
are only generated by the rapidity evolution.  Because of this, and
because multiple $W$ contributions are suppressed by powers of $g_s$,
cf. \cref{eq:projectile_W_exp}, only the single-$W$ term in
\cref{eq:projectile_W_exp} contributes at LL. 
For the same reason, no perturbative corrections to the impact
factors enter in the LL approximation, \ie $\IFpre_X \to 1$.
Up to an overall -- hence immaterial -- phase $\phi^{[AB]}$ that
depends on spinor conventions, the connected $S$-matrix
element~\cref{eq:S_to_A} that we need to consider is then
\begin{equation}
\begin{aligned} \label{eq:WtoW_1}
  \mathcal{S}_\mathrm{LL} &=
  \phi^{[AB]}
  \left[2\pi \delta(p_1^+ - p_5^+) \delta_{\lambda_1\lambda_5}
    \times 2 p_1^+\right]
    \left[2\pi \delta(p_2^- - p_3^-) \delta_{\lambda_2\lambda_3}
      \times 2 p_2^-\right]
    \\
  & \times     
    \bigg\langle \timeOrd \,
    \Big(ig_s \llbracket
      W_{\eta_A}(\qp_A)\rrbracket^{c_5 c_1}_{r_A}\Big)
      a^{a_4}_{\lambda_4}(p_4)
      \Big(
      ig_s\llbracket
      \Wb_{\eta_B}(\qp_B)
      \rrbracket^{c_3 c_2}_{r_B} \Big)
    \bigg\rangle \, .
\end{aligned}
\end{equation}
To evaluate this correlator, we first use
\cref{eq:full_regge_trajectory_def} to evolve the $W_{\eta_A}(\qp_A)$
field to central rapidity
\begin{equation}
  W^a_{\eta_A}(\qp_A) = e^{\Delta\eta_{A4} \tau_g(\qp_A)} W^a_{\eta_4}(\qp_A),
  \label{eq:wevol}
\end{equation}
where at LL we only require the LO (i.e. $\mathcal O(\as)$) Regge trajectory.
It reads
\begin{equation}\label{eq:tau_one_loop}
  \left(\frac{\as}{4\pi}\right)\tau^{(1)}_g(\qp) =
  \left(\frac{\as}{4\pi}\right)
   2 N_c \frac{e^{\ep\gamma_E}
       \Gamma(1-\ep)^2\Gamma(1+\ep)}{\Gamma(1-2\ep)\ep}\left(\frac{\mu^2}{\qp^2}\right)^\ep,
\end{equation}
with $\gamma_E$ the Eulero-Mascheroni constant, $\gamma_E \approx 0.577216$.
At small $x$, the rapidity difference can be written as
\begin{equation}
  \Delta\eta_{ij} = \eta_i-\eta_j =
  \ln\frac{|s_{ij}|}{|\pp_i| |\pp_j|} + \mathcal O(x).
\end{equation}
We remind the reader that $\eta_A = 1/2 \ln p_5^+/p_5^-$ and
$\eta_B = 1/2 \ln p_3^+/p_3^-$.
As a next step, we apply \cref{eq:ope1_0} to compute the LO
interaction of $W^a_{\eta_4}(\qp_A)$ with the emitted gluon
\begin{equation}
  W^a_{\eta_4}(\qp_A) a_{\lambda_4}^{a_4}(p_4) \to 2 g_s \llbracket
  W\rrbracket^{a_4 a}_{\eta_4} (\qp_A+\pp_4)
  \left[\frac{\boldsymbol{\vep}^*_{\lambda_4}\cdot
      \pp_4}{\pp_4^2}+\frac{\boldsymbol{\vep}^*_{\lambda_4}\cdot
      \qp_A}{\qp_A^2}\right] + \dots,
\end{equation}
and finally evolve the resulting $W$ field from rapidity $\eta_4$ to
rapidity $\eta_B$, and compute the equal-rapidity correlator
\cref{eq:vev_3}.
Using \cref{eq:S_to_A}, we can then write a LL prediction for the
amplitude:\footnote{We note that with our definitions for the lightcone
$\mathrm d p^+ \mathrm d p^- = 2 \mathrm dp^0 \mathrm dp^z$.}
\begin{equation}
\begin{aligned}
  \mathcal{A}_\LL =
  \mathcal{A}^{(--)}_\LL =
  \phi^{[AB]}& \, 4 g_s^3 s_{12}
  \left[(T_{r_B})_{c_3 c_2}^b (T_{r_A})_{c_5 c_1}^a if^{aba_4}\right]
        \times \\ &e^{\tau_g(\qp_A) \Delta\eta_{A4}
        }
        \frac{1}{\qp_B^2}\left[\frac{\boldsymbol{\vep}^*_{\lambda_4}\cdot
            \pp_4}{\pp_4^2}+\frac{\boldsymbol{\vep}^*_{\lambda_4}\cdot
            \qp_A}{\qp_A^2}\right] e^{\tau_g(\qp_B) \Delta\eta_{4B}}.
\end{aligned}
\end{equation}
Using the explicit results for the polarisation vectors in
\cref{se:spinorproducts}, the term in the square bracket can be written as
\begin{equation}
  \left[\frac{\boldsymbol{\vep}^*_{\lambda_4}\cdot
      \pp_4}{\pp_4^2}+\frac{\boldsymbol{\vep}^*_{\lambda_4}\cdot
      \qp_A}{\qp_A^2}\right] = \frac{1}{\sqrt2}\times
  V_{\lambda_4}(\qp_A,\pp_4) \times\frac{1}{\qp_A^2},
\end{equation}
with
\begin{equation}
  V_+(\qp_A,\pp_4) = \frac{\bar q_{A,\perp}\,
    q_{B,\perp}}{p_{4,\perp}},
  \quad\quad V_-(\qp_A,\pp_4) =
  \frac{q_{A,\perp} \,\bar q_{B,\perp}}{\bar p_{4,\perp}},
  \label{eq:CEV0}
\end{equation}
where we used the complex notation $p_\perp = p^x + i p^y$, $\bar
p_\perp = p^x - i p^y$, see \cref{se:spinorproducts}. The function
$V_\lambda$ in \cref{eq:CEV0} is the LO Lipatov vertex, or central
emission vertex, which is a critical ingredient for computing
amplitudes in the MRK, see e.g. ref.~\cite{DelDuca:2022skz} and
references therein.

In terms of the Lipatov vertex, the LL amplitude
can be written in the suggestive form
\begin{equation}
  \begin{split}
    &\mathcal{A}^{(--)}_{\LL} = \phi^{[AB]} \, 2\sqrt{2}\,g_s^3 s_{12}
    \\
  &\quad\times
  (T_{r_A})_{c_5 c_1}^a
  e^{\tau_g(\qp_A) \Delta\eta_{A4} } \frac{1}{\qp_A^2}
  \left[if^{aba_4} V_{\lambda_4}(\qp_A,\pp_4)\right]
    \frac{1}{\qp_B^2} e^{\tau_g(\qp_B) \Delta\eta_{4B}}(T_{r_B})_{c_3 c_2}^b ,
  \end{split}
  \label{eq:myLL}
\end{equation}
where the $t$-channel exchange structure is apparent. Repeating the
same steps we performed here for the $2\to n$ amplitude in MRK, it is
straightforward to see that such a structure iterates at all
multiplicities.  We briefly discuss this in \cref{se:redef}, but
first we give explicit expressions for the phase factors
$\phi^{[AB]}$. These depend on the spinor conventions adopted.  To
fix them, we match \cref{eq:myLL} against \cref{eq:uv_ren_ampls},
remembering that in MRK only the signature-odd amplitude with
colour-octet exchange in both the 1--5 and 2--3 channels is non vanishing
at LO, see \cref{eq:LOH}.  Using the spinor conventions of
\cref{se:spinorproducts}, for the helicity choices in
\cref{eq:spinorfactors} we obtain 
\begin{equation}
  \phi^{[gg]} = \frac{\bar p_{3,\perp}}{p_{3,\perp}},
  \quad
  \phi^{[qg]} = -i \frac{\bar p_{3,\perp}}{p_{3,\perp}},
  \quad
  \phi^{[qQ]} = \frac{\bar p_{3,\perp}}{|p_{3,\perp}|}.
\end{equation}
The same procedure can be applied to all other helicites. Since the
$\phi^{[AB]}$ phase is immaterial, we refrain from reporting here
explicit formulas for all the other possible helicity configurations.

We now discuss the NLL generalisation of \cref{eq:myLL}.
As we have mentioned in \cref{se:review},
$W$ is a signature-odd field. As a consequence, the $(--)$
amplitudes only receives contributions
when an odd number of $W$ is emitted from both the $A$ and $B$
lines. Simple power counting then shows that only one-$W$ exchanges
arise at NLL, making the NLL generalisation of the LL results above
 straightforward.  Indeed, the factorised form of
the amplitude \cref{eq:myLL} remains basically unchanged, except for
the addition of (multiplicative) radiative corrections for the impact
factors $\IFpre$ \cref{eq:projectile_W_exp} and the $WWg$ interaction
vertex $\wilsonV_\lambda$ \cref{eq:cevW}. The NLL amplitude then reads
\begin{equation}
  \begin{split}
    &\mathcal{A}^{(--)}_{\NLL} = \phi^{[AB]} \, 2\sqrt{2}\,g_s^3 s_{12}
    \times
    (T_{r_A})_{c_5 c_1}^a \IFpre(\qp_A)
    e^{\tau_g(\qp_A) \Delta\eta_{A4} }
    \\
  &\quad\times
 \frac{1}{\qp_A^2}
 \left[if^{aba_4} V_{\lambda_4}(\qp_A,\pp_4)
 \wilsonV_{\lambda_4}(\qp_A,\pp_4)\right]
  \frac{1}{\qp_B^2} e^{\tau_g(\qp_B) \Delta\eta_{4B}}
  \IFpre(\qp_B)
  (T_{r_B})_{c_3 c_2}^b .
  \end{split}
  \label{eq:myNLL}
\end{equation}
At this order, one only needs $\mathcal O(\as)$ corrections to the
impact factors $\IFpre$ and vertex $\mathcal W$, but two-loop
$\mathcal O(\as^2)$ corrections to the gluon Regge trajectory $\tau_g$.
Both $\IFpre$ and the Regge trajectory can be extracted from the
$2\to 2$ scattering amplitude, thus allowing one to obtain the one-loop
contribution to 
$\mathcal W_\lambda$ by matching \cref{eq:myNLL} to the one-loop $2\to 3$
amplitude. To compare NLL results with the literature, we find it
convenient to perform a redefinition of the evolution variable, impact
factors $\IFpre$, and vertex $\wilsonV$. However, before elaborating on this,
we discuss NLL predictions for the other signatures at one loop.

\subsubsection{One-loop NLL predictions for the other signatures}
\label{sec:NLLother}

\begin{figure}[t!]
    \centering
    \subfigure[]{\scalebox{0.78}{\includegraphics{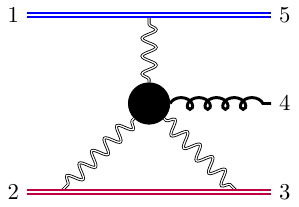}} \label{fig:WtoWW}}
    \subfigure[]{\scalebox{0.78}{\includegraphics{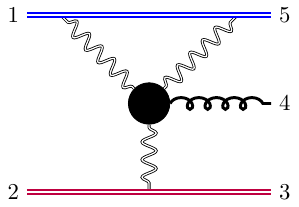}} \label{fig:WWtoW}}
    \subfigure[]{\scalebox{0.78}{\includegraphics{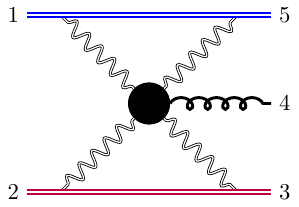}} \label{fig:WWtoWW}}
    \caption{Schematic diagrams for multi-$W$ contributions to
      the $(-+)$, $(+-)$, and $(++)$ one-loop amplitudes.
\label{fig:regge_cuts_even}}
\end{figure}

We consider one-loop amplitudes with signatures $(-+)$, $(+-)$ and
$(++)$ up to NLL, starting from the former.  The $(-+)$ amplitude only
receives contributions where an odd number of $W$ fields are emitted
from the $A$ line, and an even one from the $B$ line.  Since each
$W$ is accompanied by a $g_s$ factor without any large logarithm, see
\cref{eq:projectile_W_exp}, at NLL we need to consider the emission of
one $W$ from the $A$ and two $W$s from the $B$ line.  The rapidity
evolution is signature preserving (see the discussion around
\cref{eq:full_regge_trajectory_def}), hence the only non-vanishing
contribution at NLL comes from the $W\to WW$ transition, cf. the
second line of \cref{eq:ope1_0} and \cref{fig:WtoWW}.  Crucially, this
is $\mathcal O(g_s)$ with respect to the LL result. As a consequence,
to NLL accuracy the $W$ fields can be treated at LO. This makes NLL
predictions for the $(-+)$ signature conceptually trivial. The only
subtlety is that the two-$W$ state is not an eigenstate of the
rapidity evolution Hamiltonian, hence a simple exponentiation like in
\cref{eq:wevol} does not hold. However, the effect of the evolution
starts appearing at the two-loop ($\as^2 L$) order. Here we focus on
the one-loop amplitude, the evolution then does not play any role and
we can immediately write down the NLL result.

By combining the appropriate terms in
\cref{eq:projectile_W_exp,eq:ope1_0}, we find
\begin{equation}
\begin{aligned} \label{eq:NLLmp}
  \mathcal{S}&^{(-+)}_\mathrm{NLL} =\\
  & =\phi^{[AB]}
  \left[2\pi \delta(p_1^+ - p_5^+) \delta_{\lambda_1\lambda_5}
    \times 2p_1^+\right]
    \left[2\pi \delta(p_2^- - p_3^-) \delta_{\lambda_2\lambda_3}
      \times 2p_2^-\right]
    \\
  & \times     
    \Bigg\langle \timeOrd \,
    \Big(ig_s \llbracket
      W_{\eta_A}(\qp_A)\rrbracket^{c_5 c_1}_{r_A}\Big)
      a^{a_4}_{\lambda_4}(p_4)
      \Bigg(
      \frac{(ig_s)^2}{2!}\llbracket
      \Wb_{\eta_B}\otimes\Wb_{\eta_B}
      \rrbracket^{c_3 c_2}_{r_B}(\qp_B) \Bigg)
      \Bigg\rangle \, =
      \\
     &=
      \phi^{[AB]}
      \left[2\pi \delta(p_1^+ - p_5^+) \delta_{\lambda_1\lambda_5}\right]
      \left[2\pi \delta(p_2^- - p_3^-) \delta_{\lambda_2\lambda_3}
      \right]\times
      4 s_{12} 
      \\
      &\times
      \left[ig_s (T_{r_A}^a)_{c_5 c_1}\right]\times
      i g_s^2  \int \measF{\kp_1}
      \left[
        \frac{\boldsymbol{\vep}^*_\lambda\cdot \qp_A}{\qp_A^2} + \frac{\boldsymbol{\vep}^*_\lambda\cdot (\kp_1-\qp_A)}{(\kp_1-\qp_A)^2}
        \right]
      \\
      &\times
      \Bigg\langle \timeOrd \,
      \llbracket W_{\eta_4}(\qp_A+\pp_4-\kp_1) W_{\eta_4}(\kp_1) \rrbracket^{a_4 a}
      \Bigg(
      \frac{(ig_s)^2}{2!}\llbracket
      \Wb_{\eta_B}\otimes\Wb_{\eta_B}
      \rrbracket^{c_3 c_2}_{r_B}(\qp_B) \Bigg)
      \Bigg\rangle.
\end{aligned}
\end{equation}
The equivalent result for the amplitude, see \cref{eq:S_to_A}, reads
\begin{equation}
 \begin{split}
  \mathcal A_{\NLL}^{(-+)} & =
      4 i g_s^3 \phi^{[AB]}
      s_{12}
      \left[
        (T_{r_A})_{c_5 c_1}^a if^{a_4 a_1 e} if^{e a_2 a} (\delta_{a_1 b_1} \delta_{a_2 b_2} + \delta_{a_1 b_2} \delta_{a_2 b_1})
        (T_{r_B}^{b_1} \gcdot T_{r_B}^{b_2})_{c_3 c_2}
        \right]
        \\
        &\times
        \frac{g_s^2}{4}
        \int \measF{\kp_1}
      \frac{1}{(\qp_B-\kp_1)^2} \frac{1}{\kp_1^2} 
      \left[
        \frac{\boldsymbol{\vep}^*_\lambda\cdot \qp_A}{\qp_A^2} + \frac{\boldsymbol{\vep}^*_\lambda\cdot (\kp_1-\qp_A)}{(\kp_1-\qp_A)^2}
        \right] \big(1+\mathcal O(\as)\big).
 \end{split}
 \label{eq:myANLLmp}
\end{equation}

We now look at the colour structure of \cref{eq:myANLLmp}. To do so, we
introduce the graphic notation
\begin{equation}
  \includegraphics[valign=m]{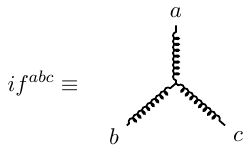} , \quad\quad \includegraphics[valign=m]{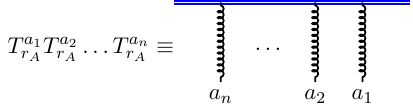}\,,
\end{equation}
in terms of which we can write the Lie algebra identity
\begin{equation}\label{eq:lie_algebra}
   [T^a_{r_A} , T^b_{r_A}] = i f^{abc} T_{r_A}^c \rightarrow \includegraphics[valign=m]{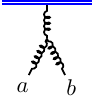} = \includegraphics[valign=m]{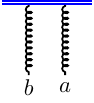} - \includegraphics[valign=m]{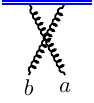}.
\end{equation}
In this notation, the colour structure of \cref{eq:myANLLmp} can be written as
\begin{equation}
        T_{r_A}^a if^{a a_2 e} if^{e a_1 a_4} (\delta_{a_1 b_1}
        \delta_{a_2 b_2} + \delta_{a_1 b_2} \delta_{a_2 b_1})
        (T_{r_B}^{b_1} \gcdot T_{r_B}^{b_2}) \rightarrow \includegraphics[valign=m]{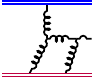} +
        \includegraphics[valign=m]{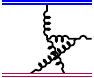} .
\end{equation}
After simple manipulations, using \cref{eq:lie_algebra} we obtain
\begin{equation} 
\begin{aligned}
    \includegraphics[valign=m]{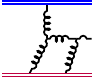} 
        +
    \includegraphics[valign=m]{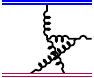}
    &=
    -
    \includegraphics[valign=m]{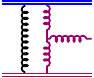}
    +
    \includegraphics[valign=m]{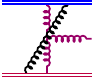}
    +
    \includegraphics[valign=m]{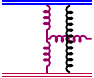}
    -
	\includegraphics[valign=m]{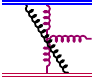}     
     \\
    &= 
    -\boldsymbol{\mathcal{T}}_{+-} \,  \includegraphics[valign=m]{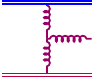},
\end{aligned}
\label{eq:colourMP}
\end{equation}
where we have coloured in purple the underlying tree-level colour
structure and identified the remaining black gluons in terms of the
colour operators ${\bf T}^{\pm}_{ij}$ in \cref{eq:titjpm}. We then
combined these into $\boldsymbol{\mathcal{T}}_{+-}$, defined in
\cref{eq:colcombinations}.

Using \cref{eq:colourMP}, we can write the amplitude \cref{eq:myANLLmp}
as
\begin{equation}
  \begin{split}
  \mathcal A_{\NLL}^{(-+)} & = -i \pi
  \left[\frac{g_s^2 S_\ep}{(4\pi)^2}\right]
  \twoDint^{(1)}_{\{1,2\}} \boldsymbol{\mathcal{T}}_{+-}
  \mathcal A^{(0)}\left(1 + \mathcal O(\as)\right) =
  \\
  & =
  - i\pi \left(\frac{\as}{4\pi}\right)
  \twoDint^{(1)}_{\{1,2\}} \boldsymbol{\mathcal{T}}_{+-}
  \mathcal A^{(0)}
  \left(1 + \mathcal O(\as)\right),
  \end{split}
\end{equation}
or
\begin{equation}
  \mathcal A_{\NLL}^{(1),(-+)}  =
  - i\pi 
    \twoDint^{(1)}_{\{1,2\}} \boldsymbol{\mathcal{T}}_{+-}
    \mathcal A^{(0)}.
    \label{eq:myNLLmp}
\end{equation}
In these equations, $S_\ep = (4\pi)^{\ep} e^{-\ep \gamma_E}$ is the
standard $\MSbar$ renormalisation factor,
\begin{equation}
  \twoDint^{(1)}_{\{1,2\}} =
  \left[\twoDint^{(0)}_{\{1,1\}}\right]^{-1}
  \int 
  \frac{\measJ{\kp_1}}{\kp_1^2 (\qp_B-\kp_1)^2}
  \left[
    \frac{\boldsymbol{\vep}^*_\lambda\cdot \qp_A}{\qp_A^2} + \frac{\boldsymbol{\vep}^*_\lambda\cdot (\kp_1-\qp_A)}{(\kp_1-\qp_A)^2}
    \right],
\end{equation}
with $\mathfrak{D}p = (\mu^{2\ep} e^{\ep \gamma_E}
/\pi^{1-\ep})\mathrm{d}^{2-2\ep}p$, and
\begin{equation}
  \twoDint^{(0)}_{\{1,1\}} =
  \frac{1}{\qp_B^2}\left[\frac{\boldsymbol{\vep}^*_{\lambda_4}\cdot
    \pp_4}{\pp_4^2}+\frac{\boldsymbol{\vep}^*_{\lambda_4}\cdot
    \qp_A}{\qp_A^2}\right].
\end{equation}
We note that all the functions $\twoDint^{(l)}_{\{i,j\}}$ depend on
the gluon polarisation $\lambda_4$. We have left this dependence
implicit. We discuss these  functions in more
detail in \cref{se:twodimintegrals}; here we limit ourselves to say
that they are pure (i.e. they have no rational prefactors), of uniform transcendental 
weight, and that the $\lambda_4$ dependence starts at $\mathcal
O(\ep)$.  Up to finite terms, for both positive and negative gluon
helicities we obtain
\begin{equation}
  \twoDint^{(1)}_{\{1,2\}} = -\frac{1}{\ep} +
  \ln\left(\frac{\qp_B^2\pp_4^2}{\qp_A^2\mu^2}\right) + \mathcal O(\ep).
  \label{eq:K1_12_ep0}  
\end{equation}
Generic expressions at higher orders in $\ep$ can be found in
\cref{se:twodimintegrals}.

We can repeat the same procedure to obtain the $(+-)$ one-loop
amplitude, with the important difference that the rapidity evolution
and $W$/gluon expansion have to be performed from $B$ to $A$,
\cref{fig:WWtoW}.  We remind the reader that the formulas for
$W$/gluon interactions in \cref{se:review} have been obtained by
making a reference choice for the emitted gluon such that its
polarisation vector does not have components on the large lightcone
direction, $\vep_4\cdot p_2 = 0$ for the evolution from $A$ to $B$. If
we evolve instead from $B$ to $A$, we can use the same formulas but
have now to impose $\vep_4\cdot p_1 = 0$.  To differentiate
polarisation vectors with different reference choices, we refer to the
transverse components in the $\vep\cdot p_2=0$ case by
${\boldsymbol{\vep}}$ and to those in the $\vep\cdot p_1=0$ by
$\boldsymbol{\vepT}$, see \cref{se:spinorproducts} for more details and
explicit representations. 

Apart from this caveat the calculation proceeds
exactly like in the $(-+)$ case, and we obtain
\begin{equation}
  \mathcal A^{(1),(+-)}_{\NLL} = -i\pi \twoDint^{(1)}_{\{2,1\}}
  \boldsymbol{\mathcal{T}}_{-+}
  \mathcal A^{(0)},
  \label{eq:myNLLpm}
\end{equation}
with now
\begin{equation}
  \twoDint^{(1)}_{\{2,1\}} =
  \left[\widetilde{\twoDint}^{(0)}_{\{1,1\}}\right]^{-1}
  \int 
  \frac{\measJ{\kp_1}}{\kp_1^2 (\qp_A+\kp_1)^2}
  \left[
    -\frac{\boldsymbol{\vepT}^*_\lambda\cdot \qp_B}{\qp_B^2} + \frac{\boldsymbol{\vepT}^*_\lambda\cdot (\kp_1+\qp_B)}{(\kp_1+\qp_B)^2}
    \right],
  \label{eq:K1_21}
\end{equation}
and
\begin{equation}
  \widetilde{\twoDint}^{(0)}_{\{1,1\}} =
  \frac{1}{\qp_A^2}\left[\frac{\boldsymbol{\vepT}^*_{\lambda_4}\cdot
    \pp_4}{\pp_4^2}-\frac{\boldsymbol{\vepT}^*_{\lambda_4}\cdot
    \qp_B}{\qp_B^2}\right].
\end{equation}
Up to finite terms, \cref{eq:K1_21} reads
\begin{equation}
  \twoDint^{(1)}_{\{2,1\}} = -\frac{1}{\ep} +
  \ln\left(\frac{\qp_A^2\pp_4^2}{\qp_B^2\mu^2}\right) + \mathcal O(\ep).
  \label{eq:K1_21_ep0}
\end{equation}
We note that the result \cref{eq:myNLLpm} can also be immediately
obtained from \cref{eq:myNLLmp} by simply exchanging the two
lightcones.  As discussed at the end of \cref{se:multireggekin}, this
amounts to the kinematics exchange $z\leftrightarrow 1-\zbar$ (which
in \cref{eq:K1_12_ep0,eq:K1_21_ep0} simply amounts to the $\pp_3
\leftrightarrow \pp_5$ exchange), followed by an appropriate
permutation in the colour operators.

The last signature we need to consider is $(++)$, see \cref{fig:WWtoWW}.
We write the connected $S$-matrix for the emission of two $W$ from each of
the $A$ and $B$ lines as
\begin{equation}
\begin{aligned} \label{eq:NLLpp}
  \mathcal{S}&^{(++)}_\mathrm{NLL} = \\
  &
  = \phi^{[AB]}
  \left[2\pi \delta(p_1^+ - p_5^+) \delta_{\lambda_1\lambda_5}
    \times 2p_1^+\right]
    \left[2\pi \delta(p_2^- - p_3^-) \delta_{\lambda_2\lambda_3}
      \times 2 p_2^-\right]
    \\
  & \times     
    \Bigg\langle \timeOrd \,
    \Bigg(
      \frac{(i g_s)^2}{2!}\llbracket
      W_{\eta_A}\!\otimes\!W_{\eta_A}
      \rrbracket^{c_5 c_1}_{r_A}(\qp_A) \Bigg)
      a^{a_4}_{\lambda_4}(p_4)
      \Bigg(
      \frac{(i g_s)^2}{2!}\llbracket
      \Wb_{\eta_B}\!\otimes\!\Wb_{\eta_B}
      \rrbracket^{c_3 c_2}_{r_B}(\qp_B) \Bigg)
      \Bigg\rangle \, =
      \\
     &=
      \phi^{[AB]}
      \left[2\pi \delta(p_1^+ - p_5^+) \delta_{\lambda_1\lambda_5}\right]
      \left[2\pi \delta(p_2^- - p_3^-) \delta_{\lambda_2\lambda_3}
      \right]\times
      4 s_{12} 
      \\
      &\times
      \left[\frac{(i g_s)^2}{2!} \left(T_{r_A}^{(d_1}\gcdot T_{r_A}^{d_2)}
        \right)_{c_5 c_1} \right]\times
      2g_s  \int \measF{\kp_1}
      \left[
        \frac{\boldsymbol{\vep}^*_\lambda\cdot \pp_4}{\pp_4^2} + \frac{\boldsymbol{\vep}^*_\lambda\cdot(\qp_A-\kp_1)}{(\qp_A-\kp_1)^2}
        \right]
      \\
      &\times
      \Bigg\langle \timeOrd \,
       \llbracket W_{\eta_4} (\qp_A+\pp_4-\kp_1) \rrbracket^{a_4 d_1}W_{\eta_4}^{d_2}(\kp_1) 
      \Bigg(
      \frac{(i g_s)^2}{2!}\llbracket
      \Wb_{\eta_B}\otimes\Wb_{\eta_B}
      \rrbracket^{c_3 c_2}_{r_B}(\qp_B) \Bigg)
      \Bigg\rangle,
\end{aligned}
\end{equation}
where we used the notation $T_r^{(i_1}\gcdot T_r^{i_2)} =
T_r^{i_1}\gcdot T_r^{i_2} + T_r^{i_2}\gcdot T_r^{i_1}$.
This leads to the one-loop amplitude
\begin{equation}
  \begin{split}
    \mathcal A^{(++)}_{\NLL} 
    & =
      4 i g_s^3 \phi^{[AB]}
      s_{12}
      \left[
        \left(T_{r_A}^{(d_1}\gcdot T_{r_A}^{d_2)}\right)_{c_5 c_1}
        i f^{a_4 b_1 d_1}
        \left(T_{r_B}^{(b_1}\gcdot T_{r_B}^{d_2)}\right)_{c_3 c_2}
        \right]
        \\
        &\times
        \frac{g_s^2}{4}
        \int \measF{\kp_1}
      \frac{1}{(\qp_B-\kp_1)^2} \frac{1}{\kp_1^2} 
      \left[
        \frac{\boldsymbol{\vep}^*_\lambda\cdot \pp_4}{\pp_4^2} + \frac{\boldsymbol{\vep}^*_\lambda\cdot (\qp_A-\kp_1)}{(\qp_A-\kp_1)^2}
        \right] \big(1+\mathcal O(\as)\big).
 \label{eq:myANLLpp}
  \end{split}
\end{equation}
The colour algebra now reads
\begin{align}
  \left(T_{r_A}^{(d_1}\gcdot T_{r_A}^{d_2)}\right)_{c_5 c_1}
  i f^{a_4 b_1 d_1}
  \left(T_{r_B}^{(b_1}\gcdot T_{r_B}^{d_2)}\right)_{c_3 c_2}
  &=  -\includegraphics[valign=m]{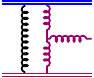}
    +
    \includegraphics[valign=m]{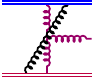}
    -
    \includegraphics[valign=m]{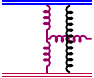}
    +
    \includegraphics[valign=m]{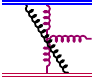} \notag\\
    &= -\boldsymbol{\mathcal{T}}_{--} \includegraphics[valign=m]{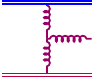} \,,
\end{align}
which allows us to write
\begin{equation}
  \mathcal A_{\NLL}^{(1),(++)} = -i\pi \twoDint^{(1)}_{\{2,2\}}
  \boldsymbol{\mathcal{T}}_{--} \,\mathcal A^{(0)},
  \label{eq:myNLLpp}
\end{equation}
with now
\begin{equation}
  \begin{split}
    \twoDint^{(1)}_{\{2,2\}} & =
    \left[\twoDint^{(0)}_{\{1,1\}}\right]^{-1}
    \int
    \frac{\measJ{\kp_1}}{\kp_1^2(\qp_B-\kp_1)^2} 
    \left[
      \frac{\boldsymbol{\vep}^*_\lambda\cdot \pp_4}{\pp_4^2} + \frac{\boldsymbol{\vep}^*_\lambda\cdot (\qp_A-\kp_1)}{(\qp_A-\kp_1)^2}
      \right] = \\
    & = -\frac{1}{\ep} + \ln\left(\frac{\qp_A^2\qp_B^2}{\pp_4^2\mu^2}\right)
      +\mathcal O(\ep).
  \end{split}
\end{equation}
We note that the result \cref{eq:myNLLpp} is symmetric
under $A$ and $B$ exchange, as expected. Results for $\twoDint^{(1)}_{\{2,2\}}$
at higher orders in $\ep$ are given in \cref{se:twodimintegrals}.
This concludes our discussion of NLO amplitudes. As we will discuss in
\cref{se:results}, these predictions can be directly compared to the
one-loop results obtained in \cref{se:amplexpansion}.

\subsection{NLL generalisation to $n$-point scattering and
  evolution redefinition}
\label{se:redef}
Starting from NLL, we find it convenient to slightly redefine the
evolution variable $\Delta\eta_{ij}$, the impact factors $\IFpre$, and
the vertex correction $\wilsonV_\lambda$.  Such a redefinition is immaterial for
the final result, but it allows us to make contact with the Regge
formalism. To explain how this comes about, we first note that also at NLL the
$t$-channel structure of \cref{eq:myNLL} iterates, 
for the same reasons
discussed in the previous subsection.
\begin{figure}[t!]
    \centering
    \subfigure{\scalebox{0.9}{\includegraphics[valign=m]{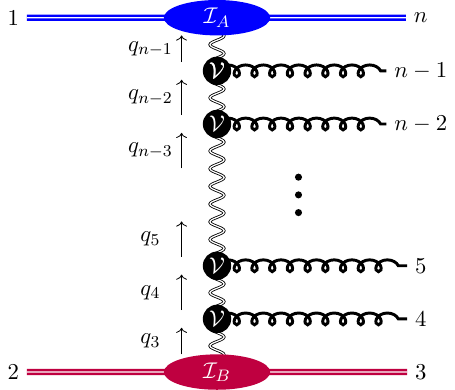}} \label{fig:TWOtoN}}
    \hspace{2mm}
    \subfigure{\scalebox{1}{\includegraphics[valign=m]{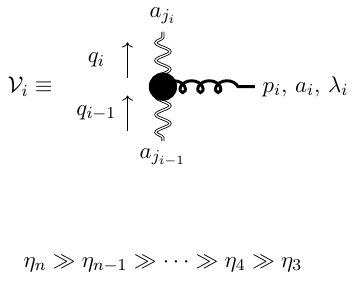}} \label{fig:vertexDef}}
\caption{Schematic depiction of $n$-point MRK odd signature scattering amplitudes.  
\label{fig:regge_pole_2ton}}
\end{figure}
Schematically, the signature-odd $n$-point NLL amplitude reads (see \cref{fig:regge_pole_2ton})
\begin{equation}
  \begin{split}
    \label{eq:myNLLn}
    \mathcal A^{(--)}_{\NLL} \propto & \quad (T_{r_A})_{c_n c_1}^{a_{n-1}} \IFpre_A(\qp_{n-1}) \times\\
  & 
  \frac{e^{\tau_{n-1}\Delta\eta_{n,n-1}}}{\qp_{n-1}^2}
    \mathcal V_{n-1}
    \frac{e^{\tau_{n-2}\Delta\eta_{n-1,n-2}}}{\qp_{n-2}^2}\cdots
      \frac{e^{\tau_{4}\Delta\eta_{5,4}}}{\qp_{4}^2}
        \mathcal V_{4}
        \frac{e^{\tau_{3}\Delta\eta_{4,3}}}{\qp_{3}^2}
        \times
        \IFpre_B(\qp_3) (T_{r_B})_{c_3 c_2}^{a_3} , 
  \end{split}
\end{equation}
where we have omitted irrelevant overall prefactors and used the shorthands
\begin{equation} \label{eq:short_tau_V}
    \tau_i = \tau_g(\qp_i), \quad\quad 
    \mathcal V_i = i f^{a_{j_i}
    a_{j_{i-1}} a_i}  V_{\lambda_i} (\qp_{i},\pp_i)
    \wilsonV_{\lambda_i} (\qp_{i},\pp_i).
\end{equation}
In order to improve readability, from this point onwards we will drop 
the dependence on the two-dimensional momenta from the impact factors and the vertex correction.
This implicit dependence can be easily and unambiguously inferred from their indices.

We now write the rapidity differences as
\begin{equation}\label{eq:fromETAtoL}
  \begin{split}
    \Delta\eta_{ij} = \ln\frac{|s_{ij}|}{|\pp_i||\pp_j|} &=
    \ln\frac{|s_{ij}|}{\mu^2_{ij}} + \frac{1}{2} \ln\frac{\mu_{ij}^2}{\pp_i^2}
    + \frac{1}{2} \ln\frac{\mu_{ij}^2}{\pp_j^2} =
    \\
    &=
    L_{ij} + \frac{1}{2} \left(\ln\frac{\mu_{ij}^2}{\pp_i^2} + \frac{i\pi}{2}\right)
    + \frac{1}{2} \left(\ln\frac{\mu_{ij}^2}{\pp_j^2} + \frac{i\pi}{2}\right),
  \end{split}
\end{equation}
where $\mu_{ij}$ is a $\mathcal O(1)$ scale and 
where we have introduced the signature-even logarithm
\begin{equation}
  L_{ij} = \frac{1}{2}\left(\ln\frac{s_{ij}}{\mu^2_{ij}} +
    \ln\frac{-s_{ij}}{\mu^2_{ij}}\right)=
  \ln\frac{|s_{ij}|}{\mu^2_{ij}} - \frac{i\pi}{2}.
\end{equation}
Our goal is to redefine all quantities in \cref{eq:myNLLn} so that the rapidity evolution is expressed in terms of exponentials of the type  $e^{\tau_{i}L_{i+1,i}}$. The leftover terms in \cref{eq:fromETAtoL} then need to be reabsorbed in $\IFpre$ and $\wilsonV_\lambda$. 
In practice we redefine the impact factors and
vertex corrections as
\begin{align}
\label{eq:redefinition_nll}
    & \IFpre_A\, e^{\frac{\tau_{n-1}}{2}\left(
      \ln\frac{\mu_{n,n-1}^2}{\pp_n^2} + \frac{i\pi}{2}\right)} \cos^{-\frac{1}{2}}\left(\frac{\pi\tau_{n-1}}{2} \right)
    \equiv \Bar{\IFpre}_A,
    \notag\\
    & \IFpre_B\, e^{\frac{\tau_{3}}{2}\left(
      \ln\frac{\mu_{4,3}^2}{\pp_3^2} + \frac{i\pi}{2}\right)}
      \cos^{-\frac{1}{2}}\left(\frac{\pi \tau_{3}}{2} \right)
    \equiv \Bar{\IFpre}_B,
    \\
    &
    \cos^{\frac{1}{2}}\left( \frac{\pi \tau_{i}}{2}\right)
    e^{\frac{\tau_{i}}{2}
      \left(\ln\frac{\mu^2_{i+1,i}}{\pp_i^2} + \frac{i\pi}{2}\right)}
    \wilsonV_{\lambda_i}
    e^{\frac{\tau_{i-1}}{2}\left(\ln\frac{\mu^2_{i,i-1}}{\pp_{i}^2} + \frac{i\pi}{2}\right)}
    \cos^{-\frac{1}{2}}\left( \frac{\pi \tau_{i-1}}{2}\right)
    \equiv \overline {\wilsonV}_{\lambda_i}, \notag
\end{align}
which also prompts us to introduce $\overline { \mathcal V}_i = i f^{a_{j_i}
    a_{j_{i-1}} a_i}  V_{\lambda_i}
    \overline{\wilsonV}_{\lambda_i}$.
The redefinition above introduces a dependence on the arbitrary factorisation scales $\mu_{ij}$ in the new quantities $\Bar{\IFpre}$ and $\overline{\wilsonV}$. 
In particular they satisfy the RGE-like
equations
\begin{equation}\label{eq:RGE_mu}
  \Bar{\IFpre}(\mu') = \Bar{\IFpre}(\mu) \left(\frac{\mu'}{\mu}\right)^{\tau_i},
  \quad\quad
  \overline{\wilsonV}_{\lambda_i}(\mu'_1,\mu'_2) =
  \left(\frac{\mu_1'}{\mu_1}\right)^{\tau_{i-1}}
  \overline{\wilsonV}_{\lambda_i}(\mu_1,\mu_2)
  \left(\frac{\mu_2'}{\mu_2}\right)^{\tau_i}.
\end{equation}
For simplicity, we set all scales $\mu_{ij} = \rho$ and keep $\rho$ implicit.
If needed, one can easily reconstruct the full scale dependence using \cref{eq:RGE_mu}.
We also point out that the cosine factors in \cref{eq:redefinition_nll} are immaterial at $\NLL$ and can be replaced by $1$ at this order. 
The reason why we introduced them will become clear in the next section. 
The NLL result \cref{eq:myNLLn} then becomes
\begin{equation}
  \mathcal A_{\NLL} \propto (T_{r_A})_{c_n c_1}^{a_{n-1}}\Bar{\IFpre}_A
  \frac{e^{\tau_{n-1}L_{n,n-1}}}{\qp_{n-1}^2}
  \overline{\mathcal V}_{n-1}
  \frac{e^{\tau_{n-2}L_{n-1,n-2}}}{\qp_{n-2}^2}\cdots
  \frac{e^{\tau_{4}L_{5,4}}}{\qp_{4}^2}
  \overline{\mathcal V}_{4}
  \frac{e^{\tau_{3}L_{4,3}}}{\qp_{3}^2}
  \Bar{\IFpre}_B (T_{r_B})_{c_3 c_2}^{a_3} .
  \label{eq:myNLLnNEW}
\end{equation}

We conclude this section by presenting explicit results for the $2\to2$ and $2\to 3$ amplitudes.
These can be used to extract the
$\Bar{\IFpre}$ impact factors
and the $\overline{\wilsonV}$ vertex, 
respectively. The amplitudes read
\begin{align}
    \mathcal A_{2\to 2,\NLL}^{(--)} &=
    \Bar{\IFpre}_A
    e^{\tau_s L_{s} } \cos \left( \frac{\pi \tau_s}{2}\right)
    \Bar{\IFpre}_B\times \mathcal A_{2\to 2}^{(0)},
    \label{eq:2to2_factorised} \\
    \mathcal{A}^{(--)}_{2\to 3,\NLL} &=
    \Bar{\IFpre}_A
    e^{\tau_A L_A }
    \overline{\wilsonV}_{\lambda_4}
    e^{\tau_B L_B }
    \Bar{\IFpre}_B \times \mathcal A^{(0)}_{2\to 3},
   \label{eq:2to3_factorised}
\end{align}
where for the $2\to 2$ case we have defined
$\qp = q_3$, $\tau_s = \tau_3$, 
$L_s = L_{34}$, and for the $2\to 3$ case
\begin{equation}
    L_A = L_{45}, \quad L_B = L_{34}, \quad \qp_A = \qp_4, \quad \qp_B = \qp_3, \quad \tau_A = \tau_4, \quad \tau_B = \tau_3,
\end{equation}
in the notation of \cref{fig:regge_pole_2ton}.

\subsection{Bridging with Regge-pole factorisation}
\label{se:comparison_to_regge}
To motivate our choices in the previous section,
we now compare \cref{eq:2to2_factorised,eq:2to3_factorised} to predictions based on
unitarity and Regge-pole factorisation.
We refer the reader to refs.~\cite{BARTELS1980365,Bartels:2008ce} for a detailed discussion of the latter. A Regge-pole
contribution to the $2\to 2$ scattering
amplitudes reads
\begin{equation}
\begin{split}
  \mathcal A_{2\to 2, {\rm pole}}^{(--)} &= \IF_A \IF_B\times 
    \frac{1}{2}
    \left[
      \left(\frac{s_{34}}{\rho^2}\right)^{\tau_s}
      +
      \left(\frac{-s_{34}}{\rho^2}\right)^{\tau_s}
      \right] \times \mathcal{A}^{(0)}_{2\to 2}
      = 
      \\
      &=
     \IF_A
    e^{\tau_s L_{s} } \cos \left( \frac{\pi \tau_s}{2}\right)
    \IF_B\times \mathcal A_{2\to 2}^{(0)},   
      \label{eq:2to2_analytic} 
\end{split}   
\end{equation}       
where we introduced impact factors $\IF_X$, which are
in principle different from the ones in the
previous section.
We see from this that 
\cref{eq:2to2_factorised} has the correct form
for an odd-
signature Regge-pole exchange at all logarithmic orders. This motivates our redefinition, 
despite the single-$W$ amplitude alone 
only corresponding to part of the full Regge-pole 
exchange, see e.g. refs.~\cite{Fadin:2017nka,Falcioni:2021dgr}.

The Regge-pole structure for the $2\to 3$
amplitude can be written as
\begin{equation}
\begin{split}
    \mathcal A_{2\to 3, ~{\rm pole}}^{(--)} &= \IF_A \,  \IF_B
    \times\left[
    (T_{r_A})_{c_5 c_1}^a i f^{a b a_4} 
    (T_{r_B})_{c_3 c_2}^b \right]
    \times  \\
    &  \Bigg\lbrace 
    \frac{1}{4}\left[
    \left(\frac{s_{45}}{\rho^2}\right)^{\tau_A-\tau_B} + 
    \left(\frac{-s_{45}}{\rho^2}\right)^{\tau_A-\tau_B}\right] 
    \Bigg[
    \left(\frac{s_{12}}{\rho^2}\right)^{\tau_B} + 
    \left(\frac{-s_{12}}{\rho^2}\right)^{\tau_B}\Bigg]
    R_{\lambda_4} + \label{eq:2to3_analytic} \\
    & \;\;\; \frac{1}{4}\Bigg[
    \left(\frac{s_{34}}{\rho^2}\right)^{\tau_B-\tau_A} + 
    \left(\frac{-s_{34}}{\rho^2}\right)^{\tau_B-\tau_A}\Bigg] 
    \Bigg[
    \left(\frac{s_{12}}{\rho^2}\right)^{\tau_A} + 
    \left(\frac{-s_{12}}{\rho^2}\right)^{\tau_A}\Bigg]
    L_{\lambda_4}\Bigg\rbrace.
\end{split}
\end{equation}
The dependence on the helicity of the centrally-emitted
gluon $\lambda_4$ is carried only by the left and right vertices, $L_{\lambda_4}$ and $R_{\lambda_4}$ respectively,
which are defined through $R_{\lambda_4} = R_\mu \varepsilon^\mu_{\lambda_4}(p_4)$ and 
$L_{\lambda_4} = L_\mu \varepsilon^\mu_{\lambda_4}(p_4)$, with $R^\mu$ and $L^\mu$ 
real-valued functions. They are typically referred
to as the right and left reggeon-reggeon-gluon
vertices, respectively. We rewrite
\cref{eq:2to3_analytic} in the factorised form
\begin{equation}
\label{eq:2to3_pole_factorised}
 \mathcal{A}^{(--)}_{2\to 3,\mathrm{pole}} =
    \IF_A\,
    e^{\tau_A L_A }\,
    \mathcal L_{\lambda_4}\,
    e^{\tau_B L_B }\,
    \IF_B \times \mathcal A^{(0)}_{2\to 3},
\end{equation}
where $L$, $R$, and $\mathcal L$ are related by
\begin{equation}
\label{eq:rlto_cev_nll}
\begin{split} 
    L_{\lambda} &= 
    \frac{-i e^{\tau_A \ln\frac{\pp_4^2}{\rho^2}}}
    {\sin\left(\pi(\tau_A-\tau_B)\right)
    \cos\left(\frac{\pi\tau_A}{2}\right)} \left( e^{-i\tau_A \frac{\pi}{2}} \mathcal{L}_{\lambda} - e^{i\tau_A \frac{\pi}{2}} \mathcal{L}^*_{-\lambda} \right) V_\lambda, \\
    R_{\lambda} &= 
    \frac{-i e^{\tau_B \ln\frac{\pp_4^2}{\rho^2}}}
    {\sin\left(\pi(\tau_B-\tau_A)\right)
    \cos\left(\frac{\pi\tau_B}{2}\right) } \left( e^{-i\tau_B \frac{\pi}{2}} \mathcal{L}_{\lambda} - e^{i\tau_B \frac{\pi}{2}} \mathcal{L}^*_{-\lambda} \right) V_\lambda.
\end{split}
\end{equation}
We will use these expressions in \cref{se:results} when connecting our
result with the one-loop results for the Lipatov vertex in
ref.~\cite{Fadin:2023roz}.

%%%%%%%%%%%%%%%%%%%%%%%%%%%%%%%%%%%%%%%%%%%%%%%%%%%%%%%%%%%%%%%%%%%%%%%%%%%%%%%%%%%%%

\subsection{Two-loop NNLL predictions for the $(--)$ signature}
\label{se:mmNNLL}

\begin{figure}[t!]
    \centering
    \subfigure[]{\scalebox{0.78}{\includegraphics[valign=m]{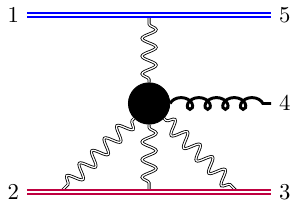}} \label{fig:WtoWWW}}
    \subfigure[]{\scalebox{0.78}{\includegraphics[valign=m]{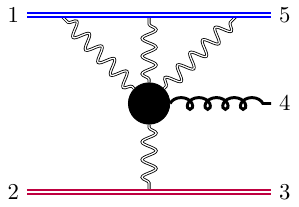}} \label{fig:WWWtoW}}
    \subfigure[]{\scalebox{0.78}{\includegraphics[valign=m]{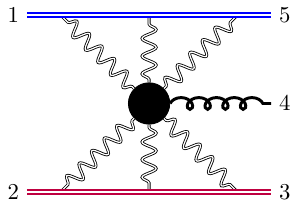}} \label{fig:WWWtoWWW}}
\caption{Schematic diagrams for multi-$W$ contributions to the $(--)$
  $\NNLL$ amplitude at two-loops.
\label{fig:regge_cuts_odd} }
\end{figure}

In this section we investigate NNLL predictions for the two-loop
$(--)$ amplitude, which involves two-loop corrections to the $WWg$
vertex $\wilsonV$. At this order however, there is a significant
differences with respect to the results discussed so far. Indeed, at
NLL each definite-signature amplitude receives contributions from a
single configuration of $W$ fields emitted from the $A$ and $B$
lines. In the case that we consider now, this is no longer
true. Indeed, at NNLL the $(--)$ amplitude can be written as
\begin{equation}
  \mathcal A_{\NNLL}^{(--)} =
  \mathcal A_{\NNLL,\{1,1\}}^{(--)} +
  \mathcal A_{\NNLL,\{1,3\}}^{(--)} +
  \mathcal A_{\NNLL,\{3,1\}}^{(--)} +
  \mathcal A_{\NNLL,\{3,3\}}^{(--)},
  \label{eq:myNNLLsplit}
\end{equation}
where $A_{\NNLL,\{i,j\}}^{(--)}$ represents the contribution coming
from a configuration with $i$ fields $W$ emitted from the $A$ line
and $j$ fields $\Wb$ emitted from the $B$ line.
The single-$W$ term is formally identical to the one given in
\cref{eq:2to3_factorised}, but with the impact factors
$\Bar{\IFpre}_X$ and vertex $\overline{\wilsonV}_\lambda$ expanded to
two-loops and the Regge trajectory $\tau_g$ expanded to three loops
(though the latter only enters starting from the three-loop
level). Contributions involving the exchange of multiple $W$s are
depicted in \cref{fig:regge_cuts_odd}.  Similarly to what we discussed in
\cref{se:LL&NLL}, in order to compute the full $\NNLL$ $(--)$
amplitude one should account for the rapidity evolution of the single-
and triple-$W$ intermediate states. However, in analogy with the NLL case
at one loop, the evolution starts at three loops so it does not affect
our two-loops analysis. The computation of the multi-$W$ contributions then
proceeds along lines which are very similar to the one described in
\cref{se:LL&NLL} for one-loop multi-$W$ contributions at
NLL.  Crucially, all the $W$ and gluon interactions can be computed at
LO, thus making the calculation tedious but straightforward. Before
reporting our results, however, we stress that the mixing of single- and
multi-$W$ contributions makes the extraction of a universal vertex
function problematic at this order. We postpone the discussion of this
issue to \cref{se:results}, and now focus on the calculation of the
multi-$W$ terms in \cref{eq:myNNLLsplit}.

We start by discussing the $\mathcal A^{(--)}_{\NNLL,\{1,3\}}$,
see \cref{fig:WtoWWW}. We need to consider the connected $S$-matrix
contribution
\begin{equation}
\begin{aligned} \label{eq:NNLL_Wto3W}
  &\mathcal{S}^{(--)}_{\NNLL,\{1,3\}}  =\\
  &=\phi^{[AB]}
  \left[2\pi \delta(p_1^+ - p_5^+) \delta_{\lambda_1\lambda_5}
    \times 2p_1^+\right]
    \left[2\pi \delta(p_2^- - p_3^-) \delta_{\lambda_2\lambda_3}
      \times 2 p_2^-\right]
    \\
  & \times     
    \Bigg\langle \timeOrd \,
    \Big(ig_s \llbracket
      W_{\eta_A}(\qp_A)\rrbracket_{r_A}\Big)
      a^{a_4}_{\lambda_4}(p_4)
      \Bigg(
      \frac{(ig_s)^3}{3!}\llbracket
      \Wb_{\eta_B}\otimes\Wb_{\eta_B}\otimes\Wb_{\eta_B}
      \rrbracket_{r_B}(\qp_B) \Bigg)
      \Bigg\rangle \, =
      \\
     &=
      \phi^{[AB]}
      \left[2\pi \delta(p_1^+ - p_5^+) \delta_{\lambda_1\lambda_5}\right]
      \left[2\pi \delta(p_2^- - p_3^-) \delta_{\lambda_2\lambda_3}
      \right]\times
      4 s_{12}
      \times
      \left[ig_s (T_{r_A}^a)_{c_5c_1}\right]
      \\
      &\times
      g_s^3 \int \measF{\kp_1}\measF{\kp_2}
      \Bigg\langle \timeOrd \llbracket W_{\eta_4}(\pp_4 +\qp_A-\kp_1) W_{\eta_4}(\kp_1-\kp_2)W_{\eta_4}(\kp_2)\rrbracket^{a_4 a}
      \\
      &\times
      \Bigg(
      \frac{(ig_s)^3}{3!}\llbracket
      \Wb_{\eta_B}\otimes\Wb_{\eta_B}\otimes\Wb_{\eta_B}
      \rrbracket^{c_3 c_2}_{r_B}(\qp_B) \Bigg)
      \Bigg\rangle
      \\
        &\times\bigg[
      \frac{1}{6}\left(\frac{\boldsymbol{\vep}^*_{\lambda_4}\cdot (\kp_1-\qp_A)}{(\kp_1-\qp_A)^2}\right)
      -\frac{1}{2}\left(\frac{\boldsymbol{\vep}^*_{\lambda_4}\cdot (\kp_2-\qp_A)}{(\kp_2-\qp_A)^2}\right)
      -\frac{1}{3}\left(\frac{\boldsymbol{\vep}^*_{\lambda_4}\cdot \qp_A}{\qp_A^2}\right)
      \bigg],
\end{aligned}
\end{equation}
which leads to the two-loop amplitude
\begin{equation}
\begin{split}
  &\mathcal A_{\NNLL,\{1,3\}}^{(--)} = 4 g_s^3 \phi^{[AB]} s_{12}
  \times (i\pi)^2\times  \bigg[\frac{4}{3} (T^a_{r_A})_{c_5 c_1}\,
      i f^{a_4 x_1 c_1} \,if^{c_1
        x_2 c_2} \,if^{c_2 x_3 a}
  \\
  & \quad\times
 \left(T_{r_B}^{(x_1}\gcdot
    T^{x_2}_{r_B}\gcdot T^{x_3)}_{r_B}\right)_{c_3 c_2} \bigg]
  \times
  \left[\frac{g_s^2 S_\ep}{(4\pi)^2}\right]^2
  \int \frac{\measJ{\kp_1}\measJ{\kp_2}}{\kp_2^2 (\kp_1-\kp_2)^2
    (\qp_B-\kp_1)^2}
  \\
  &\quad\times
  \bigg[
      \frac{1}{6}\left(\frac{\boldsymbol{\vep}^*_{\lambda_4}\cdot (\kp_1-\qp_A)}{(\kp_1-\qp_A)^2}\right)
      -\frac{1}{2}\left(\frac{\boldsymbol{\vep}^*_{\lambda_4}\cdot (\kp_2-\qp_A)}{(\kp_2-\qp_A)^2}\right)
      -\frac{1}{3}\left(\frac{\boldsymbol{\vep}^*_{\lambda_4}\cdot \qp_A}{\qp_A^2}\right)
      \bigg]\left(1+\mathcal O(\as)\right),
\end{split}
\label{eq:myANNLL13}
\end{equation}
where $T_{r_B}^{(x_1}\gcdot
T^{x_2}_{r_B}\gcdot T^{x_3)}_{r_B} = \sum_{\sigma\in S_3}
T_{r_B}^{x_{\sigma(1)}}\gcdot
T^{x_{\sigma(2)}}_{r_B}\gcdot T^{x_{\sigma(3)}}_{r_B}$. Using
the colour formalism of \cref{se:LL&NLL}, the term in the first square
bracket becomes
\begin{equation}
  \begin{split}
    \frac{4}{3} (T^a_{r_A})_{c_5 c_1}\,
    & i f^{a_4 x_1 c_1} \,if^{c_1
      x_2 c_2} \,if^{c_2 x_3 a}
    \left(T_{r_B}^{(x_1}\gcdot
    T^{x_2}_{r_B}\gcdot T^{x_3)}_{r_B}\right)_{c_3 c_2} =
    \\
    & -\left[2\boldsymbol{\mathcal{T}}_{+-}^2  +
      \frac{2}{3}\boldsymbol{\mathcal{T}}_{++}^2  +
      \frac{2N_c}{3}\boldsymbol{\mathcal{T}}_{++}\right]
    \left[(T_{r_B})_{c_3 c_2}^b (T_{r_A})_{c_5 c_1}^a if^{aba_4}\right],
  \end{split}
  \label{eq:mycol13}
\end{equation}
where in the rhs we have highlighted the tree-level colour structure. In doing this
calculation, we have found the following Lie-algebra graphic rule
useful:
\begin{equation}
  \includegraphics[valign=m]{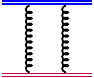} - \includegraphics[valign=m]{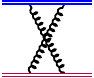} = - \frac{N_c}{2} \includegraphics[valign=m]{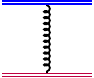} .
\end{equation}
Using \cref{eq:mycol13}, we can write the two-loop result \cref{eq:myANNLL13}
as
\begin{equation}
  \mathcal A^{(2),(--)}_{\NNLL,\{1,3\}} =\, \pi^2
  \twoDint^{(2)}_{\{1,3\}}
  \left[2\boldsymbol{\mathcal{T}}_{+-}^2  +
      \frac{2}{3}\boldsymbol{\mathcal{T}}_{++}^2  +
      \frac{2N_c}{3}\boldsymbol{\mathcal{T}}_{++}\right]
  \mathcal A^{(0)},
  \label{eq:myNNLL2_13}
\end{equation}
with
\begin{equation}
  \begin{split}
    \twoDint^{(2)}_{\{1,3\}} &=
    \left[\twoDint^{(0)}_{\{1,1\}}\right]^{-1}
    \int \frac{\measJ{\kp_1}\measJ{\kp_2}}{\kp_2^2 (\kp_1-\kp_2)^2
      (\qp_B-\kp_1)^2}
    \\
    &\times\bigg[
      \frac{1}{6}\left(\frac{\boldsymbol{\vep}^*_{\lambda_4}\cdot (\kp_1-\qp_A)}{(\kp_1-\qp_A)^2}\right)
      -\frac{1}{2}\left(\frac{\boldsymbol{\vep}^*_{\lambda_4}\cdot (\kp_2-\qp_A)}{(\kp_2-\qp_A)^2}\right)
      -\frac{1}{3}\left(\frac{\boldsymbol{\vep}^*_{\lambda_4}\cdot \qp_A}{\qp_A^2}\right)
      \bigg].
  \end{split}
  \label{eq:Ktwo13}
\end{equation}
Results for $\twoDint^{(2)}_{\{1,3\}}$ are reported in
\cref{se:twodimintegrals}. Here we stress that, as in the one-loop case,
these functions are pure, single-valued and of uniform transcendentality.
The other multi-$W$ contributions in \cref{eq:myNNLLsplit} can be obtained
by a completely analogous procedure to what discussed in
\cref{se:LL&NLL} and above. For this reason, here we limit ourselves
to presenting the final results. They read
\begin{equation}
  \begin{split}
    &\mathcal A^{(2),(--)}_{\NNLL,\{3,1\}} =\, \pi^2
    \twoDint^{(2)}_{\{3,1\}}
    \left[2\boldsymbol{\mathcal{T}}_{-+}^2  +
      \frac{2}{3}\boldsymbol{\mathcal{T}}_{++}^2  +
      \frac{2N_c}{3}\boldsymbol{\mathcal{T}}_{++}\right]
    \mathcal A^{(0)},
    \\
    &
    \mathcal A^{(2),(--)}_{\NNLL,\{3,3\}} =\, -\pi^2
    \twoDint^{(2)}_{\{3,3\}}
    \left[    \frac{1}{2} \boldsymbol{\mathcal{T}}_{--}^2  + \frac{1}{6}\boldsymbol{\mathcal{T}}_{+-}^2  + 
        \frac{1}{6}\boldsymbol{\mathcal{T}}_{-+}^2  + \frac{\boldsymbol{\mathcal{T}}_{++}^2}{18}  + \frac{2N_c}{9}\boldsymbol{\mathcal{T}}_{++} \right]
    \mathcal A^{(0)}.
  \end{split}
  \label{eq:myNNLL2_31&33}
\end{equation}
Once again, results for the $\twoDint^{(2)}_{\{i,j\}}$ functions
are reported in \cref{se:twodimintegrals}.

\begin{figure}[t!]
    \centering
    \subfigure[]{\scalebox{0.9}{\includegraphics[valign=m]{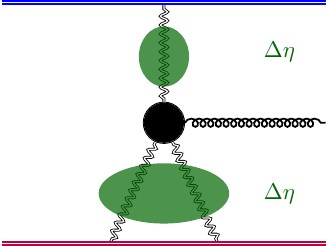}}}
    \hspace{9mm}
    \subfigure[]{\scalebox{0.9}{\includegraphics[valign=m]{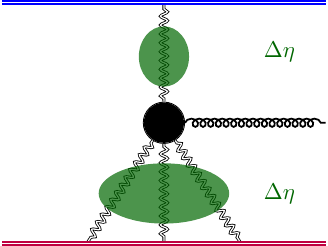}}}
    \caption{ Examples of multi-$W$ contributions involving rapidity
      evolution.  Diagram (a) corresponds to the all-orders $\NLL$
      $\mathcal A^{(-+)}_{\NLL}$ amplitude,
      diagram (b) to the all-orders $A^{(--)}_{\NNLL,\{1,3\}}$ amplitude.
      Similar diagrams would arise for all other
      transitions.  The black dot stands for the leading-order
      vertices, while the green blobs are associated with the
      leading-order rapidity evolution of the $W$ states.  While the
      evolution of a single $W$ is multiplicative and captured by the
      one-loop Regge trajectory ($\tau_{g}^{(1)}$), the evolution of
      two or more $W$s requires the computation of
      new 2d Feynman integrals at each perturbative order.  }
    \label{fig:multiW_with_evolution}
\end{figure}

We conclude this section by stressing that predicting the amplitude at
NNLL beyond two loops would require computing the rapidity evolution
of multi-$W$ states, see \cref{fig:multiW_with_evolution}(b). As we
mentioned before, these are not eigenstates of the rapidity evolution,
hence this step, while not presenting conceptual challenges,
would require the
calculation of higher-loop 2d integrals. This is not within the scope
of this paper, and we defer it to the future.
We also observe that in \cref{eq:myNNLL2_13,eq:myNNLL2_31&33}
the diagonal operator $\boldsymbol{\mathcal{T}}_{++}$ generates a
contribution only in the octet-octet exchange and it is leading in
$N_c$. Indeed $\boldsymbol{\mathcal{T}}_{++} \, \mathcal{A}^{(0)} =
-N_c/2 \, \mathcal{A}^{(0)}$, irrespectively of the particle types $A$
and $B$. The non-diagonal ones, $\boldsymbol{\mathcal{T}}_{+-}^2$,
$\boldsymbol{\mathcal{T}}_{-+}^2$ and
$\boldsymbol{\mathcal{T}}_{--}^2$ have a twofold effect. First, they
are responsible for a universal, representation-independent,
leading-$N_c$ contribution in the octet-octet exchange.  Second, they
are the only source of colour structures different from the tree-level
octet-octet one. Importantly these contributions are purely
sub-leading colour, and representation dependent, \ie they
differ in quark and gluon amplitudes.

%%%%%% -------
%%%%%% -------
%%%%%% -------
%%%%%% -------
%%%%%% -------
%%%%%%%%%%%%%%%%%%%%%%%%%%%%%%%%%%%%%%%%%%%%%%%%%%%%%%%%%%%%%%%%%%%%%%%%%%%%%%%%%%%%%%%%%%%%%%%%%%%%%%%%%%%%%%%%%%%%%%%%%%%%%%%%%%%%%%%%%%%%%%%%%%%%%%%%%%%%%%%%%%%%%%%%%%%%%%%%%%%%%%%%%%
%%%%%% -------
%%%%%% -------
%%%%%% -------
%%%%%% -------
%%%%%% -------

\section{Results}
\label{se:results}
%%%------------------------------------------------------------------------------------

With the results obtained so far we are ready to extract the universal vertex $\wilsonV$ up to two-loops. 
We start by collecting all shockwave amplitude predictions obtained in \cref{se:shockwave} at the different logarithmic and perturbative orders.
Comparing these results with the explicit MRK amplitudes of \cref{se:amplexpansion} will allow us to thoroughly check the validity of our predictions
in \cref{se:shockwave} at NLL and determine for the first time the two-loop QCD contribution to the $WWg$ vertex.
The $2\to 3$ MRK amplitudes computed in ref.~\cite{Caron-Huot:2020vlo} will also allow us to compute the same quantity in $\mathcal{N}=4$ super Yang--Mills theory, and explicitly verify the maximal transcendentality principle in
this context. 

The main results of this publication are also distributed through the ancillary files of this publication.  All results in electronic format are also available at~\cite{results:url}.  
The interested reader will find a $\texttt{README}$ file with a detailed explanation of their content and the notation adopted.

\subsection{Summary and extraction of the gluon emission vertex}
At tree level the only contribution which corresponds to LL accuracy comes from the single--$W$ amplitude in \cref{eq:myLL}, which yields
\begin{equation}
    \begin{aligned}
        \mathcal{A}^{(0)}_{\text{LL}} &= 
        {\includegraphics[valign=m]{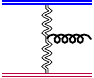}} = \mathcal{A}^{(0)} .
    \end{aligned}  \label{eq:A0_ll}
\end{equation}
At one-loop one has instead
\begin{align}
    \mathcal{A}^{(1)}_\text{LL} &= 
    {\includegraphics[valign=m]{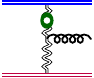}} + 
    {\includegraphics[valign=m]{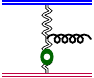}} = \left[  L_A \tau_A^{(1)} + L_B \tau_B^{(1)}  \right] \mathcal{A}^{(0)}  ,
    \label{eq:A1_ll}
    \\
    \mathcal{A}^{(1),(--)}_{\text{NLL}} &= 
    {\includegraphics[valign=m]{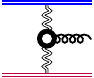}} +
    {\includegraphics[valign=m]{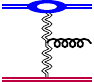}} +
    {\includegraphics[valign=m]{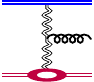}} 
    =
    \left[ \Bar{\IFpre}_A^{(1)} + \Bar{\IFpre}_B^{(1)} + \overline{\wilsonV}_{\lambda_4}^{(1)} \right] \mathcal{A}^{(0)} ,
    \label{eq:A1_nll_mm}
    \\
    \mathcal{A}^{(1),(+-)}_{\text{NLL}} &= 
    {\includegraphics[valign=m]{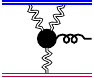}} 
    =  i \pi 
    B^{(1)}_{-+}  \boldsymbol{\mathcal{T}}_{-+}
    \mathcal{A}^{(0)} ,
    \label{eq:A1_nll_pm}
    \\
    \mathcal{A}^{(1),(-+)}_{\text{NLL}} &= 
    {\includegraphics[valign=m]{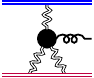}} 
    =  i \pi 
    B^{(1)}_{+-}  \boldsymbol{\mathcal{T}}_{+-}
    \mathcal{A}^{(0)} ,
    \label{eq:A1_nll_mp}
    \\
    \mathcal{A}^{(1),(++)}_{\text{NLL}} &= 
    {\includegraphics[valign=m]{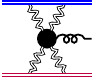}} 
    =  i \pi 
    B^{(1)}_{--}  \boldsymbol{\mathcal{T}}_{--}
    \mathcal{A}^{(0)} ,
    \label{eq:A1_nll_pp}
\end{align}
where we introduced the multi-$W$ coefficients
\begin{equation}
        B^{(1)}_{-+} = -\twoDint^{(1)}_{\{2,1\}}, 
        \quad\quad  
        B^{(1)}_{+-} = -\twoDint^{(1)}_{\{1,2\}}, \quad\quad 
        B^{(1)}_{--} = -\twoDint^{(1)}_{\{2,2\}} .
\end{equation}
In \cref{eq:A1_ll} $\tau^{(1)}_{A,B}$ is the one-loop Regge trajectory given by 
\cref{eq:tau_one_loop,eq:short_tau_V}, whereas $\Bar{\IFpre}^{(1)}_{A,B}$ and $\overline{\wilsonV}_{\lambda_4}^{(1)}$ in \cref{eq:A1_nll_mm} are, respectively, the impact factors and the central-emission vertex at one-loop accuracy. 

We stress that the formulas given so far provide a complete prediction of the one-loop MRK amplitude,
\ie for all possible signatures.
Moreover, the expressions are universal across all partonic channels, 
with the only representation-dependent components being the impact factors $\Bar{\IFpre}_X$, and the effect of the non-diagonal colour operators.
We report the explicit result of the action of such colour operators on the tree-level amplitude for the various partonic channels in \cref{se:appendixcolour}.

At two loops we restrict our attention to the odd-odd part of the amplitude where the various logarithmic orders read
\begin{align}
        \mathcal{A}^{(2)}_{\text{LL}} &= 
        {\includegraphics[valign=m]{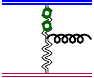}} +
        {\includegraphics[valign=m]{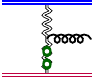}} +
        {\includegraphics[valign=m]{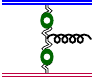}} = 
        \frac{1}{2}(L_A \tau_A^{(1)}  + L_B \tau_B^{(1)} )^2 \mathcal{A}^{(0)} ,  \label{eq:A2_ll}
        \\[8pt]
        \mathcal{A}^{(2),(--)}_{\text{NLL}} &= 
        {\includegraphics[valign=m]{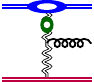}} + 
        {\includegraphics[valign=m]{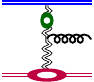}} + 
        {\includegraphics[valign=m]{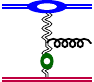}} + 
        {\includegraphics[valign=m]{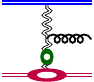}} 
        \notag\\& +
        {\includegraphics[valign=m]{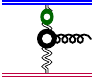}} +
        {\includegraphics[valign=m]{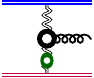}} +
        {\includegraphics[valign=m]{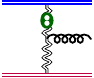}} + 
        {\includegraphics[valign=m]{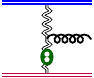}} 
        \label{eq:A2_nll}\\ 
        &= \Big[ 
          (L_A \tau_A^{(2)}  + L_B \tau_B^{(2)} ) +
          (L_A \tau_A^{(1)}  + L_B \tau_B^{(1)} )
         ( \Bar{\IFpre}_A^{(1)} + \Bar{\IFpre}_B^{(1)} + \overline{\wilsonV}_{\lambda_4}^{(1)} )  \Big] \mathcal{A}^{(0)} , \notag\\[8pt]
        \mathcal{A}^{(2),(--)}_{\NNLL} &= 
        {\includegraphics[valign=m]{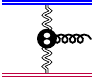}} +
        {\includegraphics[valign=m]{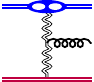}} +
        {\includegraphics[valign=m]{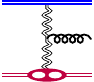}} +
        {\includegraphics[valign=m]{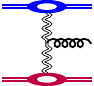}} +
        \notag\\& +
        {\includegraphics[valign=m]{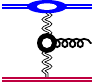}} +
        {\includegraphics[valign=m]{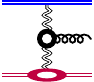}} +
        {\includegraphics[valign=m]{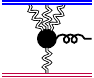}} +
        {\includegraphics[valign=m]{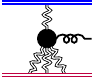}} +
        {\includegraphics[valign=m]{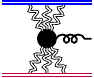}} 
         \label{eq:A2_nnll} \\
        &= \Bigg[ 
           \overline{\wilsonV}^{(2)}_{\lambda_4} +
           \Bar{\IFpre}_A^{(2)} + \Bar{\IFpre}_B^{(2)} + 
            \Bar{\IFpre}_A^{(1)}\Bar{\IFpre}_B^{(1)}  +  
            \overline{\wilsonV}_{\lambda_4}^{(1)}( \Bar{\IFpre}_A^{(1)} + \Bar{\IFpre}_B^{(1)}) 
             \notag\\
        &  \quad\quad +  (i \pi)^2  
            \Big(
            B^{(2)}_{+-}\boldsymbol{\mathcal{T}}_{+-}^2
            + 
            B^{(2)}_{--} \boldsymbol{\mathcal{T}}_{--}^2
            + 
            B^{(2)}_{-+}\boldsymbol{\mathcal{T}}_{-+}^2
            -
            B^{(2)}_{d} \frac{N_c^2}{4}            
            \Big) 
        \Bigg]\mathcal{A}^{(0)} . \notag
\end{align}
The multi-$W$ kinematic coefficients at the two-loop level are 
\begin{equation}
    \begin{gathered}
        B^{(2)}_{+-} =  -2\twoDint^{(2)}_{\{1,3\}}  + \frac{\twoDint^{(2)}_{\{3,3\}}}{6}, \quad\quad
        B^{(2)}_{-+} =  -2\twoDint^{(2)}_{\{3,1\}} + \frac{\twoDint^{(2)}_{\{3,3\}}}{6}, \quad\quad
        B^{(2)}_{--} =  \frac{\twoDint^{(2)}_{\{3,3\}}}{2}, \\
        B^{(2)}_{d}  = -\frac{2}{3}\left( \twoDint^{(2)}_{\{1,3\}}+\twoDint^{(2)}_{\{3,1\}} \right) 
        +
        \frac{7}{18}\twoDint^{(2)}_{\{3,3\}} .
    \end{gathered}
\end{equation}
We point out that the contributions from the diagonal operators $\boldsymbol{\mathcal{T}}_{++}$ and $\boldsymbol{\mathcal{T}}^2_{++}$ appearing in \cref{eq:myNNLL2_13,eq:myNNLL2_31&33}
have been absorbed in the coefficient $B^{(2)}_{d}$.
As in the one-loop case, the superscript $(2)$ refers to two-loop corrections to the MRK building blocks.

We now outline our approach for the extraction of the universal coefficients $\overline{\wilsonV}^{(1)}$ and $\overline{\wilsonV}^{(2)}$. While the former is already present in the literature, the latter is in fact unknown.
We begin by noting that by using the Balitsky-JIMWLK formalism, the Regge trajectory as well as the quark and gluon impact factors can be extracted by studying the Regge limit of $2\to2$ scattering amplitudes.
Because of this we will regard $\tau_g^{(\ell)}$ and $\Bar{\IFpre}_X^{(\ell)}$ as known quantities. 
Starting at LL accuracy, the only relevant contribution is the LO gluon Regge trajectory. 
It is straightforward to check that \cref{eq:A1_ll,eq:A2_ll}, combined with \cref{eq:tau_one_loop}, indeed match the amplitudes computed in \cref{se:amplexpansion} at this logarithmic order.

At NLL, the prediction for the odd-odd component includes for the first time a perturbative correction to the impact factors and to the $WWg$ vertex $\overline{\wilsonV}_\lambda$. 
We can then extract $\overline{\wilsonV}^{(1)}$ from \cref{eq:A1_nll_mm},
by matching with the one-loop MRK amplitudes obtained in \cref{se:amplexpansion}.
This was indeed the procedure followed in ref.~\cite{DelDuca:1998cx} to obtain the one-loop corrections to the Lipatov vertex at $\mathcal{O}(\ep^0)$.
In addition, we can now fully predict the two-loop NLL odd-odd component $\mathcal{A}^{(2),(--)}_{\text{NLL}}$ via \cref{eq:A2_nll}, which serves as a strong check of the formalism up to this logarithmic accuracy. 
As far as the components involving even signatures are concerned, there are no unknown quantities involved, thus no matching to explicit amplitudes is required and eqs.~\eqref{eq:A1_nll_pm}, \eqref{eq:A1_nll_mp} and \eqref{eq:A1_nll_pp} directly reproduce the results obtained in \cref{se:amplexpansion}.
Taking into account the various ingredients described above, we find full agreement.

Finally, at $\NNLL$ we focus on the odd-odd amplitude, which receives contributions from various effects: the two-loop impact factors $\Bar{\IFpre}^{(2)}_X$, the two-loop correction to $\overline{\wilsonV}$ and the $W\to3W$, $3W\to W$ and $3W\to3W$ transitions.
Our direct calculation of the multi-$W$ coefficients $B^{(2)}_{\sigma_1,\sigma_2}$ then leaves only $\overline{\wilsonV}^{(2)}$ undetermined and allows us to compute it by matching the corresponding explicit UV-renormalised MRK amplitude. 

\subsection{Collecting the universal contributions}
The results obtained in the previous section are consistent with the effective approach described in \cref{se:shockwave}. However, we point out again that the multi-$W$ contributions found at $\NNLL$ contain universal leading-$N_c$ terms in the
octet-octet colour channel. In particular, at large $N_c$
\begin{equation}
    \boldsymbol{\mathcal{T}}_{-+}^2 \approx \boldsymbol{\mathcal{T}}_{+-}^2 \approx \boldsymbol{\mathcal{T}}_{--}^2 \approx \frac{N_c^2}{4}, 
\end{equation}
so we find that in leading-colour approximation the multi-$W$ contributions are given by
\begin{equation}
 {\includegraphics[valign=m]{tikz-fig/fig_66.pdf}} \!+\!
        {\includegraphics[valign=m]{tikz-fig/fig_67.pdf}} \!+\!
        {\includegraphics[valign=m]{tikz-fig/fig_68.pdf}} \approx
    (i \pi)^2 \frac{N_c^2}{4}  
            \Big(
            B^{(2)}_{+-}
            \!+\! 
            B^{(2)}_{--} 
            \!+\! 
            B^{(2)}_{-+}
            \!-\!
            B^{(2)}_{d}            
            \Big) 
       \mathcal{A}^{(0)} . 
\end{equation}
Therefore, the only part which distinguishes between the representations of the projectiles is given by the sub-leading $N_c$ coefficients in the 
multi-$W$ exchanges. 

As discussed in ref.~\cite{Falcioni:2021dgr}, it is natural to
isolate the universal components of the multi-$W$ interactions and reabsorb it into a redefinition of the radiative corrections of the single-$W$ amplitude.
For consistency, this has to be done both for the $2\to2$ and for $2\to 3$ amplitudes.
In the $2 \to 2$ case this amounts to defining new impact factors and Regge-trajectory so that up to  $\NNLL$ 
\begin{equation}\label{eq:universal_2to2}
    \mathcal A_{2\to 2,\NNLL}^{(--)} =
    \IFU_A
    e^{\tauU_s L_{s} } \cos \left( \frac{\pi \tauU_s}{2}\right)
    \IFU_B\times \mathcal A_{2\to 2}^{(0)} + [\text{multi $W$ exchanges}]_{\mathrm{SLC}}.
\end{equation}
While up to two-loops the Regge trajectory remains unchanged (hence the slight abuse of notation in the equation above), the $\mathcal{O}(\as^2)$ correction to the impact factors is modified with respect to the one in \cref{se:shockwave}.
It is straightforward to obtain the new impact factors $\IFU_X$ from the results of ref.~\cite{Falcioni:2021dgr}, after having taken into account the additional cosine factor in the factorisation formula~\eqref{eq:universal_2to2}. The explicit formulae for these impact factors are given in the ancillary files. 

Coming to the $2 \to 3$ case, we define a modified $WWg$ vertex coefficient $\wilsonU_{\lambda}$, which up to two loops reads
\begin{equation}\label{eq:W_lipatov}
    {\wilsonU}_{\lambda}^{(0)} = 1, \quad
    {\wilsonU}_{\lambda}^{(1)} = \overline{\wilsonV}^{(1)}_{\lambda}, \quad 
    {\wilsonU}_{\lambda}^{(2)} = \overline{\wilsonV}^{(2)}_{\lambda} + \frac{N_c^2}{4} \left( B^{(2)}_{+-} + B^{(2)}_{--} + B^{(2)}_{-+}-B^{(2)}_{d}\right) .
\end{equation}
This allows us to rewrite the $\NNLL$ odd-odd amplitude prediction as follows:
\begin{equation}\label{eq:A2_nnll_universal}
        \begin{aligned}
        &\mathcal{A}^{(2),(--)}_{\NNLL} = \Bigg\lbrace
           \wilsonU_{{\lambda_4}}^{(2)} +
           \IFU_A{}^{(2)} + \IFU_B{}^{(2)} + 
            \IFU_A{}^{(1)}\IFU_B{}^{(1)}  +  
            \wilsonU_{{\lambda_4}}^{(1)}( \IFU_A{}^{(1)} + \IFU_B{}^{(1)}) 
             \\
        &   +  (i \pi)^2  
            \Bigg[
            B^{(2)}_{+-} \left(\boldsymbol{\mathcal{T}}_{+-}^2 \!-\! \frac{N_c^2}{4} \right)
            + B^{(2)}_{--} \left(\boldsymbol{\mathcal{T}}_{--}^2 \!-\! \frac{N_c^2}{4} \right)
            + B^{(2)}_{-+} \left(\boldsymbol{\mathcal{T}}_{-+}^2 \!-\! \frac{N_c^2}{4} \right)
                \Bigg] \Bigg\rbrace\mathcal{A}^{(0)} ,
    \end{aligned} 
\end{equation}
where the first line contains all universal contributions while the second one is manifestly sub-leading-colour.
In the following sections, we will present our results in terms of the universal coefficient $\wilsonU$. 

Before presenting any result, we point out that, since we are working with UV-renormalised amplitudes, the vertex corrections ${\wilsonU}^{(1)}_\lambda$ and ${\wilsonU}^{(2)}_\lambda$ obtained from the procedure above still contain IR poles and are regularisation scheme dependent. Furthermore they are affected by spurious kinematic singularities which obscure their simplicity.
For these reasons, we find it convenient to express our results in terms of finite remainders, which we define in the next section.
We will see that, in addition to being free of IR poles, they are also free of spurious singularities when expressed in terms of the single-valued functions defined in \cref{se:finite_remainders}. This is to be expected since the same properties are seen in the finite amplitudes defined in \cref{se:amplexpansion}.
%%%---------------------------------------------------------------------
\subsection{Finite remainders}
\label{se:hard_functions}
%%%---------------------------------------------------------------------
In order to define finite remainders for the various building blocks of MRK factorisation, we look more closely at the IR anomalous dimension \cref{eq:soft_anomalous_dimension}.
We highlight the different contributions by rewriting it as follows
\begin{equation}
\begin{aligned} \label{eq:regge_soft_anomalous_dimension}
    {\bf \Gamma}_{IR} &= 
    \gamma_K \mathcal{C}_A \ln\frac{-s_{51}}{\mu^2} -\frac{\gamma_K}{2} \ln\frac{-s_{51}}{\rho^2}({\bf T}^{15}_{+})^2 + 2\gamma_A
    \\
    &+
    \gamma_K \mathcal{C}_B\ln\frac{-s_{23}}{\mu^2} -\frac{\gamma_K}{2} \ln\frac{-s_{23}}{\rho^2}({\bf T}^{23}_{+})^2 + 2\gamma_B
    \\ 
    &+ \gamma_K 
    L_A ({\bf T}^{15}_{+})^2 + \gamma_K  
    L_B ({\bf T}^{23}_{+})^2   \\ 
    &+\frac{\gamma_K}{2} 
    \left(
       -\mathcal{C}_4 \ln \frac{\mu^2}{{\bf p}_4^2}  + \ln\frac{\rho^2}{{\bf p}_4^2} ({\bf T}^{15}_{+})^2 + \ln\frac{\rho^2}{{\bf p}_4^2}  ({\bf T}^{23}_{+})^2 - i \pi \, \boldsymbol{\mathcal{T}}_{++} \right)  + \gamma_4
        \\
    &+\frac{\gamma_K}{2} 
    \times i \pi \left( 
        \boldsymbol{\mathcal{T}}_{+-}+
        \boldsymbol{\mathcal{T}}_{--}+
        \boldsymbol{\mathcal{T}}_{-+}
        \right).
\end{aligned}
\end{equation}
The first two lines contain the collinear and soft singularities of the projectiles; these are the only terms that depend on the particle types $A$ and $B$.
The third line is the only source of large logarithms and is therefore associated with the rapidity evolutions in the $s_{51}$ and $s_{23}$ channels.
The fourth lines captures the dependence on the central gluon momentum, colour charge and collinear anomalous dimension and can therefore be associated with the $WWg$ interaction, at least at one loop. 
Finally, the fifth line contains the only non-diagonal and signature mixing operators and is purely imaginary, thus it is connected with the multi--$W$ exchange contributions.  

We can then proceed by writing the IR renormalisation matrix as
\begin{equation} \label{eq:Z_IR_split}
    {\bf Z}_{IR} = \exp \left( \z_A + \z_B + \z_{\tau_A} + \z_{\tau_B} + \z_{+} + \z_{-} \right) ,
\end{equation}
where the quantities in the exponent correspond to the different components of \cref{eq:regge_soft_anomalous_dimension} after the scale integration of \cref{eq:Z_exponentiation} has been carried out. 
Explicitly they read
\begin{gather}
    \z_A =  K \left( \mathcal{C}_A\ln\frac{-s_{51}}{\mu^2} -\frac{1}{2}\ln \frac{-s_{51}}{\rho^2} ({\bf T}^{15}_{+})^2\right)  + 2G_A  -2K' \mathcal{C}_A , \notag \\
    \z_B = K \left( \mathcal{C}_B\ln\frac{-s_{23}}{\mu^2} -\frac{1}{2}\ln \frac{-s_{23}}{\rho^2} ({\bf T}^{23}_{+})^2\right) + 2G_B -2K' \mathcal{C}_B ,  \notag \\ 
    \z_{\tau_A} = K L_A ({\bf T}^{15}_{+})^2  , \quad
    \z_{\tau_B} = K L_B ({\bf T}^{23}_{+})^2  ,  \label{eq:zetas_regge}\\
    \z_{-} =  \frac{i \pi}{2} K  \left( 
        \boldsymbol{\mathcal{T}}_{+-}+
        \boldsymbol{\mathcal{T}}_{--}+
        \boldsymbol{\mathcal{T}}_{-+}
        \right),  
        \notag\\
    \z_{+} = \frac{K}{2} 
    \Big[-\mathcal{C}_4 \ln \frac{\mu^2}{{\bf p}_4^2}  + \ln\frac{\rho^2}{{\bf p}_4^2} ({\bf T}^{15}_{+})^2 + \ln\frac{\rho^2}{{\bf p}_4^2}  ({\bf T}^{23}_{+})^2 - i \pi \, \boldsymbol{\mathcal{T}}_{++}
        \Big]  + G_4 - K' \mathcal{C}_4,    \notag
\end{gather}
with the additional definitions 
\begin{equation}\label{eq:KKG}
    K =  \int_\mu^\infty \frac{d\lambda}{\lambda} \gamma_K(\alpha(\lambda)),
    \quad
    K' = \int_\mu^\infty \frac{d\lambda}{\lambda} K(\alpha(\lambda)),
    \quad
    G_i = \int_\mu^\infty \frac{d\lambda}{\lambda} \gamma_i(\alpha(\lambda)),
\end{equation}
whose perturbative expansions up to the required order are given in \cref{se:beta_and_anomalous}.

We now get to the definition of finite remainders for the various building blocks of MRK factorisation.
We start from the Regge trajectory and impact factors. Based on the results of ref.~\cite{Falcioni:2021buo}, we know that, at least up to two loops, the quantities
\begin{equation} \label{eq:tau_and_I_finite}
    \taufin_g(\qp) = {\tau_g}(\qp)  - K N_c, \quad\quad \IFfin_X(\qp) = e^{-\zeta_X}  \IFU_X(\pp)
\end{equation}
are finite. The finite corrections to the Regge trajectory read
\begin{align}
    \taufin^{(1)}(\qp;\mu)  &= 2 N_c \ln \left(\frac{\mu ^2}{\qp^2}\right)
    ,\\
    \taufin^{(2)}(\qp;\mu)  &= N_c \left[\beta_0 \, \ln^2\left(\frac{\mu ^2}{\qp^2}\right) + 
    \frac{\gamma_K^{(2)}}{2}\ln \left(\frac{\mu ^2}{\qp^2}\right) + N_c\left(\frac{404}{27}-2\zeta_3\right) -\frac{56}{27} N_f \right]\,.
\end{align}
The perturbative coefficients of the impact factors instead are
\begin{align}
    \IFfin_g^{(1)} &= \left( 4\zeta_2-\frac{67}{18}\right) N_c+\frac{5 N_f}{9}, \\
    \IFfin_q^{(1)} &= \left(\frac{13}{18}+\frac{7 \zeta_2}{2}\right) N_c+\frac{8-\zeta_2}{2N_c}-\frac{5 N_f}{9}, \\
    \IFfin_g^{(2)} &=  \left(\frac{88 \zeta _3}{9}-\frac{3\zeta_4}{2}+\frac{335 \zeta_2}{18}-\frac{26675}{648}\right) N_c^2 + \left(\frac{2 \zeta _3}{9}-\frac{25 \zeta_2}{9}+\frac{2063}{216}\right) N_c N_f  \notag\\
    & -\frac{25 N_f^2}{162} +\left(2 \zeta _3-\frac{55}{24}\right)\frac{
   N_f}{N_c}, \\ 
    \IFfin_q^{(2)} &= \left(\frac{41 \zeta _3}{9}-\frac{35 \zeta_4}{16}+\frac{87 \zeta_2}{4}+\frac{22537}{2592}\right) N_c^2 -  \left(\frac{23 \zeta _3}{9}+ 4\zeta_2 +\frac{650}{81}\right) N_c N_f  \notag\\
   & + \frac{25
   N_f^2}{54} -\left(\frac{19 \zeta _3}{9}+\zeta_2+\frac{505}{81}\right)\frac{N_f}{N_c}-\frac{205 \zeta _3}{18}-\frac{47 \zeta_4}{8}+\frac{19 \zeta_2}{2}+\frac{28787}{648}  \notag\\
   &+ \left(-\frac{15 \zeta _3}{2}-\frac{83 \zeta_4}{16}+\frac{21\zeta_2}{4}+\frac{255}{32} \right) \frac{1}{N_c^2} .
\end{align}
where we set the renormalisation and factorisation scales to $\mu^2 = \rho^2 = \qp^2$. 
We provide results for generic scales as well as their UV renormalised counterparts in the ancillary files of this publication.

We now move to the determination of finite remainders for the cut coefficients and the $WWg$ vertex. 
Our strategy to do so consists in defining a finite scattering amplitude\footnote{
We note that $\mathcal{F}$ can be written in terms of the finite quantities of \cref{se:amplexpansion} as
$$
  \mathcal F^{[AB]}_{\boldsymbol{\lambda}}=
  g_s^3 \,2 \sqrt{2} \, \Phi^{[AB]}_{\boldsymbol{\lambda}}
  \sum_{n} \mathcal C_n^{[AB]}\mathcal H_{n,{\boldsymbol{\lambda}}}^{[AB]},
$$
where we made the helicity and partonic channel indices explicit.
}
\begin{equation}\label{eq:myF}
  \mathcal{F}=
  \lim_{\ep\to0} \,
  \mathbf{Z}_{IR}^{-1}\,
  {\mathcal A},
\end{equation}
inserting the formal expressions of eqs.~\eqref{eq:A0_ll}---\eqref{eq:A1_nll_pp},  \eqref{eq:A2_ll}---\eqref{eq:A2_nll} and \eqref{eq:A2_nnll_universal} into $\mathcal{A}$ and writing the Regge trajectory and the impact factors in terms of their finite remainders in \cref{eq:tau_and_I_finite}. 
We then compare the various signature components on the two sides of the equation at different logarithmic and perturbative orders.
Since the components of $\mathcal{F}$ are finite we can simply read off the finite remainders for $B_{\sigma_1,\sigma_2}^{(\ell)}$ and $\wilsonU$. We now describe in detail how we do this.

Starting from the one- and two-loop LL amplitudes, we simply find 
\begin{equation}
    \mathcal{F}^{(1),(--)}_{\mathrm{LL}} = \left[  L_A \hat{\tau}_A^{(1)} + L_B \hat{\tau}_B^{(1)} \right] \mathcal{A}^{(0)}, 
    \quad\quad
    \mathcal{F}^{(2),(--)}_{\mathrm{LL}} = \frac{1}{2}  \left[  L_A \hat{\tau}_A^{(1)} + L_B \hat{\tau}_B^{(1)}  \right]^2 \mathcal{A}^{(0)},
\end{equation}
which is consistent with the finiteness of $\hat{\tau}_g$ and does not provide any further information. 
Moving to NLL, the odd-odd one-loop component reads
\begin{equation}
    \mathcal{F}^{(1),(--)}_{\mathrm{NLL}} 
    =
    \left[ \IFfin_A^{(1)} + \IFfin_B^{(1)} + \left(\wilsonU^{(1)}_{\lambda_4}  -\z_+^{(1)} \right)\right] \mathcal{A}^{(0)} ,
\end{equation}
so that the one-loop finite remainder for the vertex is given by
\begin{equation}
    \hat{\wilsonU}^{(1)}_{\lambda} = \wilsonU^{(1)}_{\lambda}  -\zeta_+^{(1)},
\end{equation}
where $\zeta_+$ is defined (at all orders) as the eigenvalue  $\z_+ \mathcal{A}^{(0)} = \zeta_+ \mathcal{A}^{(0)}$.
Moving to components involving even signatures, we get
\begin{equation}
\begin{split}
\mathcal{F}^{(1),(-+)}_\NLL 
        =  i \pi 
       (
       B^{(1)}_{+-} - \frac{K^{(1)}}{2} )\boldsymbol{\mathcal{T}}_{+-}
        \mathcal{A}^{(0)},\\
\mathcal{F}^{(1),(+-)}_\NLL
        =  i \pi 
        (
        B^{(1)}_{-+} - \frac{K^{(1)}}{2} )\boldsymbol{\mathcal{T}}_{-+}
        \mathcal{A}^{(0)},\\
\mathcal{F}^{(1),(++)}_\NLL
        =  i \pi 
       (
       B^{(1)}_{--} - \frac{K^{(1)}}{2} )\boldsymbol{\mathcal{T}}_{--}
        \mathcal{A}^{(0)}  , 
\end{split}
\end{equation}
which prompts us to define the finite remainders for the one-loop multi-$W$ coefficients as
\begin{equation}
    \hat{B}^{(1)}_{\sigma_1 \sigma_2} = B^{(1)}_{\sigma_1 \sigma_2} - \frac{K^{(1)}}{2} .
\end{equation}
Finally, at $\NNLL$ we organise the odd-odd two-loop amplitude according to its colour structure and find:
\begin{equation}\label{eq:F2_nnll_universal}
        \begin{aligned}
        &\mathcal{F}^{(2),(--)}_{\NNLL} = \Bigg\lbrace
           \hat{\wilsonU}_{{\lambda_4}}^{(2)} +
           \IFfin_A{}^{(2)} + \IFfin_B{}^{(2)} + 
            \IFfin_A{}^{(1)}\IFfin_B{}^{(1)}  +  
            \hat{\wilsonU}_{{\lambda_4}}^{(1)}( \IFfin_A{}^{(1)} + \IFfin_B{}^{(1)}) 
             \\
        &   +  (i \pi)^2  
            \Bigg[
            \hat{B}^{(2)}_{+-} \left(\boldsymbol{\mathcal{T}}_{+-}^2 \!-\! \frac{N_c^2}{4} \right)
            + \hat{B}^{(2)}_{--} \left(\boldsymbol{\mathcal{T}}_{--}^2 \!-\! \frac{N_c^2}{4} \right)
            + \hat{B}^{(2)}_{-+} \left(\boldsymbol{\mathcal{T}}_{-+}^2 \!-\! \frac{N_c^2}{4} \right)
                \Bigg] \Bigg\rbrace\mathcal{A}^{(0)} ,
    \end{aligned} 
\end{equation}
where we have defined the finite remainders for two-loop vertex correction and the two-loop multi-$W$  coefficients as
\begin{equation}
    \begin{aligned}
        \hat{\wilsonU}^{(2)}_{\lambda_4} &=
        \wilsonU^{(2)} - \zeta_+^{(1)} \wilsonU^{(1)} + \frac{1}{2}  (\zeta_+^{(1)})^2 - \zeta_+^{(2)}  
        \\ &
           \quad\quad
           -(i \pi)^2 \frac{N_c^2}{4} \left(
            \frac{K^{(1)}}{2} (B^{(1)}_{+-}+B^{(1)}_{-+}+B^{(1)}_{--}) - \frac{3}{8} (K^{(1)})^2  \right), \\
        \hat{B}^{(2)}_{\sigma_1 \sigma_2}  &= 
            B^{(2)}_{\sigma_1 \sigma_2} - \frac{K^{(1)}}{2} B^{(1)}_{\sigma_1 \sigma_2} + \frac{1}{8} (K^{(1)})^2  . 
    \end{aligned}
\end{equation}
In particular, the subtraction of the leading-colour terms from the operators $\boldsymbol{\mathcal{T}}^2_{\sigma_1 \sigma_2}$ arising from the definition \cref{eq:A2_nnll_universal} causes the appearance of the second line in of the definition of $\hat{\wilsonU}^{(2)}$. 

We highlight the fact that, thanks to our definition of the finite remainders $\taufin$, 
$\IFfin_X$, $\hat{\wilsonU}$ and $\hat{B}^{(\ell)}_{\sigma_1 \sigma_2}$, 
the finite amplitude has exactly the same MRK-factorised form of the UV renormalised one.
In other words, the components of the finite amplitude can be obtained by replacing  each quantity in eqs.~\eqref{eq:A0_ll}---\eqref{eq:A1_nll_pp}, \eqref{eq:A2_ll}---\eqref{eq:A2_nll} and \eqref{eq:A2_nnll_universal} by its corresponding finite remainder, as defined above.
In principle this allows us to perform the extraction of the finite quantities directly from $\mathcal{F}$, without ever working with the UV-renormalised but still IR-divergent amplitude $\mathcal{A}$.

\subsection{Finite results in QCD and $\mathcal{N}=4$ sYM}
\label{se:results_vertex}
We now present our results for the $WWg$ vertex coefficient in QCD and $\mathcal{N}=4$ sYM.
In particular we provide the one- and two-loop corrections, which enter at $\NNLL$ accuracy in the MRK regime.
While reminding the reader of the perturbative expansion
\begin{equation}
\label{eq:uvert_exp}
	\hat\wilsonU_{\lambda_4} = 1 + \left( \frac{\as}{4\pi} \right) \hat\wilsonU_{\lambda_4}^{(1)} + \left( \frac{\as}{4\pi} \right)^2 \hat\wilsonU_{\lambda_4}^{(2)} + \dots \,    ,
\end{equation} 
we note that one-loop correction is not new, however we report its finite remainder here for completeness.
The two-loop term instead appears here for the first time. 
We present both written in terms of the single-valued functions $h_{i,j}$ and the rational functions $r_{i}$ defined in \cref{se:amplexpansion}.

The QCD results are
\begin{align} 
    \hat{\wilsonU}&_{+,QCD}^{(1)} = 
     \frac{N_c}{2} \left(5 \zeta _2- h_{1,2} \left(h_{1,2}+3 r_3\right)-i \pi  
   h_{1,1}\right)
   - 
   \frac{N_c-N_f}{3} \left(r_1 h_{1,2}+r_2\right),
    \label{eq:U_QCD_1}\\
    \hat{\wilsonU}&_{+,QCD}^{(2)} = 
   N_c^2 
   \Bigg[
   \frac{1}{144} i \pi  \bigg(-72 \zeta _3+h_{1,1}
   \left(-36 \zeta _2+9 h_{1,2} \left(3 r_3+4 h_{1,2}\right)-456\right)+464 
   \notag\\
   &\quad\quad
   -27 r_3 \left(h_{1,3} h_{1,4}+8 h_{2,2}-8
   h_{2,3}\right)\bigg)+\frac{1}{432} \bigg(216 r_2-1809 \zeta _4+216 r_1 h_{1,2}-2872
   \notag\\
   &\quad\quad
   +36 \zeta _2 \left(-18
   h_{1,1}^2+3 \left(9 r_3-7 h_{1,2}\right) h_{1,2}+209\right) -9 \big(-6 h_{1,2}^4+98 h_{1,2}^2
   +9 r_3 \big(2 h_{1,2}^3
   \notag\\
   &\quad\quad
   +3 \left(\left(h_{1,1}-4\right) h_{1,1}+24\right) h_{1,2}+h_{1,1} \left(h_{1,3} h_{1,4}+8 h_{2,2}-8
   h_{2,3}\right)+64 h_{3,6}\big)\big)\bigg)
   \Bigg] 
   \notag\\
      &+
   N_c \left(N_c-N_f\right) 
   \Bigg[
   \frac{1}{216} i \pi 
   \bigg(36 r_4+36 r_2 \left(h_{1,1}-1\right)+108 r_3 h_{1,2}+3 h_{1,1} \left(3 r_1 h_{1,2}-40\right)
   \notag\\
   &\quad\quad
   -9 r_1 \left(12
   h_{1,2}+h_{1,3} h_{1,4}+8 h_{2,2}-8 h_{2,3}\right)+112\bigg)+\frac{1}{648} \bigg(
   36 \zeta _2 \left(9 r_1 h_{1,2}+55\right)
   \notag\\
   &\quad\quad 
   +36 \left(3 \left(5r_3+r_6\right)-113 r_1\right) h_{1,2}
   +36 r_2 \left(3 h_{1,2}^2-15 \zeta
   _2+6 h_{1,1}-137\right)-9 \big(9 r_1 h_{1,2} h_{1,1}^2
   \notag\\
   &\quad\quad 
   -3 \big(4 r_4-12 r_3 h_{1,2}+r_1 \left(36 h_{1,2}-h_{1,3}
   h_{1,4}-8 h_{2,2}+8 h_{2,3}\right)-4\big) h_{1,1}
   +2 \big(3 r_1 h_{1,2}^3
   \notag\\
   &\quad\quad 
   +\left(6 r_5+2\right) h_{1,2}^2-18
   \left(r_1-r_3\right) \left(h_{1,3} h_{1,4}+8 h_{2,2}-8 h_{2,3}\right)+96 r_1
   h_{3,6}\big)\big)-260\bigg)
   \Bigg] \notag\\
    &
    + N_c \beta ^{\text{(0)}} 
    \Bigg[
    \frac{1}{8} i \pi  \bigg(h_{1,1}^2+2 h_{1,2}^2+4 \zeta _2-8 h_{2,1}\bigg)
   +\frac{1}{48} \bigg(-h_{1,4}^3-3
   h_{1,1}^2 h_{1,4}+3 h_{1,2}^2 h_{1,4}
    \notag\\
   &\quad\quad 
   -9 h_{1,2} \left(h_{1,3} h_{1,4}+8 h_{2,2}-8 h_{2,3}\right)-48 \left(2 \zeta _2 h_{1,4}-2
   h_{3,4}+2 h_{3,5}+h_{3,7}\right)
   \notag\\
   &\quad\quad
   +3 h_{1,3}^2 h_{1,4}+232 \zeta _3+3 h_{1,1} \left(5 h_{1,2}^2+2 h_{1,3}
   h_{1,2}-16 h_{2,1}\right)
   \bigg)
   \Bigg] \notag\\
   &+
   \frac{\left(N_c-N_f\right){}^2}{54} \!
   \Bigg[
   \left( r_2 \!+\! r_1 h_{1,2} \right) \left(6 h_{1,1}\!-\!20\right)+3h_{1,2}  h_{1,2}
   \Bigg]
   \!+\!
   \frac{N_f}{2 N_c} \!
   \Bigg[
   r_2+\left(r_1\!-\!2 r_3\right)h_{1,2}
   \Bigg] ,
    \label{eq:U_QCD_2}
\end{align}
where we have set $\mu^2 = \rho^2 = \pp_4^2$ and selected the $\lambda_4 = +1$ helicity of the emitted gluon. The vertex coefficients for $\lambda_4 = -1$ can be obtained by simply swapping $z \leftrightarrow \zbar$ in the equations above.
The results with full dependence on the scales $\mu$ and $\rho$,  as well as the higher orders in $\ep$ of the one-loop correction, are given in the ancillary files.

Two comments are now in order:
\begin{itemize}
    \item the one-loop result is purely leading colour\footnote{Here by leading colour we refer to the QCD planar limit $N_c \to \infty$ with $N_f/N_c$ held fixed.}. The two-loop correction is also leading colour, except for a sub-leading $N_f/N_c$ term. This is in exact analogy with the corrections to the Regge trajectory. There, both the LL and NLL contributions are leading colour, but the NNLL term, extracted in refs.~\cite{Falcioni:2021dgr,Caola:2021izf},  has a single sub-leading colour term proportional to $N_f/N_c$;
    \item as anticipated above, the finite remainders of \cref{eq:U_QCD_1,eq:U_QCD_2} are free of the spurious kinematic singularities associated with the letters $1-z-\zbar$ and $z - \zbar$ discussed in \cref{se:amplexpansion};
\end{itemize}

In order to extract the same quantity in $\mathcal{N}=4$, we take advantage of the five-point  amplitudes in the MRK limit provided in ref.~\cite{Caron-Huot:2020vlo}. 
These are given at the level of the finite remainders,  which the authors define identically to us (cf.  eqs.~(3.26) and (3.28) in ref. \cite{Caron-Huot:2020vlo} with \cref{eq:regge_soft_anomalous_dimension,eq:myF}  in this paper).
We notice that the finite remainder of the MRK  $\mathcal{N}=4$ amplitude is equal to the leading-transcendental weight component of the QCD one,  both at one and two loops. 
Since the same holds for the Regge trajectory $\taufin_g$ and the impact factors $\IFfin_{q,g}$,  and the multi-$W$ coefficients $\hat{B}^{(1)}_{\sigma_1\sigma_2}$ and $\hat{B}^{(2)}_{\sigma_1\sigma_2}$ are universal across gauge theories,  the maximal transcendentality principle must hold for $\wilsonU^{(1)}$ and $\wilsonU^{(2)}$ as well. 
As a consequence we simply find the $\mathcal{N}=4$ results via
\begin{equation}
	\hat\wilsonU^{(1)}_{\mathcal{N}=4} = \left. \hat\wilsonU^{(1)}_{QCD} \right|_\mathrm{LT},  \quad\quad \hat\wilsonU^{(2)}_{\mathcal{N}=4} = \left. \hat\wilsonU^{(2)}_{QCD} \right|_\mathrm{LT},  
\end{equation}
where $\mathrm{LT}$ stands for projection on the leading-transcendental component.
Explicitly, we find the remarkably simple result
\begin{align} 
    \hat{\wilsonU}&_{\mathcal{N}=4}^{(1)} = \frac{N_c}{2}  \left(-h_{1,2}^2-i \pi  h_{1,1}+5 \zeta _2\right),
    \label{eq:U_N4_1}\\
    \hat{\wilsonU}&_{\mathcal{N}=4}^{(2)} = 
    -\frac{N_c^2}{4}  \left( h_{1,2}^2 \left(7 \zeta _2\!-\!i \pi  h_{1,1}\right)\!+\! \zeta _2 h_{1,1} \left(6 h_{1,1}\!+\!i \pi \right)\!-\! \frac{h_{1,2}^4}{2} \!+\!2 i \pi
    \zeta _3\!+\!\frac{67}{4} \zeta _4\right).
    \label{eq:U_N4_2}
\end{align}
Note that we removed the helicity index since the $\mathcal{N}=4$ corrections are identical for $\lambda_4 = \pm 1$.
Here we point out that the two-loop $\mathcal{N}=4$ correction only contains transcendental weight 1 functions $h_{1,n}$. 
We emphasize that this is due to the definition of $\wilsonU$, which incorporates the universal component of the multi-$W$ terms, thereby eliminating a function of transcendental weight-2 ($D_2(z,\zbar)$) that would otherwise appear.

\subsection{Checks and validation}
\label{se:checks}

We now report on a series of checks we performed to validate our results.
The first non-trivial one is at the amplitude-level, where we compared
our QCD results of \cref{se:generalaspects} to those in $\mathcal{N}=4$ sYM presented in ref.~\cite{Caron-Huot:2020vlo}.
In particular we compare the leading-transcendental component of the gluon-gluon QCD scattering amplitude to the $\mathcal{N}=4$ one. 
Though this comparison is limited to a single partonic channel, it provides a strong validation of the MRK expansion procedure described in
\cref{se:amplexpansion} which is common to all partonic channels.
The main difference wrt ref.~\cite{Caron-Huot:2020vlo} is in the conventions adopted. In the current paper, particle $p_5 (p_1)$ travels along the positive
light-cone direction and $p_3 (p_2)$ travels along the negative one, whereas ref.~\cite{Caron-Huot:2020vlo} adopts the opposite convention. 
As discussed at the end of \cref{se:multireggekin}, this can be easily adjusted with the transformations $s_1 \leftrightarrow s_2$ and $z \leftrightarrow 1-\zbar$.
Once this is taken into account, together with the appropriate normalisations of the colour-basis elements, 
we find full agreement at one and two loops between the $\mathcal{N}=4$ sYM results and the leading-transcendental part of the $gg$-scattering amplitude. Although not new, in the ancillary files we provide the $\mathcal{N}=4$ sYM results in our kinematics and colour conventions, which we described in \cref{se:generalaspects}.

The second important check is the comparison of our one-loop QCD corrected Lipatov vertex to the results presented in ref.~\cite{DelDuca:1998cx} and more recently in ref.~\cite{Fadin:2023roz} to $\mathcal{O}(\epsilon^2)$ accuracy. The comparison against ref.~\cite{DelDuca:1998cx} is straightforward since 
the authors adopt a factorised expression for the $2\to3$ scattering amplitude, which is identical to our \cref{eq:2to3_factorised}. We find 
agreement with their results up to $\mathcal{O}(\epsilon^0)$. 
In ref.~\cite{Fadin:2023roz}, the bare Lipatov vertex was extracted through a diagrammatic calculation and presented in an analytic and gauge-invariant form equivalent to \cref{eq:2to3_analytic}.
The result in ref.~\cite{Fadin:2023roz} is expressed in terms of dimensionally-regulated one-loop triangle, box and pentagon integrals.
Instead we obtained an equivalent result from the one-loop helicity amplitudes computed to $\mathcal{O}(\epsilon^2)$.
As explained in \cref{se:comparison_to_regge}, we can make contact with the Regge-pole analytic form and, thanks to \cref{eq:rlto_cev_nll}, relate the absorptive ($R_\lambda+L_\lambda$) and dispersive ($R_\lambda-L_\lambda$) parts of the amplitude to $\wilsonU_\lambda$ at
$\mathcal{O}(\alpha_s)$ accuracy via
\begin{equation}
\label{eq:rpl_rml}
\begin{aligned}
& R_\lambda + L_\lambda = \frac{\wilsonU_\lambda  + \wilsonU^{*}_{-\lambda} }{2}V_\lambda,\\
& i\pi(\tau_A-\tau_B)\left(R_\lambda - L_\lambda\right) = 
i \pi (\tau_A+\tau_B)\frac{\wilsonU_\lambda  + \wilsonU^{*}_{-\lambda}}{2}V_\lambda - 2\left(\wilsonU_\lambda  - \wilsonU^{*}_{-\lambda}\right)V_\lambda\,.
\end{aligned}
\end{equation}
We remind the reader that, at one-loop order,  the Lipatov vertex $\mathcal{L}_\lambda$ of~\cref{eq:rlto_cev_nll} is identical to the universal $WWg$ vertex  $\wilsonU_\lambda$ introduced in~\cref{eq:W_lipatov}.
 We provide the complete one-loop result for $\wilsonU_\lambda$ to $\mathcal{O}(\epsilon^2)$ written in terms of Goncharov polylogarithms in the ancillary files, including the $N_f$ contributions which were omitted in the literature. Using this result, we readily obtain the $R_\lambda \pm L_\lambda$ combinations which agree with the results of ref.~\cite{Fadin:2023roz} to $\mathcal{O}(\epsilon^2)$.
We just mention that, in our conventions, $R_\lambda + L_\lambda = V_\lambda + \mathcal{O}(\alpha_s)$ whereas $R_\lambda - L_\lambda = \mathcal{O}(\alpha_s)$, and we stress again that~\cref{eq:rpl_rml} is strictly valid only up to next-to-leading order.

The last check regards the computation of the multi-$W$ contributions described in \cref{se:shockwave}. At one-loop, the results in \cref{sec:NLLother} provide a complete prediction of the even-signature amplitudes. Starting from these equations, combining the action of the signature operators given in \cref{se:colour} for the various partonic channels with the expressions for the 2d one-loop integrals presented in \cref{se:twodimintegrals} gives the complete result to $\mathcal{O}(\epsilon^2)$. We checked that we have full agreement with the corresponding expansion of the one-loop helicity amplitudes.
At the two-loop level, a strong check of the multi-$W$ transitions in the odd-odd amplitude is provided by their presence in sub-amplitudes that do not feature a double-octet exchange, thus where single- and multi-$W$ terms do not mix.
This allows for a direct comparison against the two-loop amplitudes 
presented in \cref{se:amplexpansion}. Also in this case we find perfect
agreement. In particular, we established that the sub-leading colour
contributions coming from multi-$W$ exchanges fully capture the
differences between the various partonic channels, after the effect of the
impact factors has been accounted for. Therefore, we have demonstrated
that multi-$W$ exchanges correctly describe the violation of universality
at NNLL. 

Finally, we note that an independent calculation of similar quantities is
ongoing, see ref.~\cite{Abreu:2024mpk}. We have corresponded with the authors
and found agreement for all quantities we have compared, namely the
results for multi-$W$ exchanges.

%%%%%% -------
%%%%%% -------
%%%%%% -------
%%%%%% -------
%%%%%% -------
%%%%%%%%%%%%%%%%%%%%%%%%%%%%%%%%%%%%%%%%%%%%%%%%%%%%%%%%%%%%%%%%%%%%%%%%%%%%%%%%%%%%%%%%%%%%%%%%%%%%%%%%%%%%%%%%%%%%%%%%%%%%%%%%%%%%%%%%%%%%%%%%%%%%%%%%%%%%%%%%%%%%%%%%%%%%%%%%%%%%%%%%%%
%%%%%% -------
%%%%%% -------
%%%%%% -------
%%%%%% -------
%%%%%% -------

\section{Conclusions}
\label{se:conclusions}

In this paper, we have presented an analysis of two-loop $2 \to 3$ QCD
scattering amplitudes in the high-energy limit. To do so, we extended
the results of ref.~\cite{Caron-Huot:2013fea} to deal with all the
contributions required for predicting the signature-odd two-loop
$2\to3$ MRK amplitude, and compared against a direct expansion of
full-colour QCD amplitudes. 
This was made possible thanks to the recent publication of the
latter~\cite{Agarwal:2023suw,DeLaurentis:2023izi,DeLaurentis:2023nss}. 
Our calculation allowed us to test MRK universality at the NNLL level
for the first time. We did so by separating contributions which are
independent of the flavour of the partons participating in the
scattering, from others which are flavour dependent but that can still
be described in terms of universal functions.  While doing so, we were
also able to test predictions for quantities which are closely related
to Regge cuts. We found that, besides the well-understood flavour
dependence of the impact factors, the only source of non-universality
arises from subleading-colour terms, which are accurately captured by
multiple exchanges of the $W$ field of \cref{se:shockwave}. These can
in turn be expressed in terms of single-valued, pure, and
uniform-transcendental quantities describing the kinematics in the
transverse plane. The form of these terms is also universal, but
it involves colour operators which act on the tree-level, and hence
depend on the colour representation of the scattering particles. 

As an important product of our investigations, we were able to extract
for the first time the vertex describing the emission of a
central-rapidity gluon, both in QCD and $\mathcal N=4$ sYM, at two-loop
accuracy. We 
verified that the latter is given by the leading-transcendental
contribution of the former.  This quantity, closely related to the
Lipatov vertex, was the last missing ingredient required to
calculate NNLL signature-odd amplitudes in MRK, to any multiplicity.
Indeed, once higher-multiplicity two-loop amplitudes will become available
in the literature, our results could provide a strong check of them.
We provide our main results through the ancillary files published alongside
this paper. 

Our study opens several interesting avenues for further investigation.
In this paper, we have limited ourselves to the study of
signature-even amplitudes at the one-loop level only. It would be
interesting to push this to higher orders. It would be equally
interesting to relax the strong rapidity ordering requirement and
study the next-to-MRK limit, and to explore the connections between
our results and the ongoing analysis within the SCET
framework~\cite{Rothstein:2016bsq,Rothstein:2024fpx,Moult:2022lfy,Gao:2024qsg}.
%%%
More broadly, it would be interesting to frame our study in the broader
context of Glauber physics. For example, very recently it has been pointed
out that Glauber modes are a key element for \emph{restoring} PDFs
factorisation in gap-between-jets cross sections~\cite{Becher:2024kmk}.
That analysis showed tantalising similarities with some of the features
of multi-$W$ exchanges contributions described in this paper, which would
be interesting to investigate further.

Also, we observed intriguing connections between
the Regge and IR structures of the theory. In the past, studying these
connections served as a guide towards a better understanding of both. We
anticipate that investigations in this direction may help elucidate
the relation between the quantities described in this paper and the ones
arising in the standard BFKL literature. 
This could potentially pave the way towards a better understanding of
the structure of Regge poles and Regge cuts in QCD, along the lines
reviewed in ref.~\cite{Fadin:2024eyf}. This would put on a solid
ground the precise relation between universal objects appearing in the
framework~\cite{Caron-Huot:2013fea} that we used in this paper, and
analogous ones in QCD Regge theory, like the Lipatov vertex.

Although
it is natural to imagine a close connection between the two along the
lines of ref.~\cite{Falcioni:2021dgr}, a rigorous proof of their precise
relation is still lacking. Establishing this could require studying
the evolution and the structure of radiative corrections of the
quantities described in \cref{se:shockwave,se:results}. 
This would in turn be crucial for
a generalisation of BFKL theory in QCD beyond NLL. We look forward
to pursuing these research avenues in the future.

%%%%%% -------
%%%%%% -------
%%%%%% -------
%%%%%% -------
%%%%%% -------
%%%%%%%%%%%%%%%%%%%%%%%%%%%%%%%%%%%%%%%%%%%%%%%%%%%%%%%%%%%%%%%%%%%%%%%%%%%%%%%%%%%%%%%%%%%%%%%%%%%%%%%%%%%%%%%%%%%%%%%%%%%%%%%%%%%%%%%%%%%%%%%%%%%%%%%%%%%%%%%%%%%%%%%%%%%%%%%%%%%%%%%%%%
%%%%%% -------
%%%%%% -------
%%%%%% -------
%%%%%% -------
%%%%%% -------

\acknowledgments We thank Andreas von Manteuffel
and Lorenzo Tancredi for discussions at the beginning of this project
and for encouraging us to pursue this work. We also thank Vittorio
Del Duca for collaborating during the initial stages of this work and
for several insightful discussions. 
Furthermore, we are grateful to Lorenzo Tancredi for many
valuable comments on this article.
We extend our gratitude to Samuel Abreu, Giulio Falcioni, Einan Gardi,
Giuseppe de Laurentis, Calum Milloy and Leonardo Vernazza for sharing 
their independent results on the 
multi-$W$ contributions and for conducting a thorough comparison. 
FC would like to thank Matthias Neubert for many interesting
discussions on the relation between the topics of this paper and
Glauber effects in the context of factorsation-breaking studies.
This research was partly supported
by the European Research Council (ERC) under the European Union’s
research and innovation programme grant agreements ERC Starting Grant
949279 \textsc{HighPHun} and ERC Starting Grant 804394
\textsc{hipQCD}.
FD is supported by the United States Department of Energy, Contract DE-AC02-76SF00515. 
FC performed this work in part at the Aspen
Center for Physics, which is supported by the National Science Foundation
Grant PHY-2210452.
We finally thank the CERN theory department for their
hospitality during the RAS workshop in August 2024 where part of this
work was conducted.

\appendix

%%%---------------------------------------------------------------------
\section{Spinor products and polarisation vectors}
\label{se:spinorproducts}
%%%---------------------------------------------------------------------
In order to explicitly define spinor-helicity variables, we work in the Weyl basis, where Dirac spinors and gamma matrices take the form
\begin{equation}
    \psi = 
    \begin{pmatrix} 
      \lambda_\alpha \\ 
      \tilde\lambda^{\dot\alpha}
    \end{pmatrix},\quad\quad
    \gamma^\mu =
    \begin{pmatrix}
    0 & \bar\sigma^{\mu}_{\alpha\dot\alpha} \\
    \sigma^{\mu,\dot\alpha\alpha} & 0
    \end{pmatrix},\quad\quad
    \gamma_5 = \begin{pmatrix}
    1 & 0 \\
    0 & -1
    \end{pmatrix},\quad\quad
\end{equation}
with $\sigma^\mu = (1,\vec\sigma)$, $\bar \sigma^\mu = (1,-\vec\sigma)$ and
$\sigma^i$ are the Pauli matrices. 
Requiring $\psi$ to be a solution of the massless Dirac equation $\slashed{p} \psi = 0$, we find that $ \lambda_\alpha$ and $\tilde\lambda^{\dot\alpha}$ must be solutions of the left- and right-handed Weyl equations respectively.
We write them in term of lightcone and complex transverse variables, defined
as
\begin{equation}
  p_{\pm} \equiv p^0 \pm p^z,
  \quad\quad p_\perp \equiv p^x + i p^y,
  \quad\quad
  \bar p_\perp \equiv p^x-i p^y,
\end{equation}
with
\begin{equation}
  p_+ p_- = p_\perp \bar p_\perp = \pp^2.
\end{equation}
The $\lambda_\alpha$ and $\tilde\lambda^{\dot\alpha}$ solutions then read
\begin{equation} \label{eq:explicit_lambdas}
  \lambda_\alpha = \begin{pmatrix}
    \sqrt{p_+}\\\sqrt{p_-}\frac{p_\perp}{|p_\perp|}\end{pmatrix},
  \quad\quad
  \tilde\lambda^{\dot\alpha} =
  \begin{pmatrix}
    \sqrt{p_-}\frac{\bar p_\perp}{|p_\perp|} \\ -\sqrt{p_+}\end{pmatrix},
\end{equation}
where we have chosen the normalisation for positive-energy states
$|\lambda|^2 = |\tilde\lambda|^2 = 2p^0 = p_+ + p_-$ and fixed the
overall complex phase for each spinor. Any other phase choice is also
valid.  With this we define the standard spinor products as
\begin{equation}
  \langle ij \rangle = \epsilon^{\alpha\beta}i_\alpha j_\beta,
  \quad\quad [ij] = \epsilon_{\dot\alpha\dot\beta}\,\tilde i^{\dot
    \alpha}\tilde j^{\dot\beta},
  \label{eq:spinprod}
\end{equation}
with $\epsilon^{12} = - \epsilon_{12} = -1$.
This immediately leads us to
\begin{equation}
  \begin{split}
    \langle ij\rangle = -\sqrt{p_i^+}\sqrt{p_j^-}\frac{\pP{j}}{|p_{\perp,j}|} +
    \sqrt{p_i^-}\sqrt{p_j^+} \frac{\pP{i}}{|p_{\perp,i}|},
    \\
         [ij] = -\sqrt{p_i^-}\sqrt{p_j^+} \frac{\pPc{i}}{|p_{\perp,i}|} +
         \sqrt{p_i^+}\sqrt{p_j^-}\frac{\pPc{j}}{|p_{\perp,j}|},
  \end{split}
  \label{eq:genspinors}
\end{equation}
which for real momenta implies $\langle ij \rangle^* = -[ji]$.
We define the analytic continuation to negative energy starting from \cref{eq:genspinors} and choose the standard branch of the square root: $\sqrt{z} = \sqrt{e^{i \theta} |z|} = e^{i \theta/2} \sqrt{|z|}$ with $\theta = \arg(z) \in (-\pi,\pi]$. As a result we find
\begin{equation}
  \begin{gathered}
    \langle (-p_i) p_j \rangle =
    \langle p_i (-p_j) \rangle = i \langle p_i p_j\rangle,\quad\quad
    \langle (-p_i) (-p_j) \rangle = - \langle p_i p_j \rangle, \\
    [ (-p_i) p_j ] =
    [ p_i (-p_j) ] = i [ p_i p_j],\quad\quad
    [ (-p_i) (-p_j) ] = - [ p_i p_j ], \\
  \end{gathered}
  \label{eq:ancont_spinors}
\end{equation}
with $p_{i,j}^{0} > 0$. 
From now on we then assume that all energies
are positive. 
For positive energies, when the associated momentum lies on the light cone $p=p_\pm$, we need to specify the azimuthal angle, \ie the ratio $p/|p|$. 
If $p = p_+$, we decide to set the azimuthal angle to zero ($p/|p| =1$), while if $p=p_-$ we set the azimuthal
angle to $\pi$ ($p/|p| = -1$). 
This completely fixes our notation.

At leading power in MRK, for the spinor products $\langle ij \rangle$ we then
have
\begin{gather}
  \langle12\rangle = -\sqrt{p^{-}_{3}p^{+}_{5}},
  \quad
  \langle13\rangle = -i\sqrt{\frac{p^{+}_{5}}{p^{+}_{3}}} \pP{3},
  \quad
  \langle14\rangle = -i\sqrt{\frac{p^{+}_{5}}{p^{+}_{4}}} \pP{4},
  \quad
  \langle15\rangle = -i p_{\perp,5}, \nonumber \\
  \langle23\rangle = -i \sqrt{p^{-}_{3} p^{+}_{3}},\quad
  \langle24\rangle = -i\sqrt{p^{-}_{3}p^{+}_{4}},\quad
  \langle25\rangle = -i\sqrt{p^{-}_{3}p^{+}_{5}}, \label{eq:spin_angle}\\
  \langle34\rangle = \sqrt{\frac{p^{+}_{4}}{p^{+}_{3}}} \pP{3},\quad
  \langle35\rangle = \sqrt{\frac{p^{+}_{5}}{p^{+}_{3}}} \pP{3},\quad
  \langle45\rangle = \sqrt{\frac{p^{+}_{5}}{p^{+}_{4}}} \pP{4}, \nonumber
\end{gather}
while for $[ij]$ we get
\begin{gather}
  [12] = \sqrt{p^{-}_{3}p^{+}_{5}}, \quad
  [13] = i\sqrt{\frac{p^{+}_{5}}{p^{+}_{3}}} \pPc{3}, \quad
  [14] = i\sqrt{\frac{p^{+}_{5}}{p^{+}_{4}}} \pPc{4}, \quad
  [15] = i \bar p_{\perp,5}, \nonumber \\
  [23] = i \sqrt{p^{-}_{3} p^{+}_{3}},\quad
  [24] = i\sqrt{p^{-}_{3}p^{+}_{4}},\quad
  [25] = i\sqrt{p^{-}_{3}p^{+}_{5}}, \label{eq:spin_square}\\
  [34] = -\sqrt{\frac{p^{+}_{4}}{p^{+}_{3}}} \pPc{3},\quad
  [35] = -\sqrt{\frac{p^{+}_{5}}{p^{+}_{3}}} \pPc{3},\quad
  [45] = -\sqrt{\frac{p^{+}_{5}}{p^{+}_{4}}} \pPc{4}. \nonumber
\end{gather}
Eqs.~\eqref{eq:spin_angle} and \eqref{eq:spin_square} allow us to express all spinor products in terms of lightcone momenta. However we can go further and replace the components $p_i^{\pm}$ and $p_{i,\perp}$ with their expressions in terms of the variables $s$, $s_{1,2}$ and $z(\zbar)$. The resulting spinor products can then be directly plugged into the amplitudes.

We start by writing the momenta as
\begin{equation}
    p_{1,2} = \frac{\sqrt{s_{12}}}{2}\left(1,0,0,\pm 1\right)
    \quad\quad \mathrm{and} \quad\quad
    p_{i} = |\pp_i|\left(\cosh y_i, \cos\phi_i,\sin\phi_i,\sinh y_i \right)\,.
\end{equation}
The parity odd invariant can the be written as
\begin{equation}
\label{eq:tr5_three_ids}
    \mathrm{tr}_5 =
    2 i s_{12} |\pp_3||\pp_4|\sin(\phi_3-\phi_4) = 
    2 i s_{12} |\pp_3||\pp_5|\sin(\phi_5-\phi_3) =
    2 i s_{12} |\pp_4||\pp_5|\sin(\phi_4-\phi_5) \,,
\end{equation}
with $\phi_i \in [0,2\pi)$.
We then exploit the azimuthal invariance in the transverse plane to fix $\phi_4 = \pi$.
Recalling then from \cref{se:multireggekin} that in MRK we have 
\begin{equation}
    \mathrm{tr}_5 = \frac{s_1 s_2}{x^2} (z-\zbar),\;\;\;
    |\pp_3| = \sqrt{\frac{s_1 s_2}{s}} |z|, \;\;\;
    |\pp_4| = \sqrt{\frac{s_1 s_2}{s}}, \;\;\;
    |\pp_5| = \sqrt{\frac{s_1 s_2}{s}} |1-z|\,,
\end{equation}
we can easily solve the first and last identities in~\cref{eq:tr5_three_ids} to get
\begin{equation}
\label{eq:sin_sol}
    \sin(\phi_3) = \frac{\Im(\zbar)}{|z|}
    \quad\quad \mathrm{and} \quad\quad
    \sin(\phi_5) = \frac{\Im(1-\zbar)}{|1-z|}\,,
\end{equation}
which immediately give
\begin{equation}
\label{eq:cos_sol}
    \cos(\phi_3) = \lambda_3 \frac{\Re(z)}{|z|}
    \quad\quad \mathrm{and} \quad\quad
    \cos(\phi_5) = \lambda_5 \frac{\Re(1-z)}{|1-z|}\,,
\end{equation}
where both $\lambda_3$ and $\lambda_5$ can either be $+1$ or $-1$.
The second relation in~\cref{eq:tr5_three_ids}, when combined with~\cref{eq:sin_sol,eq:cos_sol}, finally gives
\begin{equation}
    \lambda_3 + (\lambda_3 - \lambda_5)\Re(z) = 1,
\end{equation}
which in order to be valid for any value of $z$ implies $\lambda_3 = \lambda_5 =1$.
These relations completely determine the transverse components of $p_{3}$, $p_4$ and $p_5$ which are thus given by
\begin{equation}
    p_{4,\perp} = -\sqrt{\frac{s_1 s_2}{s}},\quad\quad
    p_{3,\perp} =  \sqrt{\frac{s_1 s_2}{s}}\zbar\,,\quad\quad
    p_{5,\perp} =  \sqrt{\frac{s_1 s_2}{s}}(1-\zbar)\,,
\end{equation}
which in turn yield \cref{eq:zzb_to_qab}.

Let us now discuss the polarisation of the centrally-emitted gluon.
In the main text, we have used $\vep^*_{\lambda_4}$ to denote the
polarisation vector of the central gluon. As explained there,
\cref{eq:OPE} (and its subsequent expansions) is derived under the
assumption $\vep^*_{\lambda_4}\cdot p_2 = 0$. An equivalent formula
holds for the emission from the ``$-$'' lightcone, i.e. for emissions
from the $B$ line.  In this case though, the $\vep^*_{\lambda_4}\cdot
p_1$ must be imposed.  To distinguish the two cases, we use the symbol
$\vep^*_{\lambda_4}$ to denote a polarisation vector that satisfies
the condition $\vep^*_{\lambda_4}\cdot p_2 = 0$ (i.e. the one that
should be used in \cref{eq:OPE} and its expansion), and
$\tilde\vep^*_{\lambda_4}$ for the one which satisfies
$\tilde\vep^*_{\lambda_4}\cdot p_1 = 0$ (i.e. the one that should be used
for the equivalent of \cref{eq:OPE} for emissions from the $B$ line).
In both cases, we denote the transverse components of these vectors by
the bold symbols $\boldsymbol{\vep} = \{\vep^x,\vep^y\}$ and
$\boldsymbol{\tilde\vep} = \{\tilde\vep^{\,x},\tilde\vep^{\,y}\}$.

To find explicit representations for the polarisation vectors, we use the
standard spinor-helicity representation\footnote{Compared to standard
spinor-helicity literature, here we use $\vep^{*}$ (rather than the
short-hand $\vep$) to make it explicit the fact that the gluon
is out-going.}
\begin{equation} 
  \vep_+^{*\mu} \equiv \frac{1}{\sqrt2}\frac{\langle q \gamma^\mu p_4]}{\langle q p_4\rangle}, 
  \quad\quad 
  \vep_-^{*\mu} \equiv -\frac{1}{\sqrt2}\frac{[ q \gamma^\mu p_4 \rangle}{[q p_4]},
\end{equation}
with $q=p_2$ for $\vep^{*}_{\lambda_4}$ and $q=p_1$ for $\tilde\vep^{*}_{\lambda_4}$. Explicitly
\begin{align}
  \label{eq:explicit_polvec}
    &\vep^*_+ = \frac{1}{\sqrt{2}}\left(\frac{\pPc{4}}{p^+_4},1,-i,-\frac{\pPc{4}}{p^+_4}\right),
    \quad
    &&\vep^*_- = \frac{1}{\sqrt{2}}\left(\frac{\pP{4}}{p^+_4},1,i,-\frac{\pP{4}}{p^+_4}\right),
    \notag\\
    &\tilde\vep^*_+ = \frac{-1}{\sqrt{2}}\left(\frac{p^+_4}{\pP{4}},\frac{\pPc{4}}{\pP{4}},i \frac{\pPc{4}}{\pP{4}}, \frac{p^+_4}{\pP{4}}\right),
    \quad
    &&\tilde\vep^*_- = \frac{-1}{\sqrt{2}}\left(\frac{p^+_4}{\pPc{4}},\frac{\pP{4}}{\pPc{4}},-i \frac{\pP{4}}{\pPc{4}}, \frac{p^+_4}{\pPc{4}}\right).
\end{align}
%
%
%
%
%
%%%---------------------------------------------------------------------
\section{Colour tensors and operators}
\label{se:appendixcolour}
%%%---------------------------------------------------------------------
We introduce below a set of tensors useful for the definition of the colour bases employed in the MRK limit. We have
\begin{align}
\label{eq:coltensors1}
\allowdisplaybreaks
    \mathcal{T}_{10-\overline{10}}^{abcd} &= \frac{i}{2} f^{acx} d^{xbd} + \frac{i}{2} d^{acx} f^{xbd}\,, \\
    \mathcal{T}_{10+\overline{10}}^{abcd} &= \frac{1}{2} \left(\delta^{ac}\delta^{bd}-\delta^{ad}\delta^{bc}\right)-\frac{f^{abx}f^{xcd}}{N_c}\,, \\
    \mathcal{T}_{27}^{abcd} &= -\frac{N_c-1}{2N_c} \mathcal{P}_{1}^{abcd} -\frac{N_c-2}{2N_c} \mathcal{P}_{8s}^{abcd}+\frac{1}{2}\mathcal{S}^{abcd} + \mathcal{W}_{-}^{abcd} + \mathcal{W}_{+}^{abcd} \,, \\
    \mathcal{T}_{0}^{abcd} &= -\frac{N_c+1}{2N_c} \mathcal{P}_{1}^{abcd} -\frac{N_c+2}{2N_c} \mathcal{P}_{8s}^{abcd} +\frac{1}{2}\mathcal{S}^{abcd} - \mathcal{W}_{-}^{abcd} - \mathcal{W}_{+}^{abcd} \,, \\
    \mathcal{T}_{10}^{abcd} &= \frac{1}{2} \left(\mathcal{A}^{abcd} - \mathcal{P}_{8a}^{abcd}\right) + \left(\mathcal{W}_{+}^{abcd} - \mathcal{W}_{-}^{abcd} \right) \,, \\
    \mathcal{T}_{\overline{10}}^{abcd} &= \frac{1}{2} \left(\mathcal{A}^{abcd} - \mathcal{P}_{8a}^{abcd}\right) - \left(\mathcal{W}_{+}^{abcd} - \mathcal{W}_{-}^{abcd} \right)\,,
\label{eq:coltensors6}
\end{align}
where we used the building blocks 
\begin{gather}
\label{eq:projectors}
    \mathcal{P}_1^{abcd} = \frac{1}{N_c^2-1} \delta^{ab}\delta^{cd}, \quad\quad
    \mathcal{P}_{8a}^{abcd} = \frac{1}{N_c} f^{abe}f^{ecd}, \quad\quad
    \mathcal{P}_{8s}^{abcd} = \frac{N_c}{N_c^2-4} d^{abe}d^{ecd}, \nonumber \\
    \mathcal{S}^{abcd} = \frac{1}{2}\left(\delta^{a c}\delta^{b d} + \delta^{a d}\delta^{b c}\right), \quad\quad
    \mathcal{A}^{abcd} = \frac{1}{2}\left(\delta^{a c}\delta^{b d} - \delta^{a d}\delta^{b c}\right), \\
    \mathcal{W}_{+}^{abcd} = \mathrm{Tr}\left( T^a T^c T^b T^d\right), \quad\quad
    \mathcal{W}_{-}^{abcd} = \mathrm{Tr}\left( T^a T^d T^b T^c\right)\,. \nonumber
\end{gather}
\begingroup
\renewcommand{\arraystretch}{1.45}
\begin{figure*}[t!]
    \hspace{0mm}
    \begin{minipage}[t]{0.30\textwidth}
        \centering
        \scalebox{0.975}{
            \begin{tabular}[t]{|c|c|}\hline
                $(r_1,r_2)$ & $\mathcal{C}^{[gg]}_i$ \\
                \hline
                          $(8_a,8_a)$   & $i f^{a_5 b a_1}\, if^{a_3 c a_2}\, i f^{b c a_4}$ \\
                          $(8_a,8_s)$   & $\frac{N_c^2}{N_c^2-4}\, i f^{a_5 b a_1}\, d^{a_3 c a_2}\,d^{b c a_4}$ \\
                          $(8_s,8_a)$   & $\frac{N_c^2}{N_c^2-4}\,d^{a_5 b a_1}\, i f^{a_3 c a_2}\, d^{b c a_4} $ \\
                          $(8_s,8_s)$   & $\frac{N_c^2}{N_c^2-4}\,d^{a_5 b a_1}\, d^{a_3 c a_2}\, i f^{b c a_4} $ \\
                          $(1,8_a)$  & $\frac{N_c^2}{N_c^2-1}\,\delta^{a_5 a_1} i f^{a_3 a_4 a_2}$ \\
                         $(8_a,1)$   & $\frac{N_c^2}{N_c^2-1}\,i f^{a_5 a_4 a_1} \delta^{a_3 a_2}$ \\
                         $(8_a,0)$   & $i f^{a_5 b a_1} \mathcal{T}_0^{a_4 b a_2 a_3}$ \\
                          $(0,8_a)$   & $\mathcal{T}_0^{a_4 b a_1 a_5} i f^{a_3 b a_2}$ \\
                         $(8_a,27)$   & $i\,f^{a_5 b a_1} \mathcal{T}_{27}^{a_4 b a_2 a_3}$ \\
                         $(27,8_a)$   & $\mathcal{T}_{27}^{a_4 b a_1 a_5} i f^{a_3 b a_2} $ \\
                         $(8_a,10)$   & $i f^{a_5 b a_1} \mathcal{T}_{10+\overline{10}}^{a_4 b a_2 a_3}$ \hspace{1mm}\\
                          \hline
            \end{tabular}
        }
    \end{minipage}
    \quad\quad\quad\;
    \begin{minipage}[t]{0.30\textwidth}
        \centering
        \scalebox{0.975}{
            \begin{tabular}[t]{|c|c|}\hline
                $(r_1,r_2)$ & $\mathcal{C}^{[gg]}_i$ \\
                \hline
                        $(10,8_a)$   & $ \mathcal{T}_{10+\overline{10}}^{a_4 b a_1 a_5} i f^{a_3 b a_2}$ \\ 
                        $(8_s,10)$   & $\frac{N_c^2}{N_c^2-4}\,d^{a_5 b a_1} \mathcal{T}_{10-\overline{10}}^{a_4 b a_2 a_3}$ \\
                        $(10,8_s)$   & $\frac{N_c^2}{N_c^2-4}\,\mathcal{T}_{10-\overline{10}}^{a_4 b a_1 a_5} d^{a_3 b a_2}$ \\
                        $(27,27)$   & $\mathcal{T}_{27}^{b c a_2 a_3} \mathcal{T}_{27}^{b e a_1 a_5}\, if^{c a_4 e}$ \\
                        $(0,0)$   & $\mathcal{T}_{0}^{b c a_2 a_3} \mathcal{T}_{0}^{b e a_1 a_5}\, if^{c a_4 e}$ \\
                        $(10,0)$   & $\mathcal{T}_{10+\overline{10}}^{b c a_1 a_5} \mathcal{T}_{0}^{b e a_2 a_3}\, if^{e a_4 c}$ \\
                        $(0,10)$   & $\mathcal{T}_{0}^{b e a_1 a_5}\mathcal{T}_{10+\overline{10}}^{b c a_2 a_3}\, if^{c a_4 e}$ \\
                        $(10,27)$   & $\mathcal{T}_{10+\overline{10}}^{b e a_1 a_5}\mathcal{T}_{27}^{b c a_2 a_3}\, if^{c a_4 e}$ \\
                        $(27,10)$   & $\mathcal{T}_{27}^{b e a_1 a_5}\mathcal{T}_{10+\overline{10}}^{b c a_2 a_3}\, if^{c a_4 e}$ \\
                        $(10,10)_1$   & $\mathcal{T}_{10+\overline{10}}^{b e a_1 a_5} \mathcal{T}_{10+\overline{10}}^{b c a_2 a_3}\, if^{c a_4 e}$ \\
                          $(10,10)_2$ & $
                        \mathcal{T}_{10+\overline{10}}^{b e a_1 a_5} \big(\mathcal{T}_{10-\overline{10}}^{b c a_2 a_3}\, d^{c a_4 e} - \mathcal{T}_{10+\overline{10}}^{b c a_2 a_3} \, if^{c a_4 e}/N_c\big)$ \\
                        \hline
            \end{tabular}
        }
    \end{minipage}
    \captionof{table}{Colour basis for $gg$ scattering. In the right column we define the basis elements, while in the left column their associated irreducible representations $(r_1,r_2)$ as defined in \cref{eq:colrep}.}
    \label{tab:colorGG}
\end{figure*}
\endgroup
%%%~~~~~~~~~~~~~~~~~~~~~~~~~~~~~~~~~~~~~~~~~~~~~~~~~~~~~~~~~~~~~
\begingroup
\begin{figure*}[h!]
    \centering
    \hspace{-0.1cm}\begin{minipage}[t]{0.27\textwidth}
    \renewcommand{\arraystretch}{1.3}
       \centering
       \scalebox{0.975}{
            \begin{tabular}[t]{|c|c|} \hline
                $(r_1,r_2)$ & $\mathcal{C}_i^{[qg]}$ \\
                \hline
                 $(8,8_a)_a$ & $T^b_{i_5 i_1} i f^{b c a_4} i f^{a_3 c a_2}$ \\
                 $(8,8_s)_s$ & $\frac{N_c^2}{N_c^2-4}\,T^b_{i_5 i_1} d^{c b a_4 } d^{a_3 c a_2}$\\
                 $(8,8_s)_a$ & $T^b_{i_5 i_1} i f^{b c a_4} d^{c a_2 a_3}$ \\ 
                 $(8,8_a)_s$ & $T^b_{i_5 i_1} d^{b c a_4} i f^{c a_2 a_3}$ \\
                 $(1,8_a)$   & $\delta_{i_5 i_1} i f^{a_4 a_2 a_3}$ \\
                 $(1,8_s)$   & $\delta_{i_5 i_1} d^{a_4 a_2 a_3}$ \\ \hline
            \end{tabular}
        }
    \end{minipage}
    \hspace{1.1cm}
    \begin{minipage}[t]{0.27\textwidth}
     \renewcommand{\arraystretch}{1.517}
        \centering
        \scalebox{0.975}{
            \begin{tabular}[t]{|c|c|} \hline
                $(r_1,r_2)$ & $\mathcal{C}_i^{[qg]}$ \\
                \hline
                 $(8,1)$     & $\frac{N_c^2}{N_c^2-1}\,T^{a_4}_{i_5 i_1} \delta^{a_2 a_3}$ \\
                 $(8,10)_1$  & $T^b_{i_5 i_1} \mathcal{T}_{10-\overline{10}}^{a_4 b a_2 a_3}$ \\
                 $(8,10)_2$  & $T^b_{i_5 i_1} \mathcal{T}_{10+\overline{10}}^{a_4 b a_2 a_3}$ \\
                 $(8,27)$    & $T^b_{i_5 i_1} \mathcal{T}_{27}^{a_4 b a_2 a_3}$ \\ 
                 $(8,0)$     & $T^b_{i_5 i_1} \mathcal{T}_{0}^{a_4 b a_2 a_3}$ \\
                          \hline
            \end{tabular}
        } \\ \vspace{2mm}\hspace{-40mm}(a)
    \end{minipage}
    \quad\quad\;
    \begin{minipage}[t]{0.27\textwidth}
     \renewcommand{\arraystretch}{1.82}
        \centering
        \scalebox{0.975}{
            \begin{tabular}[t]{|c|c|} \hline
                $(r_1,r_2)$ & $\mathcal{C}_i^{[qQ]}$ \\
                \hline
                 $(8,8)_a$ & $T^{b}_{i_5 i_1} T^{c}_{i_3 i_2} i f^{b c a_4}$ \\
                 $(8,8)_s$ & $T^{b}_{i_5 i_1} T^{c}_{i_3 i_2} d^{b c a_4}$ \\
                 $(8,1)$  & $T^{a_4}_{i_5 i_1} \delta_{i_3 i_2}$ \\ 
                 $(1,8)$  & $\delta_{i_5 i_1} T^{a_4}_{i_3 i_2}$ \\ \hline
            \end{tabular}
        } \\ \vspace{2mm} (b)
    \end{minipage}
    \captionof{table}{Definition of the colour bases for (a) the $qg$ and (b) the $qQ$ scattering channels.}
    \label{tab:colorQG}
\end{figure*}
%\twocolumngrid
\endgroup
We then use the tensors in eqs.~(\ref{eq:coltensors1}---\ref{eq:coltensors6}) to fully specify the colour bases for the $gg$, $qg$ and $qQ$ scattering channels which are given explicitly in \cref{tab:colorGG,tab:colorQG}.
Within these colour bases, the diagonal operators $({\bf T}_{15}^+)^2$ and $({\bf T}_{23}^+)^2$ for the different partonic channels are expressed as  
\begin{align}
    ({\bf T}_{15}^+)^2_{gg} &= \mathrm{diag}\big(N_c,N_c,N_c,N_c,0,N_c,N_c,2 \left(N_c-1\right),N_c,2 \left(N_c+1\right),N_c,2 N_c,N_c,2 N_c, \notag\\
    &\quad\quad\quad\;\; 2
   \left(N_c+1\right),2 \left(N_c-1\right),2 N_c,2 \left(N_c-1\right),2 N_c,2 \left(N_c+1\right),2 N_c,2 N_c\big), \notag\\
   ({\bf T}_{23}^+)^2_{gg} &= \mathrm{diag}\big(N_c,N_c,N_c,N_c,N_c,0,2 \left(N_c-1\right),N_c,2 \left(N_c+1\right),N_c,2 N_c,N_c,2 N_c,N_c, \notag\\
    &\quad\quad\quad\;\; 2
   \left(N_c+1\right),2 \left(N_c-1\right),2 \left(N_c-1\right),2 N_c,2 \left(N_c+1\right),2 N_c,2 N_c,2 N_c\big), \notag\\
   ({\bf T}_{15}^+)^2_{qg} &= \mathrm{diag}\big(N_c,N_c,N_c,N_c,0,0,N_c,N_c,N_c,N_c,N_c\big), \\
   ({\bf T}_{23}^+)^2_{qg} &= \mathrm{diag}\big(N_c,N_c,N_c,N_c,N_c,N_c,0,2 N_c,2 N_c,2 \left(N_c+1\right),2 \left(N_c-1\right)\big), \notag\\
   ({\bf T}_{15}^+)^2_{qQ} &= \mathrm{diag}\big(N_c,N_c,N_c,0\big), \notag\\
   ({\bf T}_{23}^+)^2_{qQ} &= \mathrm{diag}\big(N_c,N_c,0,N_c\big). \notag
\end{align}
With the aid of \cref{eq:tpp_diag} one can then easily derive the expressions for the diagonal operator $\boldsymbol{\mathcal{T}}_{++}$.
Additionally, the action of the signature-definite colour operators on the tree-level amplitudes in the $gg$-channel is given by
\begin{align}
    \boldsymbol{\mathcal{T}}_{--} \,\mathcal{C}_{(8a,8a)}^{[gg]} &= 
    \frac{N_c}{2} \mathcal{C}_{(8s,8s)}^{[gg]} + 4 \mathcal{C}_{(27,27)}^{[gg]} + 4 \mathcal{C}_{(0,0)}^{[gg]}, \notag\\
    \boldsymbol{\mathcal{T}}_{-+} \,\mathcal{C}_{(8a,8a)}^{[gg]} &= 
    \frac{N_c}{2} \mathcal{C}_{(8s,8a)}^{[gg]} + 2 \mathcal{C}_{(1,8a)}^{[gg]} + 2 \mathcal{C}_{(0,8a)}^{[gg]} + 2 \mathcal{C}_{(27,8a)}^{[gg]}, \notag\\
    \boldsymbol{\mathcal{T}}_{+-} \,\mathcal{C}_{(8a,8a)}^{[gg]} &= 
    -\frac{N_c}{2} \mathcal{C}_{(8a,8s)}^{[gg]} -2 \mathcal{C}_{(8a,1)}^{[gg]} - 2 \mathcal{C}_{(8a,0)}^{[gg]} - 2 \mathcal{C}_{(8a,27)}^{[gg]}, \notag\\
    \boldsymbol{\mathcal{T}}_{--}^2 \,\mathcal{C}_{(8a,8a)}^{[gg]} &= 
    \left( \frac{N_c^2}{4} + 4\right) \mathcal{C}_{(8a,8a)}^{[gg]} + N_c \mathcal{C}_{(8a,10)}^{[gg]} - N_c \,\mathcal{C}_{(10,8a)}^{[gg]} - 8 N_c\, \mathcal{C}_{(10,10)_1}^{[gg]}, \\
    \boldsymbol{\mathcal{T}}_{-+}^2 \,\mathcal{C}_{(8a,8a)}^{[gg]} &= 
    \left( \frac{N_c^2}{4} + 6\right)  \mathcal{C}_{(8a,8a)}^{[gg]} + 3 N_c\, \mathcal{C}_{(10,8a)}^{[gg]}, \notag\\
    \boldsymbol{\mathcal{T}}_{+-}^2 \,\mathcal{C}_{(8a,8a)}^{[gg]} &= 
    \left( \frac{N_c^2}{4} + 6\right)  \mathcal{C}_{(8a,8a)}^{[gg]} - 3 N_c \,\mathcal{C}_{(8a,10)}^{[gg]}, \notag
\end{align}
whereas for the $qg$-channel reads
\begin{align}    
    \boldsymbol{\mathcal{T}}_{--} \,\mathcal{C}_{(8,8a)_a}^{[qg]} &= 
    \frac{N_c}{2} \mathcal{C}_{(8,8s)_a}^{[qg]}, \notag\\
    \boldsymbol{\mathcal{T}}_{-+} \,\mathcal{C}_{(8,8a)_a}^{[qg]} &= 
    \frac{N_c}{2}
    \mathcal{C}_{(8,8a)_s}^{[qg]}
    + \mathcal{C}_{(1,8a)}^{[qg]}, \notag\\
    \boldsymbol{\mathcal{T}}_{+-} \,\mathcal{C}_{(8,8a)_a}^{[qg]} &= 
    -\frac{N_c}{2} \mathcal{C}_{(8,8s)_s}^{[qg]} 
    -2\mathcal{C}_{(8,1)}^{[qg]}
    -2 \mathcal{C}_{(8,27)}^{[qg]}
    -2 \mathcal{C}_{(8,0)}^{[qg]}, \notag\\
    \boldsymbol{\mathcal{T}}_{--}^2 \,\mathcal{C}_{(8,8a)_a}^{[qg]} &= 
    \left( \frac{N_c^2}{4} - 1\right) \mathcal{C}_{(8,8a)_a}^{[qg]}
    +N_c \,\mathcal{C}_{(8,10)_2}^{[qg]}, \\
    \boldsymbol{\mathcal{T}}_{-+}^2 \,\mathcal{C}_{(8,8a)_a}^{[qg]} &= 
    \left( \frac{N_c^2}{4} + 1\right)  \mathcal{C}_{(8,8a)_a}^{[qg]}, \notag\\
    \boldsymbol{\mathcal{T}}_{+-}^2 \,\mathcal{C}_{(8,8a)_a}^{[qg]} &= 
    \left( \frac{N_c^2}{4} + 6\right)  \mathcal{C}_{(8,8a)_a}^{[qg]}
    -3 N_c \,\mathcal{C}_{(8,10)_2}^{[qg]}, \notag
\end{align}
and finally in the $qQ$ channel
\begin{align}
    \boldsymbol{\mathcal{T}}_{--} \,\mathcal{C}_{(8,8)_a}^{[qQ]} &= 
    \frac{N_c^2-4}{2 N_c} \mathcal{C}_{(8,8)_a}^{[qQ]},  \notag \\
    \boldsymbol{\mathcal{T}}_{-+} \,\mathcal{C}_{(8,8)_a}^{[qQ]} &= 
    \frac{N_c}{2}
    \mathcal{C}_{(8,8)_s}^{[qQ]}
    +
    \mathcal{C}_{(1,8)}^{[qQ]}, \notag\\
    \boldsymbol{\mathcal{T}}_{+-} \,\mathcal{C}_{(8,8)_a}^{[qQ]} &= 
    -\frac{N_c}{2}
    \mathcal{C}_{(8,8)_s}^{[qQ]}
    -
    \mathcal{C}_{(8,1)}^{[qQ]}, \notag \\
    \boldsymbol{\mathcal{T}}_{--}^2 \,\mathcal{C}_{(8,8)_a}^{[qQ]} &= 
   \frac{(N_c^2-4)^2}{4 N_c^2} \mathcal{C}_{(8,8)_a}^{[qQ]}, 
    \\
    \boldsymbol{\mathcal{T}}_{-+}^2 \,\mathcal{C}_{(8,8)_a}^{[qQ]} &= 
    \left( \frac{N_c^2}{4} + 1\right)  \mathcal{C}_{(8,8)_a}^{[qQ]}, \notag \\
    \boldsymbol{\mathcal{T}}_{+-}^2 \,\mathcal{C}_{(8,8)_a}^{[qQ]} &= 
    \left( \frac{N_c^2}{4} + 1\right)  \mathcal{C}_{(8,8)_a}^{[qQ]}\, . \notag 
\end{align}

%%%---------------------------------------------------------------------
\section{Anomalous dimensions}
\label{se:beta_and_anomalous}
%%%---------------------------------------------------------------------
The $\beta$-function coefficients used in this paper are defined via
\begin{equation}
\frac{d \as }{d \ln \mu}  = \beta(\as) - 2\epsilon \as \; , \quad
\beta(\as) = -2 \as \sum\limits_{n=0} \beta_n \left(\frac{\as}{4 \pi}\right)^{n+1}  
\end{equation}
so that up to two-loops we have
\begin{equation}
\begin{aligned}
\beta_0 &= \frac{11}{3} C_A - \frac{2}{3}\: N_f  ,\\
\beta_1 &= \frac{34}{3} \: C_A^2-\frac{10}{3}\: C_A \:N_f - 2 \:C_F\: N_f  ,\\ 
\end{aligned}
\end{equation}
with the $SU(N_c)$ Casimir constants
\begin{equation}
    C_A = N_c, \quad  C_F = \frac{N_c^2-1}{2N_c}  .
\end{equation}
Here we also report the cusp anomalous dimension $\gamma_K$ as well as the quark and gluon collinear anomalous dimensions $\gamma_i$. They admit the series expansion in $\as$
\begin{equation}
\gamma_K = \sum\limits_{n=1} \left(\frac{\as}{4 \pi}\right)^{n}  \gamma^{(n)}_K ,  \qquad \gamma_{q,g} = \sum\limits_{n=1} \left(\frac{\as}{4 \pi}\right)^{n}  \gamma^{(n)}_{q,g}.
\end{equation}
To the  order relevant for this paper, their perturbative coefficients read
\begin{equation}
\label{eq:cusp_anomalous_coefficients}
\begin{aligned}
   \gamma_K^{(1)}&= 4 , \\
   \gamma_K^{(2)} &= \left( \frac{268}{9} 
    - \frac{4\pi^2}{3} \right) C_A - \frac{40}{9}\, N_f , \\
   \gamma_K^{(3)} & = C_A^2 \left( \frac{490}{3} 
    - \frac{536\pi^2}{27}
    + \frac{44\pi^4}{45} + \frac{88}{3}\,\zeta_3 \right) + C_A  N_f  \left(  \frac{80\pi^2}{27}- \frac{836}{27} - \frac{112}{3}\,\zeta_3 \right) \\
    &+ C_F N_f \left(32\zeta_3 - \frac{110}{3}\right) - \frac{16}{27}\, N_f^2  , \\[8pt]
   \gamma^{(1)}_q &= -3 C_F , \\
   \gamma^{(2)}_q &= C_F^2 \left( 2\pi^2 - 24\zeta_3 -\frac{3}{2}
     \right)
    + C_F C_A \left( 26\zeta_3   - \frac{11\pi^2}{6} - \frac{961}{54} 
    \right)
    + C_F  N_f \left( \frac{65}{27} + \frac{\pi^2}{3} \right) ,
    \\[8pt]
     \gamma^{(1)}_g &=   -\beta_0   , \\
   \gamma^{(2)}_g  &=    C_A^2 \left( -\frac{692}{27} + \frac{11\pi^2}{18} + 2 \zeta_3 \right)+C_A N_f \left( \frac{128}{27} - \frac{\pi^2}{9} \right) + 2 C_F N_f . \\
\end{aligned}
\end{equation}
Below we also provide the explicit perturbative expansion of the quantities defined in \cref{eq:KKG}:
\begin{equation}
\begin{aligned}
    K &=  \int_\mu^\infty \frac{d\lambda}{\lambda} \gamma_K(\alpha(\lambda)) = \left(\frac{\as}{4 \pi}\right) \frac{\gamma_K^{(1)}}{2 \epsilon} 
    +
    \left(\frac{\as}{4 \pi}\right)^2  \left( -\frac{ \beta_0 \gamma_K^{(1)}}{4 \epsilon^2} + \frac{\gamma_K^{(2)}}{4\epsilon} \right) + \mathcal{O}(\as^3), \\
    K' &=  \int_\mu^\infty \frac{d\lambda}{\lambda} K(\alpha(\lambda)) = 
    \left(\frac{\as}{4 \pi}\right) \frac{\gamma_K^{(1)}}{4\epsilon^2} 
    +
    \left(\frac{\as}{4 \pi}\right)^2  \left( -\frac{ 3\beta_0 \gamma_K^{(1)}}{16 \epsilon^3} + \frac{\gamma_K^{(2)}}{16\epsilon^2} \right) + \mathcal{O}(\as^3) , \\
    G_i & = \int_\mu^\infty \frac{d\lambda}{\lambda} \gamma_i(\alpha(\lambda))
    =
    \left(\frac{\as}{4 \pi}\right) \frac{\gamma_{i}^{(1)}}{2\epsilon} 
    +
    \left(\frac{\as}{4 \pi}\right)^2
    \left( 
    -\frac{\beta_0 \gamma_{i}^{(1)} }{4 \epsilon^2} + \frac{\gamma_{i}^{(2)}}{4\epsilon}
    \right)
    + \mathcal{O}(\as^3)
    .
\end{aligned}
\end{equation}

%%%---------------------------------------------------------------------
\section{Two-dimensional Feynman integrals}
\label{se:twodimintegrals}
%%%---------------------------------------------------------------------
In this appendix, we collect the results for the two-dimensional
integrals that we have used in \cref{se:shockwave} to describe
multi-$W$ interactions. We start from one loop. We require
the following integrals
\begin{equation}
  \begin{split}
    \twoDint^{(1)}_{\{1,2\}} & =
    \left[\twoDint^{(0)}_{\{1,1\}}\right]^{-1}
    \int 
    \frac{\measJ{\kp_1}}{\kp_1^2 (\qp_B-\kp_1)^2}
    \left[
      \frac{\boldsymbol{\vep}^*_\lambda\cdot \qp_A}{\qp_A^2} + \frac{\boldsymbol{\vep}^*_\lambda\cdot (\kp_1-\qp_A)}{(\kp_1-\qp_A)^2}
      \right],
    \\
    \twoDint^{(1)}_{\{2,1\}} & =
    \left[\widetilde{\twoDint}^{(0)}_{\{1,1\}}\right]^{-1}
    \int 
    \frac{\measJ{\kp_1}}{\kp_1^2 (\qp_A+\kp_1)^2}
    \left[
      -\frac{\boldsymbol{\vepT}^*_\lambda\cdot \qp_B}{\qp_B^2} + \frac{\boldsymbol{\vepT}^*_\lambda\cdot (\kp_1+\qp_B)}{(\kp_1+\qp_B)^2}
      \right],
    \\
    \twoDint^{(1)}_{\{2,2\}} & =
    \left[\twoDint^{(0)}_{\{1,1\}}\right]^{-1}
    \int
    \frac{\measJ{\kp_1}}{\kp_1^2(\qp_B-\kp_1)^2} 
    \left[
      \frac{\boldsymbol{\vep}^*_\lambda\cdot \pp_4}{\pp_4^2} + \frac{\boldsymbol{\vep}^*_\lambda\cdot (\qp_A-\kp_1)}{(\qp_A-\kp_1)^2}
      \right],
  \end{split}
\end{equation}
where we recall that we have defined
$\mathfrak{D}p = (\mu^{2\ep} e^{\ep \gamma_E}
/\pi^{1-\ep})\mathrm{d}^{2-2\ep}p$, and 
\begin{equation}
  \twoDint^{(0)}_{\{1,1\}} =
  \frac{1}{\qp_B^2}\left[\frac{\boldsymbol{\vep}^*_{\lambda_4}\cdot
      \pp_4}{\pp_4^2}+\frac{\boldsymbol{\vep}^*_{\lambda_4}\cdot
      \qp_A}{\qp_A^2}\right],
  \quad\quad
  \widetilde{\twoDint}^{(0)}_{\{1,1\}} =
  \frac{1}{\qp_A^2}\left[\frac{\boldsymbol{\vepT}^*_{\lambda_4}\cdot
      \pp_4}{\pp_4^2}-\frac{\boldsymbol{\vepT}^*_{\lambda_4}\cdot
      \qp_B}{\qp_B^2}\right],
\end{equation}
with the polarisation vectors defined in \cref{se:spinorproducts}.
When presenting the results we use that $\qp_A = \pp_5$ and  $\qp_B = -\pp_3$,
see \cref{eq:tchannelmomenta}.
Their calculation is straightforward. We find it convenient to express
the result in terms of two-dimensional bubble integrals $\tilde B_i$ and
a function $\tilde I_3$ which is related to the four-dimensional
off-shell scalar triangle. Specifically, they are defined as
\begin{equation}
  \tilde B_i = -\frac{2}{\ep} (\pp_i^2)^{-\ep},
  \quad\quad
    \tilde I_3 = \frac{\Gamma(1-2\ep)}{\Gamma^2(1-\ep)}
    (\pp_4^2)^{-\ep} \bar I_3,
\end{equation}
with 
\begin{equation}
  \begin{split}
    (z-\zbar)\bar I_3 & = 2 D_2(z,\zbar) + \ep\bigg[
      G_{00}(z) G_{1}(\zbar)-G_{1}(z) G_{00}(\zbar)-G_{01}(z) G_{0}(\zbar)+
      \\&
      +G_{01}(z)G_{1}(\zbar)
      -G_{0}(z) G_{01}(\zbar)-G_{1}(z) G_{01}(\zbar)+2 G_{01}(\zbar)
      G_{\zbar}(z)+
      \\&
      -G_{10}(z) G_{0}(\zbar)
      +G_{10}(z) G_{1}(\zbar)+G_{0}(z)
      G_{10}(\zbar)+G_{1}(z) G_{10}(\zbar)+
      \\&
      -2 G_{10}(\zbar)
      G_{\zbar}(z)-G_{11}(z) G_{0}(\zbar)+G_{0}(z) G_{11}(\zbar)-2 G_{1}(\zbar)
      G_{\zbar0}(z)+
      \\&
      +2 G_{0}(\zbar) G_{\zbar1}(z)-2 G_{\zbar01}(z)+2
      G_{\zbar10}(z)+G_{001}(z)-G_{010}(z)+G_{011}(z)+
      \\&
      -G_{100}(z)+G_{101}(z)-G_{110}(z)+G_{001}(\zbar)-G_{010}(\zbar)
      -G_{011}(\zbar)+
      \\&
      +G_{100}(\zbar)-G_{101}(\zbar)+G_{110}(\zbar)\bigg] + \mathcal O(\ep^2),
    \end{split}
\end{equation}
where $G$ are the standard Goncharov polylogarithms defined as
\begin{equation}
    G_{a_1\cdots a_n}(z) = \int^z_0 \frac{\mathrm{d}t}{t-a_1} G_{a_2\cdots a_n}(t), \quad\quad\quad
    G_{\vec{0}_n}(z) = \frac{1}{n!} \ln^n(z)\,.
\end{equation} 
In terms of these quantities, the $\twoDint^{(1)}_{\{i,j\}}$ functions read
\begin{equation}
  \begin{split}
    &\twoDint^{(1)}_{\{1,2\}} = \mathcal N
    \left[
    \frac{\tilde B_3 - \tilde B_5 + \tilde B_4}{2} -\lambda_4 \ep (z-\zbar)\tilde I_3\right],
    \\
    &\twoDint^{(1)}_{\{2,1\}} = \mathcal N
    \left[
      \frac{\tilde B_5 - \tilde B_3 + \tilde B_4}{2} -\lambda_4 \ep (z-\zbar)\tilde I_3\right],\\
    &
    \twoDint^{(1)}_{\{2,2\}} =
    \mathcal N\left[
    \frac{\tilde B_3 + \tilde B_5 - \tilde B_4}{2} +\lambda_4 \ep (z-\zbar)\tilde I_3\right],
  \end{split}
\end{equation}
where $\lambda_4$ is the gluon helicity and $\mathcal N = e^{\ep \gamma_E}
\Gamma^2(1-\ep)\Gamma(1+\ep)/\Gamma(1-2\ep)$.

We now move to the two-loop case. The integrals we need to consider are
\begin{equation}
  \begin{split}
    \twoDint^{(2)}_{\{1,3\}} & =
    \left[{\twoDint}^{(0)}_{\{1,1\}}\right]^{-1}
    \int \frac{\measJ{\kp_1}\measJ{\kp_2}}{\kp_2^2 (\kp_1-\kp_2)^2
    (\qp_B-\kp_1)^2}
  \\
  &\quad\times
  \bigg[
      \frac{1}{6}\left(\frac{\boldsymbol{\vep}^*_{\lambda_4}\cdot (\kp_1-\qp_A)}{(\kp_1-\qp_A)^2}\right)
      -\frac{1}{2}\left(\frac{\boldsymbol{\vep}^*_{\lambda_4}\cdot (\kp_2-\qp_A)}{(\kp_2-\qp_A)^2}\right)
      -\frac{1}{3}\left(\frac{\boldsymbol{\vep}^*_{\lambda_4}\cdot \qp_A}{\qp_A^2}\right)
      \bigg],
  \\
  %%%%%%
  \twoDint^{(2)}_{\{3,1\}} & =
    \left[\widetilde{\twoDint}^{(0)}_{\{1,1\}}\right]^{-1}
    \int \frac{\measJ{\kp_1}\measJ{\kp_2}}{\kp_2^2 (\kp_1-\kp_2)^2
    (\qp_A+\kp_1)^2}
  \\
  &\quad\times
  \bigg[
      \frac{1}{6}\left(\frac{\boldsymbol{\vepT}^*_{\lambda_4}\cdot (\kp_1+\qp_B)}{(\kp_1+\qp_B)^2}\right)
      -\frac{1}{2}\left(\frac{\boldsymbol{\vepT}^*_{\lambda_4}\cdot (\kp_2+\qp_B)}{(\kp_2+\qp_B)^2}\right)
      +\frac{1}{3}\left(\frac{\boldsymbol{\vepT}^*_{\lambda_4}\cdot \qp_B}{\qp_B^2}\right)
      \bigg],
  %%%%%%%%%%
  \\
  \twoDint^{(2)}_{\{3,3\}} & =
  \left[{\twoDint}^{(0)}_{\{1,1\}}\right]^{-1}
  \int \frac{\measJ{\kp_1}\measJ{\kp_2}}{\kp_2^2 (\kp_1-\kp_2)^2
    (\qp_B-\kp_1)^2}
  \left[
    \frac{\boldsymbol{\vep}^*_{\lambda_4}\cdot \pp_4}{\pp_4^2}+
    \frac{\boldsymbol{\vep}^*_{\lambda_4}\cdot (\qp_A-\kp_1)}{(\qp_A-\kp_1)^2}
    \right].
  \end{split}
\end{equation}
Despite being two-loop integrals, these contain a one-loop bubble
that can be integrated out, thus rendering them effectively one-loop
integrals with propagators raised to non-integer powers. This makes
their calculation simple. We expressed them in terms of the following
bubble-like functions
\begin{equation}
  \begin{split}
    &\tilde B^{(\ep)}_i = 
    -\frac{3}{2\ep}\left[\frac{\Gamma^2(1-2\ep)\Gamma(1+2\ep)}{\Gamma(1-3\ep)
        \Gamma(1-\ep)\Gamma^2(1+\ep)}\right] (\pp_i^2)^{-2\ep},
    \\
    & \tilde\Delta^{(\ep)}_3= (\pp_4)^{-2\ep}
    \left[
      -\frac{2}{\ep} + (2g_{1,4}+g_{1,5}) +
      \ep \left( -g_{1,4}^2-g_{1,4} g_{1,5} - g_{1,5}^2\right)+\mathcal O(\ep^2)
      \right],\\    
        & \tilde\Delta^{(\ep)}_4= (\pp_4)^{-2\ep}
    \left[-\frac{2}{\ep} + (g_{1,4}+g_{1,5}) +
      \ep \left(-g_{1,5}^2 + g_{1,4} g_{1,5} - g_{1,4}^2\right)+\mathcal O(\ep^2)
      \right],\\
    & \tilde\Delta^{(\ep)}_5= (\pp_4)^{-2\ep}
    \left[
      -\frac{2}{\ep} + (g_{1,4}+2g_{1,5}) +
      \ep \left( -g_{1,4}^2-g_{1,4} g_{1,5} - g_{1,5}^2\right)+\mathcal O(\ep^2)
      \right],
  \end{split}
\end{equation}
where the single-valued functions $g_{i,j}$ are defined in
\cref{se:finite_remainders}, and triangle-like functions
$\tilde I_{3,i}^{(\ep)}$. Their proper definition for our purposes
is irrelevant, since they are defined such that $\tilde I_{3,i}^{(\ep)} = 
\tilde I_{3} + \mathcal O(\ep)$. In terms of these functions
the $\mathcal K^{(2)}_{\{i,j\}}$ integrals read
\begin{equation}
  \begin{split}
    &
    \twoDint^{(2)}_{\{1,3\}} = 
    \mathcal N^2\left[-\frac{2}{\ep}\right]    
    \Bigg\{
    \frac{1}{6}
    \left[\frac{\tBep_3 - \tBep_5 + \tDep_4}{2}
      -\lambda_4 \frac{3}{2}\ep (z-\zbar)\tIep_{3,1}\right] +
    \\
    &\quad\quad\quad
    -\frac{1}{2}
    \left[\frac{\tBep_3 - \tDep_5 + \tBep_4}{2}
      -\lambda_4 \frac{3}{2}\ep (z-\zbar)\tIep_{3,3}\right]
    \Bigg\}
    ,
    \\
    &
    \twoDint^{(2)}_{\{3,1\}} = 
    \mathcal N^2\left[-\frac{2}{\ep}\right]    
    \Bigg\{
    \frac{1}{6}
    \left[\frac{\tBep_5 - \tBep_3 + \tDep_4}{2}
      -\lambda_4 \frac{3}{2}\ep (z-\zbar)\tIep_{3,1}\right] +
    \\
    &\quad\quad\quad
    -\frac{1}{2}
    \left[\frac{\tBep_5 - \tDep_3 + \tBep_4}{2}
      -\lambda_4 \frac{3}{2}\ep (z-\zbar)\tIep_{3,2}\right]
    \Bigg\},
    \\
    &
    \twoDint^{(2)}_{\{3,3\}} = 
    \mathcal N^2\left[-\frac{2}{\ep}\right]    
    \Bigg\{
    \frac{\tBep_3 + \tBep_5 - \tDep_4}{2}
    +\lambda_4 \frac{3}{2}\ep (z-\zbar)\tIep_{3,1}
    \Bigg\}.
  \end{split}
\end{equation}

\bibliographystyle{JHEP}
\bibliography{biblio.bib}

\printindex

\end{document}